\documentclass[preprint2]{aastex}

\newcommand{\HI}{H\,{\sc i}}
\newcommand{\HII}{H\,{\sc ii}}

\newcommand{\Ha}{H$\alpha$}
\newcommand{\Hb}{H$\beta$}
\newcommand{\kms}{~km\,s$^{-1}$}

\newcommand{\Mo}{$M_{\odot}$}

\newcommand{\Zo}{~Z$_{\odot}$}

\newcommand{\MSMBH}{$M_{\rm SMBH}$}

\DeclareRobustCommand{\ion}[2]{%
\relax\ifmmode
\ifx\testbx\f@series
{\mathbf{#1\,\mathsc{#2}}}\else
{\mathrm{#1\,\mathsc{#2}}}\fi
\else\textup{#1\,{\mdseries\textsc{#2}}}%
\fi}

\shorttitle{Quasar Host Galaxies and the $M_{SMBH}$ - $\sigma_{*}$ Relation}
\shortauthors{Sheinis \& L\'opez-S\'anchez}

\begin{document}

%% LaTeX will automatically break titles if they run longer than
%% one line. However, you may use \\ to force a line break if
%% you desire.

\title{Quasar Host Galaxies and the $M_{SMBH}$ - $\sigma_{*}$ Relation}

%% Use \author, \affil, and the \and command to format
%% author and affiliation information.
%% Note that \email has replaced the old \authoremail command
%% from AASTeX v4.0. You can use \email to mark an email address
%% anywhere in the paper, not just in the front matter.
%% As in the title, use \\ to force line breaks.

%\author{A. I. Sheinis\altaffilmark{1,2}, \'A. R. L\'opez-S\'anchez\altaffilmark{1,3}}
\author{Andrew I. Sheinis\altaffilmark{1,2}}
%\affil{Astronomy Department, University of California,
%    Berkeley, CA 94720}
%\author{C. D. Biemesderfer\altaffilmark{4,5}}
%\affil{National Optical Astronomy Observatories, Tucson, AZ 85719}
%\email{aastex-help@aas.org}
\and 
\author{ \'Angel R. L\'opez-S\'anchez\altaffilmark{1,3}}

%\author{M. J. Wolf\altaffilmark{4}}
%\affil{Space Telescope Science Institute, Baltimore, MD 21218}

%% Notice that each of these authors has alternate affiliations, which
%% are identified by the \altaffilmark after each name.  Specify alternate
%% affiliation information with \altaffiltext, with one command per each
%% affiliation.

\altaffiltext{1}{Australian Astronomical Observatory, PO Box 915, North Ryde, NSW 1670, Australia.}
\altaffiltext{2}{Sydney Institute for Astronomy (SIfA), School of Physics, The University of Sydney, NSW 2006, Australia.}
\altaffiltext{3}{Department of Physics and Astronomy, Macquarie University, NSW 2109, Australia.}
%\altaffiltext{4}{University of Wisconsin - Madison, Department of Astronomy, 475 N. Charter Street, Madison, WI 53706.}

%% Mark off your abstract in the ``abstract'' environment. In the manuscript
%% style, abstract will output a Received/Accepted line after the
%% title and affiliation information. No date will appear since the author
%% does not have this information. The dates will be filled in by the
%% editorial office after submission.

\begin{abstract}
We analyze the emission line profiles detected in deep optical spectra of quasars to derive the mass of their super-massive black holes (SMBH) following the single-epoch virial method. Our sample consists in 6 radio-loud quasars and 4 radio-quiet quasars. We carefully fit a broad and narrow Gaussian component for each bright emission line in both the \Hb\ (10 objects) and \Ha\ regions (5 objects). We find a very good agreement of the derived SMBH masses, $M_{\rm SMBH}$, using the fitted broad \Hb\ and \Ha\ emission lines. 
We compare our  $M_{\rm SMBH}$ results with those found by previous studies using the reverberation mapping technique, the virial method and X-ray data, as well as those derived using the continuum luminosity at 5100~\AA.
We also study the relationship between the $M_{\rm SMBH}$ of the quasar and the stellar velocity dispersion, $\sigma_{*}$, of the host galaxy. 
We use the measured  $M_{\rm SMBH}$ and  $\sigma_{*}$ to investigate the $M_{\rm SMBH}$ -- $\sigma_{*}$ relation for both the radio-loud and radio-quiet subsamples. Besides the scatter, we find a good agreement  between radio-quiet quasars and AGN+quiescent galaxies and between radio-loud quasars and AGN. The intercept in the latter case is 0.5~dex lower than in the first case.
Our analysis does not support the hypothesis of using $\sigma$([\ion{O}{iii}]~$\lambda$5007) as a surrogate for stellar velocity dispersions
in  high-mass, high-luminosity quasars. 
We also investigate the relationship between the 5\,GHz radio-continuum luminosity, $L_{\rm 5\,GHz}$, of the quasar host galaxy with both $M_{\rm SMBH}$ and $\sigma_{*}$. We do not find any correlation between $L_{\rm 5\,GHz}$ and $M_{\rm SMBH}$, although we observe a trend that galaxies with larger stellar velocity dispersions have larger 5\,GHz radio-continuum luminosities. Using the results of our fitting for the narrow emission lines of [\ion{O}{iii}]~$\lambda$5007 and [\ion{N}{ii}]~$\lambda$6583 
we estimate the gas-phase oxygen abundance of six quasars, being sub-solar in all cases.
\end{abstract}

%% Keywords should appear after the \end{abstract} command. The uncommented
%% example has been keyed in ApJ style. See the instructions to authors
%% for the journal to which you are submitting your paper to determine
%% what keyword punctuation is appropriate.

\keywords{Galaxies: Active -- Galaxies: Quasars -- Galaxies: Kinematics -- Galaxies: Super-massive black hole -- Galaxies: abundances}
\section{Introduction}

How super-massive black holes (SMBH) grow in the center of galaxies is intimately related to how galaxies are formed and evolve. Over the last decade an increasing number of both theoretical and observational studies have provided a better understanding of the physical connection between SMBH and galaxy evolution, particularly since the discovery that all galaxies with a bulge contain a SMBH \citep{kormendy95}.
However, the fundamental observational evidence is the relationship found between the SMBH mass, \MSMBH, and the host galaxy stellar velocity dispersion --bulge stellar velocity dispersion--, $\sigma_{*}$, which was first predicted by \citet{SilkRees98} and \citet{Fabian99} %Silk \& Rees (1998) and Fabian (1999) 
and later verified in both active \citep{Gebhardt+00b,Ferrarese+01,Nelson+04,Onken+04,Dasyra+07,Woo+10,Woo+13,Graham+11,Park+12} 
%(Gebhardt et al. 2000b; Ferrarese et al. 2001; Nelson et al. 2004; Onken et al. 2004; Dasyra et al. 2007; Woo et al. 2010, 2013; Graham et al. 2011; Park et al. 2012) 
and quiescent \citep{FerrareseMerritt00,Gebhardt+00a,Tremaine+02,Gultekin+09,McConnell+11,McConnellMa13}
%(Ferrarese \& Merritt 2000; Gebhardt et al. 2000a; Tremaine et al. 2002; G\"ultekin et al. 2009; McConnell et al. 2011; McConnell \& Ma 2013) 
galaxies. 
%A growing understanding of the connection between galaxies and their
%central black holes has emerged over the last decade. 
%The facts that
%all galaxies with a bulge contain supermassive black holes
%\citep{kormendy95}; that black hole mass, $M_{BH}$, is correlated with
%the host galaxy stellar velocity dispersion, $\sigma_{*}$,
%\citep{gebhardt00a}; and that 
As the bulge extends by several orders of magnitude outside the gravitational influence of the SMBH, the \MSMBH- $\sigma_{*}$ correlations suggests 
that both the bulge and SMBH co-evolve \citep{Kormendy+00}. %(Kormendy et al. 2000).
Furthermore,
the inclusion of active galactic nuclei (AGN) feedback, or an
equivalent energetic source to quench star formation above a critical
halo mass, in semi-analytic galaxy formation models
\citep{Cattaneo+06,Dekel+06} matches the galaxy demographics and
bimodality of properties observed in large surveys --SDSS:
\citet{Kauffmann+03a,Kauffmann+03b,Hogg+03,Baldry+04,Heavens+04,CidFernandes+05};
GOODS: \citet{Giavalisco+04}; COMBO-17: \citet{Bell+04}; DEEP/DEEP2: \citet{Koo+05}; %\citet{Koo+03,Koo+05}; 
MUNICS: \citet{Drory+01}; FIRES: \citet{Labbe+03};
K20: \citet{Cimatti+02}; GDDS: \citet{McCarthy+04}.
%--, all suggest that
%the growth mechanisms of the black hole and galaxy must be connected. 
However, the details of the physical processes that make %this connection, 
the connection between the growth mechanisms of the black hole and galaxy,
such as how AGN energy interacts with and is dissipated by
surrounding halo gas, are not yet known.

%Therefore, an understanding of the relationship between the central black hole mass and the host galaxy stellar velocity dispersion 
%would allow us to explore the actual role of SMBH in galaxy evolution. It is important 
We seek 
to clarify if  there is an unique \MSMBH -- $\sigma_{*}$ relationship or if it differs for different kind of objects. For example, some analyses suggest a morphological dependence of the \MSMBH -- $\sigma_{*}$ relation between low-mass --usually late-type-- and high-mass --early-type-- quiescent galaxies \citep[e.g.][]{Greene+10,McConnellMa13}.
It has been also found that both barred galaxies and galaxies hosting pseudobulges do not follow the standard \MSMBH -- $\sigma_{*}$ relation
\citep[e.g.][]{Graham08a,Graham08b,GrahamLee09,Hu+08,GadottiKauffmann09,Kormendy+11}.
Furthermore, some morphological deviations have been reported in AGN 
\citep[e.g.][]{GrahamLee09,Graham+11,Woo+10,Woo+13,Mathur+12,Park+12}. 
However, it is not clear yet if such deviations in the $M_{SMBH}$-$\sigma_{*}$ relation exist in the high-mass (and high velocity dispersion) end, as claimed by several authors 
\citep{Dasyra+07,Watson+08} %(Dasyra et al. 2007; Watson et al. 2008), 
but not  confirmed by others  \citep{Grier+13}. %(Grier et al. 2013).
Numerical models, as those presented by \citet{King10} and \citet{ZubovasKing12}, suggest that the \MSMBH -- $\sigma_{*}$ relationship is not unique and it may even depend on the environment.

The stellar velocity dispersion in the host galaxies is estimated using the stellar absorption lines in galaxy spectra of the quiescent galaxies.  This in fact was the method first used to discover the $M_{SMBH}$-$\sigma_{*}$  relation \citep{Gebhardt+00b,FerrareseMerritt00}. For bonafide quasars ($M_V < -23$), the quasar can be up to 3 magnitudes brighter than the integrated light from the host \citep{MillerSheinis03}, which complicates the extraction of the stellar absorption lines from the host. 
Thus a limited number of studies have been possible for bright quasars \citep{WolfSheinis08}.
% (cite Wolf and Sheinis 2008, Yanke 200? EtcÉ).
\citet{Shields+03} developed one method of estimating the $M_{SMBH}$-$\sigma_{*}$  relation using the results of \citet{NelsonWhittle96}, who found a correlation between  [\ion{O}{iii}] emission linewidth and $\sigma_{*}$. This method used the velocity dispersion of  [\ion{O}{iii}] as a surrogate for stellar velocity dispersion and has shown that AGN and quasars follow the $M_{SMBH}$-$\sigma_{*}$ relation at a wide range of redshifts, albeit with large scatter. However, there have been some indications that radio-loud quasars may not follow this trend  \citep{Laor00,McLureJarvis04}.

Black hole masses are derived in a number of different ways \citep[see][for recent reviews]{CzernyNikolajuk10,Shen13}. %(see Czerny \& Nikolajuk 2010 for a recent review). 
\citet{Gebhardt+00a} computed them through simulations of galaxy
stellar dynamics for quiescent galaxies. Direct $M_{SMBH}$ measurements 
should consider both spatial and spectral resolution, being only feasible for nearby galaxies. 
Recently,  \citet{ShaposhnikovTitarchuk09} introduced a new method to derive  $M_{\rm SMBH}$ based solely on X-ray spectral data. 
This technique is providing very good results in both quiescent \citep{ShaposhnikovTitarchuk09}  and active \citep{Gliozzi+11} systems.

For AGN, reverberation mapping \citep{BlandfordMcKee82} %(Blandford \& McKee 1982) 
is the most accurate method for measuring  \MSMBH. 
Assuming that the broad emission line region (BLR) is powered by photoionization
from the central source, % \citep[e.g.][]{Peterson97},
this technique uses the fact that
the continuum flux (which arises from the accretion
disc or very close to it) varies with time, and is later echoed % These variations are echoed later 
by changes in the flux of the broad emission
lines. The radius of the BLR, $R_{BLR}$, is then obtained by the cross correlation of the light curves, which provides the delay time  between the broad-line variations and the
continuum variations. The SMBH mass is then computed assuming that the BLR is virialized and the motion of the emitting clouds is dominated by the
gravitational field of the SMBH \citep[e.g.][]{Ho99,Wandel+99},  $M_{\rm SMBH}= (f\, R_{BLR}\,{V_{BLR}}^2)/G$.
$f$ is a dimensionless factor that accounts for the unknown geometry and orientation of the BLR, $V_{BLR}$ is the dispersion velocity of the gas (which is deduced from the width of the Doppler broadened emission lines).
Reverberation mapping allows to probe regions of gas that are only $\sim$0.01\,pc in extent.
However, this technique requires high-quality spectrophotometric monitoring
of an AGN over an extended period of time. 
The uncertainty of the SMBH masses derived using the reverberation mapping method typically is between 0.4 and 0.5 dex \citep{Peterson10,Shen13}.
Reverberation mapping has yielded black hole masses for
$\sim$50 AGNs thus far \citep{Kaspi+00,Peterson+04,Bentz+09b}. %(Kaspi et al. 2000; Peterson et al. 2004; Bentz et al. 2009b).
Nevertheless, few reverberation mapping measurements have well-defined velocity-resolved delay maps \citep[e.g.][]{Denney+09a,Bentz+10,Grier+13}. The main caveat of this technique is the assumption of the accretion disc morphology
\citep{Krolik01,Shen13}, which may be lead to different types of galaxies following somewhat different scaling relations between SMBH mass
and bulge properties \citep{Graham08a,Greene+08,Greene+10,Gultekin+09,McConnellMa13}, leading to accuracy within factors of 2-3 
\citep[e.g., ][]{Woo+10,Graham+11,Park+12,Grier+13}. %(Grier et al. 2013 and references within).
% \citep[e.g., ][and references within]{Woo+10,Graham+11,Park+12,Grier+13}. %(Grier et al. 2013 and references within).
Some authors \citep[e.g.][and references therein]{Pancoast+14a} have recently developed new methods to constrain the geometry and dynamics of the BLR by modeling reverberation mapping data directly. This also allows to measure SMBH masses independent of a virial coefficient. In particular, \citet{Pancoast+14a} claim they can recover the black hole mass to 0.05 -- 0.25~dex uncertainty.

The single-epoch virial method has been calibrated using reverberation mapping \citep[e.g.][]{Kaspi+00,Kaspi+05,GreeneHo05,Bentz+06a,VestergaardPeterson06,McGill+08,Bentz+09a,Bentz+13,ShenLiu12}. This technique, which assumes that the BLR gas is virialized and hence follows the virial relation, assumes a radius-luminosity relation of the form $R_{BLR} \propto L^\alpha$. The coefficients of this relation are determined from estimates of a sample of AGNs for which reverberation mapping data are available. For the case of the broad \Hb\ emission line, it has been established  $\alpha\sim0.52-0.56$. Individual black hole masses derived using reverberation mapping differ from those derived by the radius-luminosity relation of the BLR by up to 0.5 dex \citep[e.g.][]{McLureJarvis02,Peterson+04,VestergaardPeterson06,Kim+08}.

This study uses the single-epoch virial method to derive the SMBH of a sample of luminous quasars via a careful analysis of the \Hb\ and \Ha\ emission line profiles. The main objective is to investigate the  $M_{SMBH}$-$\sigma_{*}$ relation on these bright quasars. This paper is organized as follows. Section~2 describes our sample, which consists  in 6 radio-loud quasars and 4 radio-quiet quasars. Section~3 presents the analysis of the data and how the SMBH masses have been estimated. Our results are discussed in Sect.~4, which compares our mass estimations with those reported in the literature; explores the  $M_{SMBH}$-$\sigma_{*}$ and the $M_{SMBH}$-QSO radio-luminosity relations; as well as discusses the nature of the radio-loud and radio-quiet quasars. In Sect.~4 we also compare the stellar velocity dispersions with the FWHM of broad  [\ion{O}{iii}] emission and, when possible, estimate the gas-phase metallicity of the host galaxies using the narrow emission lines. Finally Sect.~5 provides the conclusions of our analysis.

\begin{deluxetable}{ c   c c  c   c c  c   c c}
\tabletypesize{\scriptsize}
%\rotate
\tablecaption{Redshifts, distances, host galaxy velocity dispersions, and radio luminosities of our QSO sample. The redshift is derived from our fit to the radial velocity of the narrow  [\ion{O}{iii}]~$\lambda$5007 emission line and have an error or $\pm$0.00008. Distances are derived from the redshift assuming a flat cosmology ($H_0$=70~km\,s$^{-1}$\,Mpc$^{-1}$, \mbox{$\Omega_M$ = 0.3}, and $\Omega_\Lambda$ = 0.7) and have an error of  $\pm$0.3\,Mpc. Host galaxy velocity dispersions are given by \citet{WolfSheinis08}. For PG~0052+251, $\sigma_{*}$ in \citet{WolfSheinis08} was an upper limit due to the spectral resolution. Radio luminosities has been extracted from \citet{Wold+10}.  \label{sigma_table}}
\tablewidth{0pt}
\tablehead{
\colhead{Quasar}  &    \colhead{Redshift}     &  \colhead{Distance}       &     \colhead{Measured $\sigma_{*}$}  &  \colhead{Avg. Observed} &       \colhead{$R_e$  }      &   \colhead{Ap-Cor $\sigma_{*}$}   &  \colhead{log ($L_{\rm 5\,GHz}$) }& \colhead{Radio} \\

 \colhead{Name}   &      &      \colhead{[Mpc]} &  \colhead{[km~s$^{-1}$] } & \colhead{Radius [arcsec]} & \colhead{[arcsec]} & \colhead{[km~s$^{-1}$]}   &    \colhead{ [erg\,s$^{-1}$]}  & \colhead{Activity} 
}
\startdata

%1 & 
PG~0052+251    &  0.154480 & 638.0  &   250 $\pm$ 53 & 3.63  &    1.8   &    279 $\pm$  59  &   39.4 & RQ\\
%2 & 
PHL~909 	         &  0.171807 & 706.5  &   150 $\pm$ 11        & 4.5   &   2.3    &    167 $\pm$   12  &  40.0 & RQ \\
%3 & 
PKS~0736+017$^a$   &  0.189136 & 774.5  &  311$\pm$ 83       & 4.5   &   3.3     &   342 $\pm$   91 &  43.0  &  RL \\
%4 & 
3C~273$^b$       & 0.157366 & 649.4  &  305 $\pm$ 57      & 4.36    &   3.7    &   334 $\pm$  62  &   44.1 & RL \\
%5 & 
PKS~1302-102   & 0.277831  & 1112.5  &  346 $\pm$  72    & 3.05    &    1.4   &   388 $\pm$   80 &  43.0 &  RL\\
%6 & 
PG~1309+355    & 0.182345 & 747.9  &  236 $\pm$ 30   & 4.05    &  2.0      &  264 $\pm$  33   &  41.3 & RQ \\
%7 & 
PG~1444+407    & 0.267252 & 1073.0  &  279 $\pm$  22    & 3.70    &    1.3   &  316  $\pm$  25 & 39.2  &RQ\\
%8 & 
PKS~2135-147   & 0.200396 & 818.3 &  278  $\pm$ 106 & 3.81    &   2.6      &  307 $\pm$  117 & 42.9  &RL \\
%9 & 
4C~31.63$^c$     & 0.334626 & 1320.4   &  290 $\pm$ 23      & 2.0    &   6.5    & 301  $\pm$  24   & 43.3 & RL \\
%   &
           "                    &       "        &   "         &  325 $\pm$ 24      & 2.5    &   6.5   & 340  $\pm$ 25   &  " &  "  \\
%   & 
        "                    &    "          &   "         &  {\it average}  &     ...      &       6.5         & 320  $\pm$  25  &  " &  " \\    
%10 & PKS~2349-014 3S   & $<$224 $\pm$ 34& 3.24    &   4.8   & 312 $\pm$ 60 \\
% 10   & 
PKS~2349-014   & 0.173844  &  714.5  & 278 $\pm$ 54  & 4.83  &   4.8  & 302 $\pm$ 59  & 42.5  & RL\\
 
\enddata
\tablenotetext{a}{Also known as [HB89]~0736+017. $^b$ Also known as PG~1226+023. $^c$ Also known as  [HB89]~2201+315.}
\end{deluxetable}

% SECTION 2: OBSERVATIONS AND DATA REDUCTION  -----------------------------------------------------------------------------------------------

\section{Data selection}\label{observation}

Here we analyse the sample of 10 nearby and luminous quasars studied by  \citet{WolfSheinis08}. 
These authors defined their sample following \citet{Bahcall+97} observations of bright ($M_V< -29.9$), low-$z$ ($z\leq0.20$), and high galactic latitude ($|b|>35^{\circ}$) quasars, with the addition of a few objects from \citet{Dunlop+03} and \citet{Guyon+06} with the same characteristics. However, this sample included only 3 radio-loud quasars, and hence 6 extra objects with redshift $0.20<z<0.30$ (3 of them being radio-loud), also drawn from \citet{Bahcall+97}, were added.
The same sample of 10 quasars considered by  \citet{WolfSheinis08} was later analysed by \citet{Wold+10} to study the connection between the stellar ages of the  host galaxies and the quasar activity.
These ten objects are
PG~0052+251, PHL~909, PKS~0736+017, 3C~273, PKS~1302-102, PG~1309+355, PG~1444+407, PKS~2135-147, 4C~31.63, and \mbox{PKS~2349-014.}

Nuclear and off-nuclear spectra of seven of the host galaxies of our sample (all but PHL~909, PKS~0736+017 and  PG~1309+355) were
obtained  with the Low Resolution Imaging Spectrograph \citep{Oke+94}
on the Keck telescope during 1996--1997, as detailed in \citet{SheinisPhD02} and \citet{MillerSheinis03}.
An image showing the approximate
off-axis slit and fibre positions with respect to each object is shown in Fig.~1 of \citet{Wold+10}.

Observed wavelength ranges covered $\sim$4500--7000~\AA\ at
a spectral resolution of $\Delta \lambda \sim $11 \AA\ (300~km\,s$^{-1}$). 
The spectroscopic data of PHL~909, PKS~0736+017 and  PG~1309+355 were obtained in 2007
using the integral field unit (IFU) Sparsepak \citep{Bershady+04,Bershady+05} 
which feeds the Bench Spectrograph of the 3.5-m WIYN Telescope\footnote{The WIYN Observatory is owned and operated by the WIYN Consortium,
Inc., which consists of the University of Wisconsin, Indiana University, Yale
University, and the National Optical Astronomy Observatory (NOAO). NOAO
is operated for the National Science Foundation by the Association of
Universities for Research in Astronomy (AURA), Inc.}. 
The configuration used provided 
an observed wavelength coverage of $\sim$4270--7130~\AA\ at a resolution
of $\Delta \lambda \sim$5~\AA\ (110~km\,s$^{-1}$).  
All spectra were corrected for Galactic
extinction using the law of \citet*{CardelliClaytonMathis89} and 
the $A_V$ values from \citet*{SchlegelFinkbeinerDavis98} as listed
in the NASA/IPAC Extragalactic Database (NED).
More details can be found in \citet{WolfSheinis08}.

Four of the chosen objects, PG~0052+251, PHL~909, PG~1309+355, and PG~1444+407, are radio-quiet (RQ) quasars. The other six objects are radio-loud (RL) quasars.
Following \citet{Wold+10}, a quasar is defined as radio-loud when $L_{\rm 5\,GHz} \gtrsim 10^{41.5}$\,erg\,s$^{-1}$. 
This relatively small sample of quasars is not representative of the local quasar population, which
consists of $\sim$10\%\ RL quasars.
While the data presented here consist of  half of the nearby and luminous
quasars known \citep[following the][final sample of 20 objects]{Bahcall+97},
and includes all the RL quasars in that sample, it is nonetheless a small
sample of of only 10 objects.  Thus we present the caveat that the initial
conclusions drawn from our analysis and comparison between the properties of
RL and RQ quasars are based on the small sample and will be better confirmed when
more objects with the same characteristics are considered in an upcoming paper.

%{\bf In any case, we have to keep in mind that we are dealing with a small sample statistics of only 10 objects, so it is difficult to draw strong conclusions from our analysis and comparison between the properties of RL and RQ quasars. Our observational trends should be confirmed when more objects with the same characteristics are considered. However, our sample consists on half of the nearby and luminous quasars known \citep[following the][final sample of 20 objects]{Bahcall+97}, and includes {\it all} the RL quasars within them.}

% SECTION 3: ANALYSIS  -----------------------------------------------------------------------------------------------

\section{Analysis}

\subsection{Stellar Velocity Dispersions \label{vd_section} }

Stellar velocity dispersions ($\sigma_{\star}$) of the host galaxies were presented in \citet{WolfSheinis08}.
$\sigma_{\star}$ were derived from stellar absorption lines in the
host galaxies by fitting a stellar template (main sequence A stars through K giants) that has been convolved
with a gaussian profile to the off-nuclear, scatter-subtracted galaxy
spectrum in pixel space using the code of Karl Gebhardt
\citep{Gebhardt+00b,Gebhardt+03}. 
The  wavelength range of 3850--4200~\AA, which contains the \ion{Ca}{ii}~H,K (3968, 3934~\AA) absorption lines,
was used for the velocity dispersion fits.

Velocity dispersion uncertainties were calculated through
Monte Carlo simulations by adding Gaussian noise to each pixel
in the final template, which has a very high S/N, at a level such
that the mean matches the noise in the initial galaxy spectrum
and the standard deviation is given by the rms of the initial fit.
The velocity dispersion was then measured for 100 noise realizations
and the mean and standard deviation of these results provide
the measured velocity dispersion and its 1$\sigma$ uncertainty.

Because velocity dispersion varies with galaxy radius, to match the
comparison data all host galaxy velocity
dispersions  were corrected from the radius at which $\sigma_{\star}$ was measured to a
radius of $R_{e}$/8 using the correction in \citet{Bernardi+03a} and \citet{Jorgensen+95},    %\citet{BruzualCharlot03}, %B03,
\begin{equation}
\sigma_{\star,corr} = \sigma_{\star,meas} \left( \frac {R_{obs}} {R_{e}/8}
\right)^{0.04} .
\end{equation}
Aperture-corrected values for the host galaxies and
average observed radii are compiled in %Table \ref{vd_table}.
Table~\ref{sigma_table}. For 4C~31.63 two $\sigma_{\star}$ values were obtained at different radii. For our analysis we will use the average value, as indicated in the table.

Eight of our objects are found in \mbox{elliptical} galaxies and two (PG~0052+251, PG~1309+355) in spirals \citep[see Table~1 in][]{WolfSheinis08}. The determination of $\sigma_{\star}$ did not consider galaxy rotation, which may induce to overestimate its real value up to $\sim$15 -- 20\% (this actually depends on the inclination angle of the rotating disk and the maximum rotation velocities) in spiral galaxies \citep{Kang+13,Woo+13}. The effect of rotation is particularly important when using the integrated flux of galaxy, however the the detailed procedure carried out by \citet{WolfSheinis08} to derive $\sigma_{\star}$ used off-nuclear regions, so we consider the effect of galaxy rotation to be small.

\subsection{Black hole masses\label{sect3.2}}

As explained in the introduction, the mass of a super massive black hole, $M_{\rm SMBH}$ can be determined via the kinematics of the ionized gas surrounding it. 
%(e.g., Peterson \& Wandel 1999; Kaspi et al. 2000; Onken et al. 2004; Peterson et al. 2004).
The virial method assumes that if the full width half-maximum (FWHM) of the broad emission lines reflects a Keplerian motion of the gas in the BLR, $M_{\rm SMBH}$ can be estimated from
\begin{eqnarray}
\label{virial} M_{\rm SMBH} = f \times \frac{R_{\rm BLR}V^2_{\rm BLR}}{G},
\end{eqnarray}
where $G$ is the gravitational constant, $R_{\rm BLR}$ is the radius of the BLR, $V_{\rm BLR}$ is the rotational velocity of the ionized gas
and $f$ is a dimensionless factor that accounts for the unknown geometry and orientation of the BLR.  
$V_{\rm BLR}$ is estimated from the FWHM of the \Ha\ or \Hb\ emission lines, while $R_{\rm BLR}$ is determined using the monochromatic continuum luminosity %$\lambda L_{\lambda}$ 
of the host galaxy at 5100~\AA, $L_{5100}$ \citep{Kaspi+00,Kaspi+05}. 
As the continuum luminosity is correlated with the luminosities of the \Ha\ and \Hb\ emission lines, the mass of the SMBH can be  estimated using both the luminosities and the FWHM of the broad \HI\ Balmer line.  
Although more recent calibrations are available \citep[i.e.][]{McGill+08,Assef+11,Park+12}   %%%%% CHECK
we prefer to use  \citet{GreeneHo05} as they provide consistent equations for \Ha, \Hb, and  $L_{5100}$. These equations are:
%provided these relations for the case of the \Ha\ and \Hb\ emission lines:
\begin{eqnarray}
\label{eqhb}
M_{\rm SMBH} =   (3.6\pm0.2)\times 10^6 \times \Big(\frac{L_{\rm H\beta}}{10^{42}\,{\rm erg\,s^{-1}}}\Big)^{(0.56\pm0.02)} \nonumber \\ \times  \Big(\frac{\rm FWHM_{H\beta}}{\rm 10^3\,km\,s^{-1}}\Big)^2 \hspace{2.8cm} [M_{\odot}],
\end{eqnarray}
and
\begin{eqnarray}
\label{eqha}
M_{\rm SMBH} =   2.0^{+0.4}_{-0.3}\times 10^6 \times \Big(\frac{L_{\rm H\alpha}}{10^{42}\,{\rm erg\,s^{-1}}}\Big)^{(0.55\pm0.02)} \nonumber \\  \times \Big(\frac{\rm FWHM_{H\alpha}}{\rm 10^3\,km\,s^{-1}}\Big)^{(2.06\pm0.06)} \hspace{1.25cm}  [M_{\odot}].
\end{eqnarray}

\citet{GreeneHo05} assumed the following virial formula for the SMBH mass:
\begin{eqnarray}
\label{eqL5100}
M_{\rm SMBH} =   (4.4\pm0.2) \times 10^6 \times \Big(\frac{\lambda L_{5100}}{10^{44}\,{\rm erg\,s^{-1}}}\Big)^{(0.64\pm0.02)} \nonumber \\ \times \Big(\frac{\rm FWHM_{H\beta}}{\rm 10^3\,km\,s^{-1}}\Big)^2 \hspace{1.25cm}  [M_{\odot}],
\end{eqnarray}
which uses the continuum luminosity at rest frame wavelength 5100\,\AA, $\lambda L_{5100}$, and the FWHM of the broad \Hb\ emission line, FWHM$_{\rm H\beta}$.

Here, we use the optical spectra of our sample of quasars to analyse the emission line profiles of the \Hb\ and [\ion{O}{iii}] $\lambda\lambda$4959,5007 emission lines, plus the \Ha\ and  [\ion{N}{ii}] $\lambda\lambda$6548,6583 emission lines if also observed, to derive their SMBH masses. We fitted a narrow and a broad Gaussian components --expressed as $G=p\times\exp[{-0.5\times(\frac{\lambda-c}{\sigma})^2}]$, being $p$ the peak of the Gaussian, $\sigma$  its width and $c$ its central wavelength-- in each of these emission lines over a continuum. When detected, we also considered other broad and narrow emission lines such as [\ion{O}{i}]~$\lambda$6300, [\ion{He}{i}]~$\lambda$6678, or [\ion{S}{ii}]~$\lambda\lambda$6717,6731. In particular, for many cases it was important to include a broad and narrow emission line component around $\lambda$4686, which may be attributed to the high-ionization \ion{He}{ii}~$\lambda$4686 emission line
and which may cause an overestimation of the continuum and the \Hb\ real flux and therefore complicates the proper measurement of the \Hb\ line dispersion 
\citep[e.g.][]{Decarli+08,Denney+09}. 
The inclusion of a broad component for the NLR lines is justified by the finding of outflows of ionized gas around AGNs and, in particular, around RL quasars \citep[e.g.][]{Nesvada+06,Holt+08,FuStockton09,Husemann+14}, with radial velocities of the ionized gas exceeding more than 500~\kms. These outflows of ionized gas have been interpreted as evidence of AGN feedback \citep{Liu+13a,Liu+13b,Husemann+14,Zakamska&Greene14}. The presence of broad components in the emission lines observed in galaxy spectra is now commonly taken into account by on-going 3D spectroscopic surveys, for example, the LZIFU code that the "Sydney-AAO Multi-IFU Galaxy Survey" \citep[SAMI,][]{Croom+12SAMI,Bryant+15} has developed \citep{Ho+14}, or in the kinematical analysis of the ionized gas in the galaxies of the "Calar-Alto Legacy Integral Field Area" \citep[CALIFA,][]{Sanchez+12CALIFA1,Garcia-Benito+15CALIFA3} Survey  \citep{Garcia-Lorenzo+15}.   
Furthermore, a broad  \ion{Fe}{ii} emission should be considered between \Hb\ and [\ion{O}{iii}]~$\lambda$4959 \citep[e.g.][]{BorosonGreen92,Park+12}. 
%Additionally, we assumed these constraints to reduce the number of free parameters in our fits:
%\vspace{-0.15cm}
%\begin{enumerate}
%\item As the spectra have been corrected for radial velocity using the peak of the [\ion{O}{iii}]~$\lambda$5007 emission, we set the wavelengths of the narrow line components to their respective laboratory wavelengths. However, to get a more accurate estimation of the distance to the quasars using the bright narrow  [\ion{O}{iii}] $\lambda$5007 emission line, we allowed this to be within $\pm$1~\AA.
%\item  We require that the FWHM of the narrow line components of the two [\ion{O}{iii}] emission lines have the same value. We imposed the same condition for the FWHM of the narrow line components of the two [\ion{N}{ii}] emission lines.
%\item As the integrated flux of the [\ion{O}{iii}] emission lines is fixed by theory \citep{OsterbrockFerland06},
%we forced that $F$[\ion{O}{iii}]~$\lambda$4959 =  0.3467$\times F$[\ion{O}{iii}]~$\lambda$5007, and $F$[\ion{N}{ii}]~$\lambda$6548 =  0.3402$\times F$[\ion{N}{ii}]~$\lambda$6583.
%\end{enumerate}
%
The broad  \ion{Fe}{ii} is a sum of several blended  \ion{Fe}{ii} lines that span from 3535 to 7530~\AA.  We use the template \ion{Fe}{ii} emission spectrum of  the narrow emission-line type Seyfert~1 galaxy I~Zw~1 provided by \citet{Veron-Cetty+04}, which was convoluted to the spectral resolution of each particular quasar spectrum, to remove the contribution of the  \ion{Fe}{ii} lines in the range 4700-5100~\AA. We note that not all the emission attributed to   \ion{Fe}{ii} lines in the quasar spectra was always removed, and that the broad [\ion{O}{iii}] lines are sometimes likely fitting these residuals.

\begin{figure*}
\centering
\includegraphics[scale=0.65,angle=90]{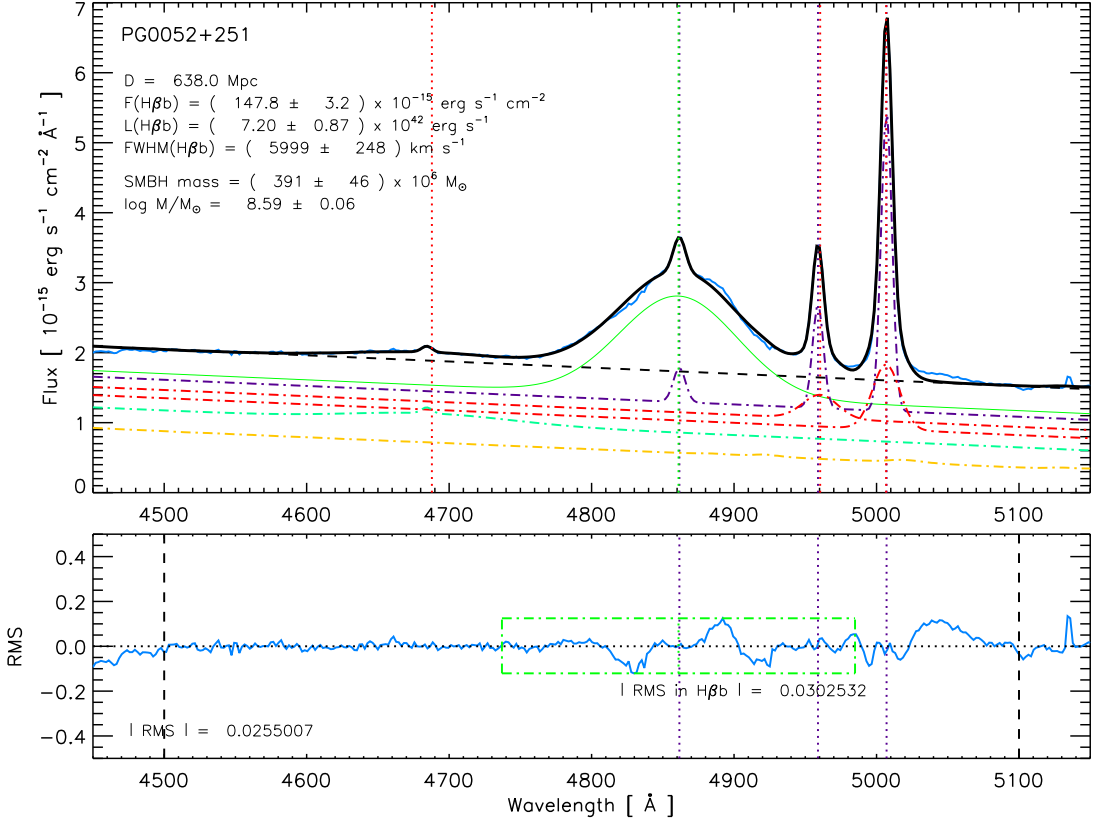} 
\caption{\small ({\it Top panel}) Fit to the emission line profiles of PG~0052+251 observed around the \Hb\ region. The blue line is the observed spectrum of the quasar, while the black line is the best fit to the data. The black dashed line is the fit to the continuum. Dotted magenta lines indicate the center of the narrow-components for \Hb\ and [\ion{O}{iii}]~$\lambda\lambda$4959, 5007. Dotted red lines indicate the center of all the broad components but \Hb. The center of the broad \Hb\ line is plotted by a green dotted line. The continuous green line shows the broad \Hb\ gaussian. The dotted-dashed magenta line represents the narrow-component spectrum. The dotted-dashed red lines correspond to the broad   [\ion{O}{iii}] lines. The dotted-dashed green line represents the broad and narrow \ion{He}{ii}~$\lambda$4686 lines.
The dotted-dashed yellow line shows the template \ion{Fe}{ii} spectrum \citep{Veron-Cetty+04} scaled for this object considering the 4500-5100~\AA\ range.
 Some derived properties of the fit are shown in the top left corner. ({\it Bottom panel}) The blue line represents the $rms$ of our fit. The fit has been done within the region between the vertical dashed lines. The average value of the $rms$ is shown in the bottom left corner. The green dashed box indicates the region used to estimate the $rms$ of the broad \Hb\ component; its average value is also shown. Dotted magenta lines indicate the center of the narrow-components for \Hb\ and [\ion{O}{iii}]~$\lambda\lambda$4959, 5007.}
\label{fits1}
\end{figure*}

\begin{figure*}
\centering
\includegraphics[scale=0.55,angle=90]{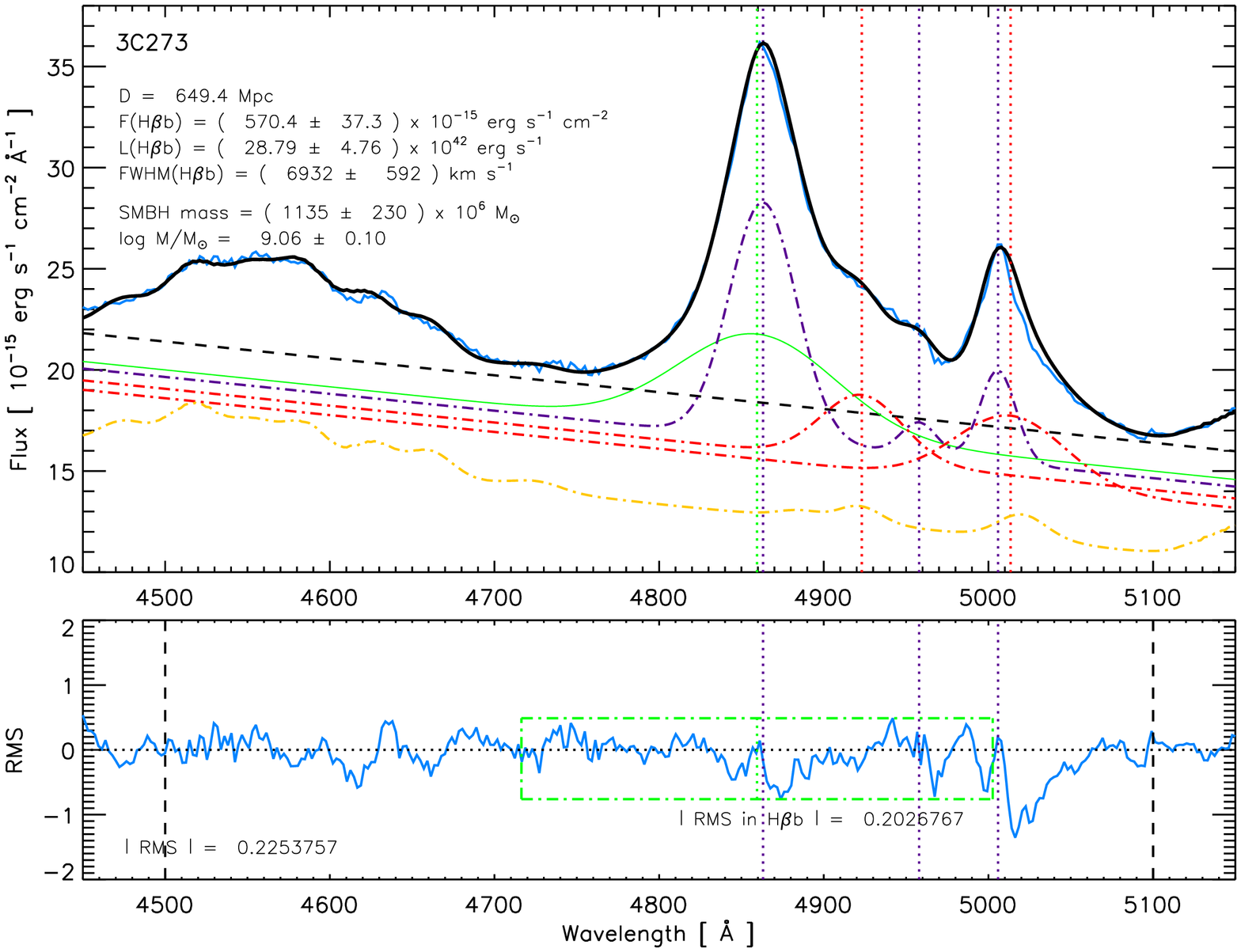}
\includegraphics[scale=0.55,angle=90]{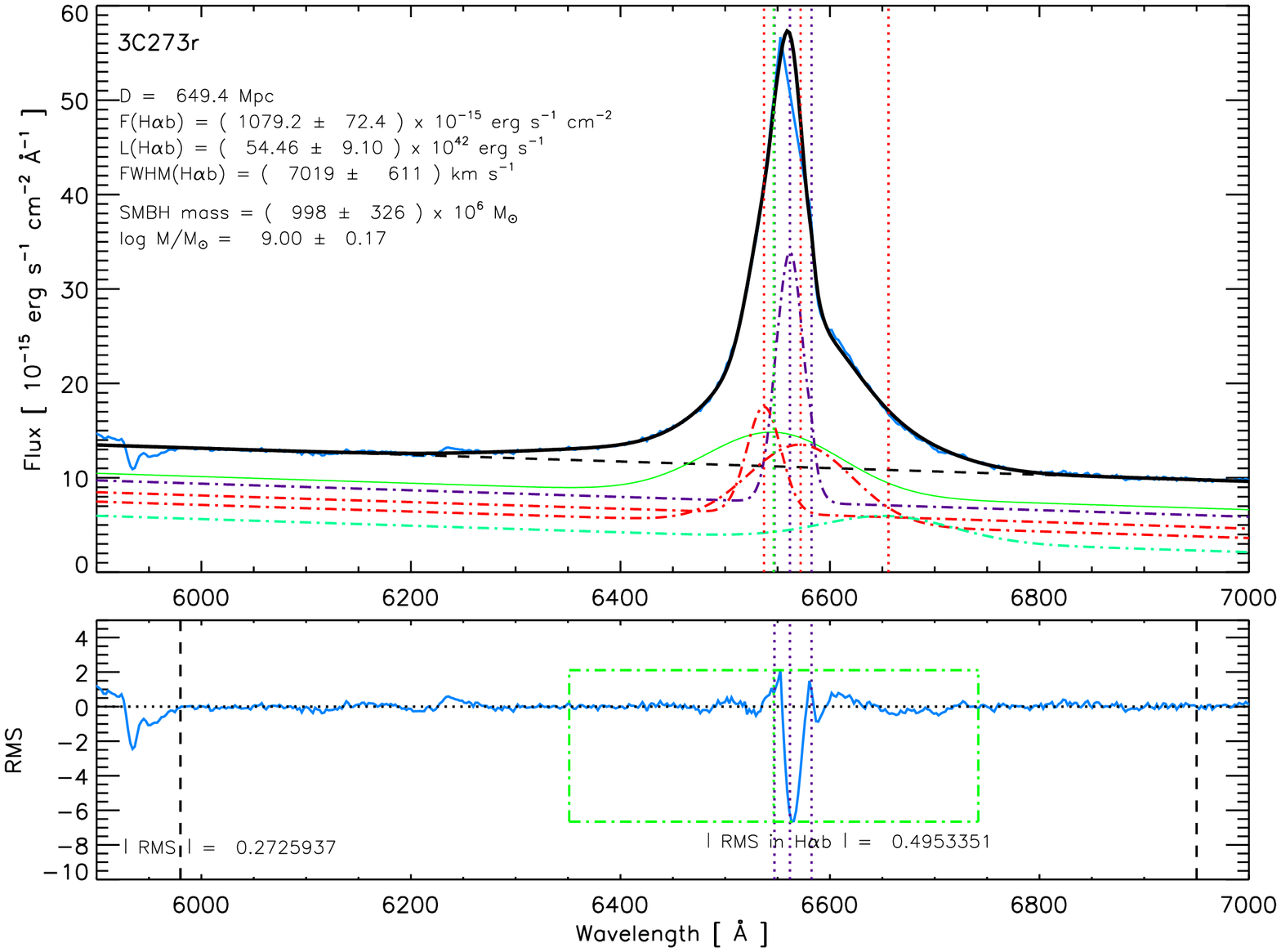} 
\caption{\small Fit to the emission line profiles of 3C 273 around the \Hb\ region (top) and the \Ha\ region (bottom).  The lines are the same described in Fig.~\ref{fits1}, following a similar description for the \Ha\ fit (the blue line is the observed spectrum, the black line is our best fit, dotted red lines are the center of the broad components but \Ha, which is plotted by a green dotted line; the continuous green line is the broad \Ha\ Gaussian, the  dotted-dashed magenta line is the narrow \Ha\ line; the dotted-dashed red lines are the broad  [\ion{N}{ii}] lines; and the dotted-dashed green line is a broad \ion{He}{i}~$\lambda$6678 line).}
\label{fits2}
\end{figure*}

\begin{deluxetable}{cc cc   ccc    c cc   cccc cc ccc}
%\begin{table}{cc cc   ccc    c cc   cccc cc ccc}
\tabletypesize{\scriptsize}
%\rotate
\tablecaption{Details of the flux contributions of the  \ion{He}{ii},  \ion{Fe}{ii} and broad \Hb\  lines to the fits in the 4500-5100~\AA\  region.   \label{detailfits} }
\tablewidth{0pt}
\tablehead{       Object  &      \ion{Fe}{ii}        &  \multicolumn{2}{c}{  \ion{He}{ii} lines }     &    &
 
 \multicolumn{2}{c}{\%   \ion{Fe}{ii} } && \multicolumn{2}{c}{\%   \ion{He}{ii} } &&  \multicolumn{2}{c}{\%  \Hb\ broad}  \\
   \noalign{\smallskip}
   \cline{3-4} \cline{6-7}\cline{9-10} \cline{12-13} 
\noalign{\smallskip}
      &      &  Broad & Narrow & & All & broad \Hb    & & All & broad \Hb  & & All & broad \Hb }
\startdata

PG~0052+251 &   VERY  FAINT$^a$ &  VERY FAINT  & YES    &&     1.2 &  0.9 &&   6.9 &  1.8 &&  56.4 & 81.0   \\
PHL~909        & FAINT   & VERY FAINT  & YES      &&          7.9 &  2.6 &&   4.3 &  0.7 &&  51.4 & 64.2  \\
PKS~0736+017 &    YES & YES & YES &  &  17.0 &  5.7 &&  15.0 &  2.7 &&  40.5 & 77.3 \\
  3C~273     &   YES   &           NO  & NO &       &  34.5 &  5.6 &&  0.0 &  0.0 &&  20.4 & 34.1 \\
PKS~1302-102       &  YES      & NO & NO & &         37.4 &  4.3 &&  0.0 &  0.0 &&  38.7 & 82.2 \\
PG~1309+355  &  YES   &           NO  & YES &&     29.2 &  2.8 &&   0.4 &  0.0 &&  42.6 & 69.9 \\
PG~1444+407  & YES   &      NO  &   NO &&        54.1 &  7.7 &&   0.0 &  0.0 &&  27.9 & 66.1 \\
PKS~2135-147 &   VERY FAINT$^a$ &  VERY FAINT  & YES   & &    4.0 &  0.8 &&   4.2 &  1.7 &&  43.6 & 46.3   \\
 4C~31.63    &   YES  &       NO  &   NO &  &        38.0 &  5.8 &&   0.0 &  0.0 &&  43.9 & 84.5  \\
PKS~2349-014  &   FAINT &    NO & VERY FAINT  &  & 8.2 &  1.7 &&   0.3 &  0.0 &&  46.8 & 75.2 \\
\enddata

\tablenotetext{a}{A fit without considering the  \ion{Fe}{ii} template essentially provides the same result.} 

\end{deluxetable}
%\end{table}

The procedure we followed for performing the fitting in the \Hb\ region was the following:
\begin{enumerate}

\item First, we fit the continuum and broad and narrow components around the \ion{He}{ii}~$\lambda$4686 emission line (if any). 

%{\bf In some cases, the broad \ion{He}{ii}~$\lambda$4686 is  clearly contaminated by several blended  \ion{Fe}{ii} lines, as seen in the template provided by \citet{Veron-Cetty+04}. Note that we do not fit those blended \ion{Fe}{ii} lines, just the overall shape of the resulting broad line.}

\item Then, we  use the template provided by \citet{Veron-Cetty+04} to scale the \ion{Fe}{ii} template spectrum  only considering the lines in the range 4500-5100~\AA\ by minimizing the fit residuals. However, as I~Zw~1 is a narrow line emission galaxy type Seyfert~1, we use the broad  \ion{Fe}{ii} lines observed in the 4500-4700~\AA\ range for providing a broadening to the  \ion{Fe}{ii} template spectrum. In any case the contribution of the \ion{Fe}{ii} lines in the broad \Hb\ range is small, typically between 1--6\% of the total flux of the derived broad \Hb\ flux (see Table~\ref{detailfits}).
%as this provides the better result over the 4500-5100~\AA\ range than a fit including the blended \ion{Fe}{ii} lines over e.g. the  4500-5100~\AA\ range.}

\item Then we fit the spectral profile between \Hb\ and  [\ion{O}{iii}] using three broad components: the broad \Hb\ and the broad [\ion{O}{iii}] lines. All these components must have sigma $> 10$\,\AA\ to assure that we are not fitting any narrow line. The scale factor for the  \ion{Fe}{ii} template spectrum is also included in the minimization of the fit residuals.

\item After that, we add the three narrow emission lines corresponding to the nebular \Hb\ and [\ion{O}{iii}] emission. We forced sigma to be the same in the three lines. As the integrated flux of the [\ion{O}{iii}] emission lines is fixed by theory %{\bf when massive stars photoionize the gas} 
\citep{OsterbrockFerland06}, we also forced that $F$[\ion{O}{iii}]~$\lambda$4959 =  0.3467$\times F$[\ion{O}{iii}]~$\lambda$5007. We then search for those three narrow Gaussians that minimized the residuals.

\item If the narrow \Hb\ emission seems to be broader than the [\ion{O}{iii}] lines, we then fit only this line to get a better solution.

\item We now search for the best combination of narrow+broad Gaussians that fits the [\ion{O}{iii}] lines, again forcing the same sigma in both narrow lines and $F$[\ion{O}{iii}]~$\lambda$4959 =  0.3467$\times F$[\ion{O}{iii}]~$\lambda$5007 in the narrow lines.

\item Finally, we slightly change the parameters of all Gaussians and search for the combination that minimizes the residuals but keeping all assumptions considered before (e.g., same sigma for narrow lines). This would be our best fit.
\end{enumerate}

%{\bf We note that, following the theory \citep{OsterbrockFerland06}, the flux of the broad [\ion{O}{iii}] lines should be also fixed. However we have not considered this assumption and let these lines to run free to account for individual variations in the  \ion{Fe}{ii} lines that cannot be considered using the standard  \ion{Fe}{ii} template of I~Zw~1 provided by  \citet{Veron-Cetty+04}, that, as we said, corresponds to a narrow emission-line type Seyfert~1 galaxy and not a quasar.}

We note that, although it is not usually considered \citep[e.g.][]{Mezcua+11,Fogarty+12,Ho+14,Wild+14,Sanchez+15}, the flux of the broad [\ion{O}{iii}] lines could be also fixed following the theory \citep{OsterbrockFerland06}. However we have considered this assumption for the most challenging case, 3C~273 and note that this assumption leads to a slightly worse fit and it does not materially change the results for the derived values (FWHM, flux, and SMBH mass).

The results of the Gaussians which are providing the best fits are listed in Table~\ref{gaussianfits}, while the plots showing these fits and their $rms$ are compiled in Figures~\ref{fits1}, \ref{fits2}, and \ref{fits3} to \ref{fits8}.
We note that some of the residuals obtained follow from some small asymmetry of the broad \Hb\ line or the broad  [\ion{O}{iii}]~$\lambda$5007 line (broad asymmetric wings). These asymmetric features can be fitted using Gauss-Hermite polynomials \citep[e.g, see][]{Park+12}, however their contribution is rather small (few \%), so we just include this discrepancy in our final error budget.

We follow a similar procedure for fitting the \Ha\ region. In this case, we force $F$[\ion{N}{ii}]~$\lambda$6548 =  0.3402$\times F$[\ion{N}{ii}]~$\lambda$6583 for the narrow emission line components. % following photoionization by massive stars} \citep{OsterbrockFerland06}. 
The spectra resolution of some of the spectra obtained around the \Ha\ region is low enough to reduce the accuracy of the fitting (particularly in 4C~31.63), but this uncertainty is included in the error budget of the broad \Ha\ line. In any case, as we will discuss below, the \Ha-based SMBH masses agree very well with the \Hb-based SMBH masses in all objects (differences smaller than 0.06~dex).

\begin{deluxetable}{ l c c c c  c c c  c c c c c c  c c c}
\tabletypesize{\scriptsize}
%\rotate
\tablecaption{Results of the best fits to the emission line profiles of our quasar spectra. For each emission line we provide the results for both the narrow (n) and the broad (b) components. We provide the $\sigma$, central wavelength, $c$, and peak, $p$, of a Gaussian fit, $G=p\times\exp\big[{-0.5\times(\frac{\lambda-c}{\sigma})^2}\big]$.  We express $p$ in units of 10$^{-16}$ erg\,s$^{-1}$\,cm$^{-2}$, $\sigma$ and $c$ in units of \AA. For some few objects we added an extra broad component (e) in \Hb\ or \Ha, which provides a most robust fit to the data.  \label{gaussianfits} }
\tablewidth{0pt}
\tablehead{
\colhead{Quasar}  &  \colhead{Comp.} & \multicolumn{3}{c}{\ion{He}{ii}~$\lambda$4686} & &  \multicolumn{3}{c}{\Hb}  & & \multicolumn{3}{c}{[\ion{O}{iii}]~$\lambda$4959} & &\multicolumn{3}{c}{[\ion{O}{iii}]~$\lambda$5007} \\
\cline{3-5} \cline{7-9} \cline{11-13} \cline{15-17}
\noalign{\smallskip}

 & &  $\sigma$  & $c$  & $p$ & &  $\sigma$  & $c$  & $p$ & &  $\sigma$  & $c$  & $p$ & &  $\sigma$  & $c$  & $p$  

}
\startdata

%Quasar & Comp. & \multicolumn{3}{c}{\ion{He}{ii}~$\lambda$4686} & &  \multicolumn{3}{c}{\Hb}  & & \multicolumn{3}{c}{[\ion{O}{iii}]~$\lambda$4959} & &\multicolumn{3}{c}{[\ion{O}{iii}]~$\lambda$5007} \\
%\cline{3-5} \cline{7-9} \cline{11-13} \cline{15-17}
%\noalign{\smallskip}

% & &  $\sigma$  & $c$  & $p$ & &  $\sigma$  & $c$  & $p$ & &  $\sigma$  & $c$  & $p$ & &  $\sigma$  & $c$  & $p$ \\ 

%\noalign{\smallskip}
% \tableline
%\noalign{\smallskip}

PG0052+251 & n  &   3.69 & 4684.1 &    0.836 &&   4.30 & 4861.4 &    4.86 &&   3.83 & 4958.7 &   14.8 &&   3.83 & 5006.8 &   42.8 \\
                      & b  &  56.1 & 4687.7 &    1.22 &&  41.3 & 4860.8 &   14.3 &&  12.2 & 4960.3 &    3.37 &&   9.54 & 5006.5 &    9.25 \\

\noalign{\smallskip}

         PHL909 & n  &   5.78 & 4687.9 &    0.482 &&   4.90 & 4861.0 &    0.842 &&   4.90 & 4958.1 &    3.32 &&   4.90 & 5006.7 &    9.58 \\
                       & b  & \nodata & \nodata &   \nodata  &&  47.0 & 4862.6 &    6.10 &&  21.8 & 4944.1 &    2.52 &&  28.1 & 4989.6 &    2.86 \\
 
\noalign{\smallskip}

PKS0736+017 & n  &   5.48 & 4687.7 &    0.276 &&   9.15 & 4863.3 &    1.05 &&   4.90 & 4958.9 &    0.779 &&   4.90 & 5006.8 &    2.49 \\
                        & b  &\nodata & \nodata &   \nodata &&  21.8 & 4866.9 &    3.81 &&  23.5 & 4939.2 &    0.625 &&  36.4 & 5029.4 &    0.442 \\
 
\noalign{\smallskip}

3C273 & n  & \nodata & \nodata &   \nodata &&  20.2 & 4864.0 &  116.6 &&  10.8 & 4958.9 &   15.7 &&  10.8 & 5006.8 &   45.4 \\
                & b  &  \nodata & \nodata &   \nodata &&  47.7 & 4860.4 &   47.7 &&  24.7 & 4924.0 &   32.3 &&  33.3 & 5014.4 &   34.2 \\
%                & e  & \nodata & \nodata& \nodata   &&   7.830 & 4864.0 &   10.0553 && \nodata &\nodata &    \nodata&&  \nodata& \nodata&   \nodata \\ 

\noalign{\smallskip}

 PKS1302-102 & n  &   \nodata & \nodata &   \nodata  &&   5.71 & 4861.6 &    1.74 &&   5.03 & 4958.0 &    2.18 &&   5.03 & 5006.1 &    5.41 \\
                         & b  & \nodata & \nodata &   \nodata &&  29.3 & 4867.9 &    5.59 &&  13.3 & 4940.6 &    1.19 &&  14.6 & 5003.7 &    2.52 \\

\noalign{\smallskip}

     PG1309+355 & n  &  10.5 & 4686.7 &    3.48 &&   4.01 & 4861.6 &   18.9 &&   4.01 & 4958.7 &   23.1 &&   4.01 & 5006.7 &   66.6 \\
                            & b  & \nodata & \nodata &   \nodata &&  43.9 & 4870.7 &   40.0 &&  14.6 & 4954.3 &   15.5 &&  12.3 & 5002.4 &   39.8 \\

\noalign{\smallskip}

PG1444+407 & n  &   \nodata & \nodata &   \nodata &&  12.4 & 4863.3 &    8.53 &&   5.14 & 4958.9 &    0.771 &&   5.14 & 5004.0 &    3.70 \\
                      & b  &    \nodata & \nodata &   \nodata  &&  29.4 & 4867.5 &   12.1 &&    15.9 & 4941.2   &    2.56 &&  28.5 & 5016.6 &    2.07 \\

\noalign{\smallskip}

PKS2135-147 & n  &   5.04 & 4686.0 &    2.26 &&   4.31 & 4861.8 &    7.53 &&   4.03 & 4959.1 &   31.1 &&   4.03 & 5006.8 &   89.7 \\
                        & b  & 125.7 & 4691.4 &    0.810 &&  58.2 & 4867.6 &   10.0 &&  13.6 & 4961.0 &    2.33 &&  12.8 & 5011.1 &    7.36 \\

\noalign{\smallskip}

4C31.63   & n  &   \nodata & \nodata &    \nodata &&   6.03 & 4870.7 &    1.73 &&   5.28 & 4960.1 &    2.52 &&   5.28 & 5006.2 &    7.85 \\
                & b  & \nodata & \nodata &   \nodata &&  26.2 & 4879.0 &   21.3 &&  13.4 & 4946.8 &    4.13 &&  39.3 & 5018.0 &    2.82 \\

\noalign{\smallskip}

%\tablebreak

PKS2349-014 & n  &   4.79  & 4686.8  &    0.168 &&    6.09 & 4863.1 &    2.01 &&   4.40 & 4959.0 &    4.55 &&   4.40 & 5006.8 &   13.1 \\
                       & b  & \nodata & \nodata &   \nodata &&  41.66 & 4862.6 &    5.13 &&   16.1 & 4956.1 &    0.990 &&  16.9 & 5004.7 &    2.52 \\

  \noalign{\medskip}
  \tableline
   \noalign{\medskip}
Object & Comp. & \multicolumn{3}{c}{[\ion{N}{ii}]~$\lambda$6548} & &  \multicolumn{3}{c}{\Ha}  & & \multicolumn{3}{c}{[\ion{N}{ii}]~$\lambda$6583} & &\multicolumn{3}{c}{\ion{He}{i}~$\lambda$6678} \\
\cline{3-5} \cline{7-9} \cline{11-13} \cline{15-17}
\noalign{\smallskip}

 & &  $\sigma$  & $c$  & $p$ & &  $\sigma$  & $c$  & $p$ & &  $\sigma$  & $c$  & $p$ & &  $\sigma$  & $c$  & $p$ \\
 
\noalign{\smallskip}
 \tableline
\noalign{\smallskip}

3C273          & n  &   2.73 & 6548.0 &    5.02 &&  13.0 & 6562.8 &  265.1 &&   2.73 & 6583.4 &   14.8 && \nodata & \nodata &  \nodata \\
                    & b  &  16.2 & 6538.0 &  113.0 &&  65.1 & 6547.3 &   66.1 &&  49.6 & 6573.0 &   83.9 &&  62.1 & 6656.9 &   26.2 \\
        %        & e  & \nodata & \nodata& \nodata   && 101.187 & 6404.1 &   12.3953 && \nodata &\nodata &    \nodata&&  \nodata& \nodata&   \nodata \\ 

   \noalign{\smallskip}

PKS1302-102    & n  &   5.69 & 6547.2 &    0.841 &&   5.69 & 6563.9 &    6.55 &&   5.69 & 6584.2 &    2.43 &&    \nodata & \nodata &  \nodata \\
                          & b  &   6.37 & 6548.6 &    3.63 &&  39.9 & 6568.9 &   11.2 &&   9.38 & 6577.2 &    1.71 &&  14.8 & 6679.1 &    1.77 \\

   \noalign{\smallskip}

PKS2135-147    & n  &   5.72 & 6547.3 &    2.70 &&   5.72 & 6562.1 &   25.4 &&   5.72 & 6582.7 &    7.78 &&    \nodata & \nodata &  \nodata \\
                          & b  &  37.6 & 6467.0 &    9.50 &&  73.2 & 6575.7 &   24.7 &&  46.3 & 6578.8 &   10.6 &&   9.51$^a$ & 6723.3$^a$ &    2.93$^a$ \\

   \noalign{\smallskip}

4C31.63       & n  &  17.42 & 6525.5 &    1.01 &&  17.4 & 6560.6 &    3.32 &&  17.4 & 6574.9 &    2.97 &&   \nodata & \nodata &  \nodata \\
                    & b  &  13.22 & 6546.4 &   10.6 &&  39.8 & 6567.5 &   27.5 &&  19.3 & 6570.6 &   15.2 &&   \nodata & \nodata &  \nodata \\
       %         & e  & \nodata & \nodata& \nodata   &&  23.420 & 6455.3 &    1.5826 && \nodata &\nodata &    \nodata&&  \nodata& \nodata&   \nodata \\ 

   \noalign{\smallskip}

PKS2349-014    & n  &   6.96 & 6548.1 &    3.20 &&   6.96 & 6562.9 &   11.2 &&   6.96 & 6582.5 &    9.39 &&   4.08$^b$ & 6300.6$^b$ &    0.881$^b$ \\
                          & b  &  16.6 & 6537.7 &    8.85 &&  68.0 & 6583.6 &    5.03 &&  25.4 & 6591.3 &    7.45 &&  11.8$^a$ & 6723.0$^a$ &    1.50$^a$ \\

\enddata
\tablenotetext{a}{The data correspond to a fit to a broad [\ion{S}{ii}] $\lambda$6717+$\lambda$6731 emission line.}
\tablenotetext{b}{The data correspond to a fit to a narrow [\ion{O}{i}] $\lambda$6300 emission line.}
\end{deluxetable}

\begin{deluxetable}{l  cc  cc    c  c   c}
\tabletypesize{\scriptsize}
%\rotate
\tablecaption{Results of the analysis of the emission line profiles of our target quasars. The second column provides the best $rms$ obtained in the fit. The third column indicates if the fit has been performed to the \Hb\ or \Ha\ region. The fourth and fifth columns compile the luminosity and FWHM of the broad \Hb\ or \Ha\ emission line. Columns sixth and seventh give the mass of the SMBH derived using our data following Eqs.~\ref{eqhb} and~\ref{eqha}.  Last column provides the stellar velocity dispersion, $\sigma_{\star}$, for the analyzed galaxies, which was obtained by  \citet{WolfSheinis08}.   \label{smbhmass} }
\tablewidth{0pt}
\tablehead{            &   &     \multicolumn{3}{c}{Broad Emission Line}         \\
\cline{3-5}
\noalign{\smallskip}
Quasar            &  $rms$          &  Line      &  $L$          & FWHM  &    $M_{\rm SMBH}$   &       $\log (M_{\rm SMBH}/M_{\odot}$)     &  log ($\sigma_{\star}$) \\
\noalign{\smallskip}
%\noalign{\smallskip}
                      &  [erg\,s$^{-1}$\,cm$^{-2}$\,\AA$^{-1}$]   &    & [10$^{42}$\,erg\,s$^{-1}$] & [km\,s$^{-1}$] & [10$^6$~\Mo]  &        & [km\,s$^{-1}$]  
}
\startdata
PG0052+251 &   0.025501 & \Hb  &    7.20 $\pm$   0.87 &   5999 $\pm$    248 &   391 $\pm$    46  &   8.59 $\pm$   0.06        & 2.45 $\pm$ 0.08  \\

        PHL909 &   0.017113 & \Hb  &    4.30 $\pm$   0.58 &   6830 $\pm$    371 &   380 $\pm$    52  &   8.58 $\pm$   0.06        &  2.22 $\pm$ 0.03 \\

PKS0736+017 &   0.008748 & \Hb  &    1.49 $\pm$   0.20 &   3157 $\pm$    160 &    44 $\pm$     5  &   7.65 $\pm$   0.06         &  2.53 $\pm$ 0.10  \\

 3C273         &   0.225376 & \Hb  &   28.8 $\pm$   4.8 &   6932 $\pm$    592 &  1135 $\pm$   230  &   9.06 $\pm$   0.10         & 2.52  $\pm$  0.07 \\
                     &   0.272594 & \Ha  &   54.5 $\pm$   9.1   &   7019 $\pm$    611 &   998 $\pm$   326  &   9.00 $\pm$   0.17          & " \\

PKS1302-102 &   0.020744 & \Hb  &    6.07 $\pm$   0.82 &   4242 $\pm$    234 &   177 $\pm$    25  &   8.25 $\pm$   0.07    & 2.59 $\pm$ 0.08 \\
                     &    0.048395 & \Ha  &    16.6 $\pm$   2.1 &   4293 $\pm$     202 &   188 $\pm$    47  &   8.28 $\pm$   0.13        & " \\ 

PG1309+355 &   0.127747 & \Hb  &   29.51 $\pm$   4.26 &   6370 $\pm$    410 &   972 $\pm$   163  &   8.99 $\pm$   0.08  & 2.42 $\pm$  0.05 \\
   
PG1444+407 &   0.034493 & \Hb  &   12.3 $\pm$   1.8 &   4261 $\pm$    270 &   265 $\pm$    42  &   8.42 $\pm$   0.08   & 2.50 $\pm$ 0.03 \\ 

PKS2135-147 &   0.036036 & \Hb  &   11.7 $\pm$   1.8 &   8440 $\pm$    624 &  1018 $\pm$   181  &   9.01 $\pm$   0.08      & 2.49 $\pm$ 0.14 \\
                     &  0.068690 & \Ha  &   36.4 $\pm$   4.3 &   7862 $\pm$    309 &  1010 $\pm$   274  &   9.00 $\pm$   0.14         &  " \\

4C31.63     &   0.040863 & \Hb  &   29.24 $\pm$   3.57 &   3795 $\pm$    160 &   343 $\pm$    45  &   8.54 $\pm$   0.06  &  2.51 $\pm$ 0.05 \\
                     &  0.039150 & \Ha  &   57.2 $\pm$   6.5 &   4275 $\pm$    145 &   369 $\pm$   103  &   8.57 $\pm$   0.14         & " \\
                  
PKS2349-014 &   0.013422 & \Hb  &    3.27 $\pm$   0.43 &   6048 $\pm$    306 &   255 $\pm$    33  &   8.41 $\pm$   0.06    &  2.48 $\pm$ 0.08 \\
                        &   0.026210 & \Ha  &    5.24 $\pm$   0.90 &   7295 $\pm$    666 &   298 $\pm$    81  &   8.47 $\pm$   0.14         & " \\

\enddata
%\tablenotetext{a}{The best fit is obtained introducing an extra, broad component for the \Hb\ or \Ha\ emission.  See text for details.}
\end{deluxetable}

Finally, we applied Eqs.~\ref{eqhb} and \ref{eqha} to derive the SMBH masses using the results provided for the best fit to the broad \Hb\ and \Ha\ lines, respectively. The luminosities of the broad \Hb\ and \Ha\ lines were derived just applying the $L=4\pi D^2 F$ relation. The distance to the quasar, $D$, was determined from the radial velocity of the narrow [\ion{O}{iii}]~$\lambda$5007 emission line, which was also provided by our fit. Both the redshift \mbox{$z=v_{rad}/c$} and the distance to our objects (estimated assuming 
a flat cosmology with $H_0$=70~km\,s$^{-1}$\,Mpc$^{-1}$, \mbox{$\Omega_M$ = 0.3}, and $\Omega_\Lambda$ = 0.7) %has been adopted throughout this paper.
%$H_0$=71\,km\,s$^{-1}$\,Mpc$^{-1}$)  
are compiled in Table~\ref{sigma_table}.
The FWHM was computed using the standard definition for a Gaussian, FWHM = 2.355$\times\sigma$. The results for the luminosity and FWHM of the broad \Hb\ and/or \Ha\ lines of each object are listed in Table~\ref{smbhmass}, which includes the $rms$ obtained in our best fit for each case.

Uncertainties were estimated from the $rms$ of the fit, which provides an error to the derived values for the flux and the $\sigma$ of the Gaussians. 
For estimating the uncertainty in luminosity we also assumed a 5\% error in the absolute flux calibration of our spectra. We propagate these errors through the equations to derive the uncertainties of the SMBH masses, taking into account the uncertainties included in Eqs.~\ref{eqhb} and \ref{eqha}. %As the uncertainties in the \Ha\ equation are much higher than those included in the \Hb\ equation, 
The \Ha-based SMBH masses have much higher uncertainties than the \Hb-based SMBH masses. 
In any case, we should keep in mind some caveats, uncertainties, and biases of SMBH masses derived using the virial method \citep[see Sect.~3 in][for a review]{Shen13}. For example, as described in the introduction, the actual value of the parameter $f$ introduced in Eq.~\ref{virial} may vary within a factor of 2--3 depending on the geometry of the BLR.
Some other uncertainties are introduced by assuming an unique line profile to infer the underlying BLR velocity structure or considering a single BLR size. The actual effects of the host starlight and the dust reddening (particularly in the blue and UV regimes) are also unknown in many cases, as well as the orientation, isotropy, radiation pressure, and velocity structure of the BLR. Finally, SMBH masses estimated using the virial method may be also affected by the AGN variability.

\begin{deluxetable}{l  c cc@{\hspace{5pt}} c@{\hspace{5pt}} ccc c c c}
\tabletypesize{\scriptsize}
%\rotate
\tablecaption{Super-massive black hole mass estimations, in the form of $\log (M_{\rm SMBH}/M_{\odot}$), for the quasars studied in this work (TW) and their estimations found in the literature using different techniques. Forth column compiles the average value of the SMBH mass derived here using the fits to both \Hb\ and \Ha\ broad emission lines. Table includes the derived $f_{5100}$ fluxes (second column and in units of 10$^{-15}\,$erg\,s$^{-1}$\,cm$^{-2}$\,\AA$^{-1}$) and the SMBH mass estimations using this value and FWHM$_{\rm H\beta}$ (third column). \label{comparison}}
\tablewidth{0pt}
\tablehead{         &      &  \multicolumn{2}{c}{TW}   & &       \multicolumn{3}{c}{Reverberation Mapping Techniques}         \\
\cline{3-4}\cline{6-8}
\noalign{\smallskip}
Quasar             &    $f_{5100}$           &   $L_{5100}$    &      Fits         & &   P04                     & K08                       & B09                      &      S03            &      K08            &   G11  
}
\startdata

PG 0052+251  & 1.50 $\pm$ 0.03      & 8.59 $\pm$  0.06 &  8.59 $\pm$  0.06  &&   8.56 $\pm$ 0.08        &  8.31 $\pm$ 0.10         &  8.57 $\pm$ 0.08      &  \nodata       &    \nodata       &  8.48 $\pm$ 0.26   \\

PHL 909          & 1.224 $\pm$ 0.014  & 8.68 $\pm$  0.06 &  8.58 $\pm$   0.06  &&  \nodata             &  \nodata                      &  \nodata                    &  \nodata               &  9.12 $\pm$ 0.20   &  \nodata    \\

PKS 0736+017& 0.430 $\pm$ 0.006 & 7.77 $\pm$  0.06 &  7.65 $\pm$   0.06   &&  \nodata                      &  \nodata                      &  \nodata                    &  \nodata              &  8.08  $\pm$ 0.20   & \nodata  \\

3C 273            & 11.1 $\pm$ 0.4  & 9.22 $\pm$  0.09 &  9.04 $\pm$   0.08   &&   8.95 $\pm$ 0.08  &  8.69 $\pm$ 0.08            &  8.95 $\pm$ 0.09       &  8.88                   &     \nodata             &  \nodata   \\
 
PKS1302-102 & 1.30 $\pm$ 0.02   & 8.52 $\pm$  0.07 &   8.26 $\pm$   0.06    && \nodata           &  \nodata                      &  \nodata                    &  \nodata                &    8.58 $\pm$ 0.30     &  \nodata        \\
                                                        
PG 1309+355  & 8.37 $\pm$ 0.05 & 9.58 $\pm$  0.08 &  8.99 $\pm$   0.09    && \nodata                      &  \nodata                      &  \nodata                    &    8.16                   &    8.38 $\pm$ 0.30    & \nodata \\
  
PG 1444+407  & 1.58 $\pm$ 0.03 & 10.11 $\pm$  0.10 & 8.42 $\pm$   0.08    && \nodata                      &  \nodata                      &  \nodata                    &  \nodata                &    8.21 $\pm$ 0.20   & \nodata  \\ 

PKS 2135-147 & 1.486 $\pm$ 0.018 & 9.03 $\pm$  0.07 &  9.01 $\pm$   0.07    && \nodata                  &  \nodata                      &  \nodata                    &  \nodata                &    8.96 $\pm$ 0.20  & \nodata     \\
                         
4C 31.63         & 1.75 $\pm$ 0.02  & 8.56  $\pm$  0.06 & 8.55 $\pm$  0.05     && \nodata                 &  \nodata                      &  \nodata                    &  \nodata                &   8.43 $\pm$ 0.30  &   \nodata    \\
                  
PKS 2349-014 & 0.620 $\pm$ 0.014 & 8.68 $\pm$  0.06 &    8.43 $\pm$   0.06    && \nodata                      &  \nodata                 &  \nodata                &  \nodata               & 8.55 $\pm$ 0.20    &   \nodata   \\

%    fv = [1.498, 1.224, 0.4299, 17.06, 1.295, 8,373, 1.583, 1.486, 1.749, 0.6119]   ; en unidades 1D-15 erg s-1 cm-1 A-1
 % efv = [0.03, 0.014, 0.006, 0.4, 0.02, 0.05, 0.03, 0.018, 0.02, 0.014]
 
 %     8.5684867       8.7235654       7.7680704       9.5473858       8.6630292       9.5841637       10.213443       9.0304008       8.5884772       8.6831054
 %    0.041961674     0.041978725     0.034486801     0.063398726     0.050356023     0.059274187     0.098916891     0.046764740     0.054527232     0.045274502

%G0052 PHL909 PKS0736 3C273 PKS1302 PG1309 PG1444 PKS2135 4C31 PKS2349
 
 %     8.5892871       8.6815306       7.7680704       9.2240135       8.5221779       9.5839942       10.113037       9.0298951       8.5604961       8.6779505
  %   0.055692868     0.060557736     0.056369455     0.089696358     0.068215030     0.080892688      0.10054402     0.074436936     0.061331181     0.060596390

\enddata
\tablerefs{(P04) \citet{Peterson+04} using reverberation mapping techniques; (K08) \citet{Kim+08} using the reverberation mapping method; (B09) \citet{Bentz+09a} using reverberation mapping techniques; (S03) \citet{Shields+03}; (K08) \citet{Kim+08}  using the virial method based on the single epoch spectrum; (G11) \citet{Gliozzi+11} using X-ray data.}
\end{deluxetable}

For completeness, we also estimated the SMBH masses using the continuum luminosity at rest frame wavelength 5100\,\AA, $\lambda L_{5100}$, and the FWHM of the broad \Hb\ emission line derived in our fit, via the virial formula provided by \citet{GreeneHo05}, see Eq.~\ref{eqL5100}. Table~\ref{comparison} compiles both the $f_{5100}$ measured in our spectra and the SMBH masses using the derived continuum luminosity.  The uncertainty of the $f_{5100}$ flux has been estimated using the $rms$ of the continuum over the $\lambda\lambda$5075-5125 range and does not include the error in the absolute flux calibration. Errors listed in the mass estimations following this method include the uncertainties in both $L_{5100}$ and FWHM$_{\rm H\beta}$ measurements as well as the uncertainties shown in Eq.~\ref{eqL5100} and the error in the absolute flux calibration.

% SECTION 4: DISCUSSION  -----------------------------------------------------------------------------------------------

\begin{figure}[t!]
\centering 
\includegraphics[scale=0.45,angle=0]{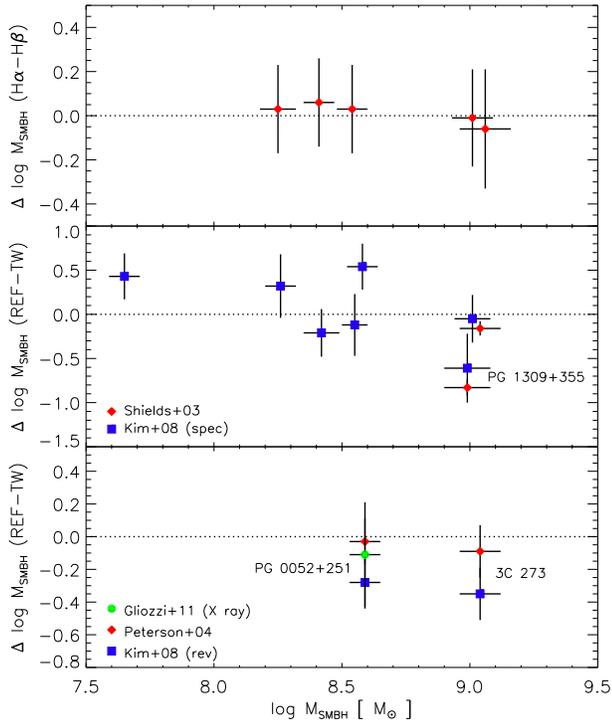} %0.425
 \caption{\small (Top) Comparison of the $M_{SMBH}$  derived here using the \Ha\ and \Hb\ profile fitting. (Middle) Comparison of our $M_{SMBH}$ estimations with the values provided by \citet{Shields+03}--red diamonds-- and \citet{Kim+08} --blue boxes--, in both cases derived following the virial method based on the single epoch spectrum. (Bottom) Comparison of our $M_{SMBH}$ estimations with the values provided by \citet{Peterson+04} --red diamonds-- and \citet{Kim+08} --blue boxes--, in both cases using reverberation mapping techniques. The value of the $M_{SMBH}$ obtained by \citet{Gliozzi+11} for PG~0052+251 using X ray data is also plotted in this panel --green circle--.
\label{masscomp} }
\end{figure}

\section{Discussion}

\subsection{Mass comparison}

We first compared our derived $M_{SMBH}$ values using the fitting to the \Ha\ and \Hb\ emission line profiles. Top panel of Fig.~\ref{masscomp} plots the difference of the logarithm of the masses found using \Ha\ and \Hb\ profile fitting ($y$-axis) versus the mass obtained from the \Hb\ profile fitting ($x$-axis) for the five quasars we have \Ha\ data. As  this panel clearly shows, the \Ha-based SMBH masses agree very well with those values derived from the analysis of the \Hb\ region. The \Ha\ spectra typically have lower spectral resolution than the \Hb\ spectra, and hence higher uncertainties, thus in these five cases we use an uncertainty-weighted average of both results to minimize the error of our SMBH mass estimations, which we compile in  Table~\ref{comparison}. The fact that both the \Ha-based and the \Hb-based SMBH mass estimations agree well in all cases reinforces the strength of the technique used here to estimate SMBH masses.

Figure~\ref{masscomp} also compares our results for the SMBH masses with those values found in the literature. Table~\ref{smbhmass} compiles the $M_{\rm SMBH}$ masses of our QSO sample found by other authors following different techniques.
We note that we checked for a consistent treatment --i.e., similar virial factor used (within 4 - 8\%) for computing SMBG masses-- when using literature data.
Only two of our quasars have an estimation of their SMBH masses using reverberation mapping techniques:  3C~273 and PG~0052+251 \citep{Kaspi+00,Peterson+04,Kim+08,Bentz+09a}. %(Kaspi et al. 2000, ApJ, 533, 631; Peterson et al. 2004, AJ, 613, 682; Kim et al. 2008; Bentz et al. 2009). 
Following Table~\ref{comparison} and the bottom panel of Fig.~\ref{masscomp}, we see that 
our results agree well within the errors for PG~0052+251: we derived $\log (M_{\rm SMBH}/M_{\odot}$) = 8.60$\pm$0.06, while \citet{Peterson+04} and \citet{Bentz+09a} estimated SMBH masses of 8.56$\pm$0.08 and 8.57$\pm$0.08, respectively. \citet{Kim+08} derived a somewhat lower mass for the SMBH in this object,  8.31$\pm$0.10.
%
%On the other hand,
Our result for 3C~273 also agrees within the errors with those values reported in the literature using reverberation mapping techniques:
%Conversely, the results differ in {\bf almost} half an order of magnitude for 3C~273:  
we derived $\log (M_{\rm SMBH}/M_{\odot}$) = 9.04$\pm$0.08 consistently using both the fitting to the \Ha\ and \Hb\ regions, while \citet{Peterson+04} and \citet{Bentz+09a}  estimated 8.95$\pm$0.08 and 8.95$\pm$0.09, respectively, for this object.
However,  \citet{Kim+08} provides again a lower mass for the SMBH within 3C~273, $\log (M_{\rm SMBH}/M_{\odot}$) = 8.69$\pm$0.08.
Previous studies of the $M_{SMBH}$  of 3C~273 using the virial method based on a single epoch spectrum also provide lower numbers that we do: \citet{Shields+03} found $\log (M_{\rm SMBH}/M_{\odot}$) = 8.88. 
We rule out a problem with the absolute flux calibration of our spectrum, as our sample considers other objects that were observed the same night and show no issues with flux calibration.
We also consider our analysis, particularly the fitting to the emission line profiles (up to 4 narrow and 5 broad components have been considered), has been performed with more detail than in previous work. 
%However, we can not discard that the activity is quickly changing in 3C~273, and hence measurements made in different epochs differ significantly between them.  
However, it is possible that the differing flux measurements of 3C~273 are due to true variations in the emission of this quasar \citep{Lloyd84,Hooimeyer+92,Ghosh+00,Soldi+2008}. 
Although following the virial hypothesis an increasing (decreasing) of the emission line luminosities will induce a decreasing (increasing) of the rotational velocity of the gas, uncorrelated physical or observational stochastic variations between line width and luminosity (or other effects as described in Sect.~\ref{sect3.2}) may also play an important role in determining the SMBH mass of 3C~273.

Middle panel of Fig.~\ref{masscomp} compares our SMBH masses with those estimations obtained by \citet{Shields+03} using the $M_{SMBH}$-$\sigma_{*}$ relation based on the [\ion{O}{iii}] sigma and \citet{Kim+08} using the virial method based on a single epoch spectrum. We see a fairly good correlation for some objects (PKS~1302-102, PKS~2135-147, 4C~31.63, PKS~2349-014) which agree well within the errors but for other objects (PHL~909, PKS~0736+017, PG~1309+355, PG~1444+407) we see poor agreement, well outside the error bars. The worst case is PG~1309+355, SMBH mass estimate disagrees by more than an order of magnitude, between the two methods. We believe our results for this object and for PHL~909 provides the most robust estimate of the MBH mass, as the $rms$ of our fits are very low in both cases. 
Furthermore, as shown in the middle panel of Fig.~\ref{masscomp} panel, there appears to be a trend in our fits providing lower SMBH masses than previously estimated in the low-mass end, while obtaining higher SMBH masses than previously estimated in the high-mass end.

\begin{figure*}
\centering 
\includegraphics[scale=0.4,angle=0]{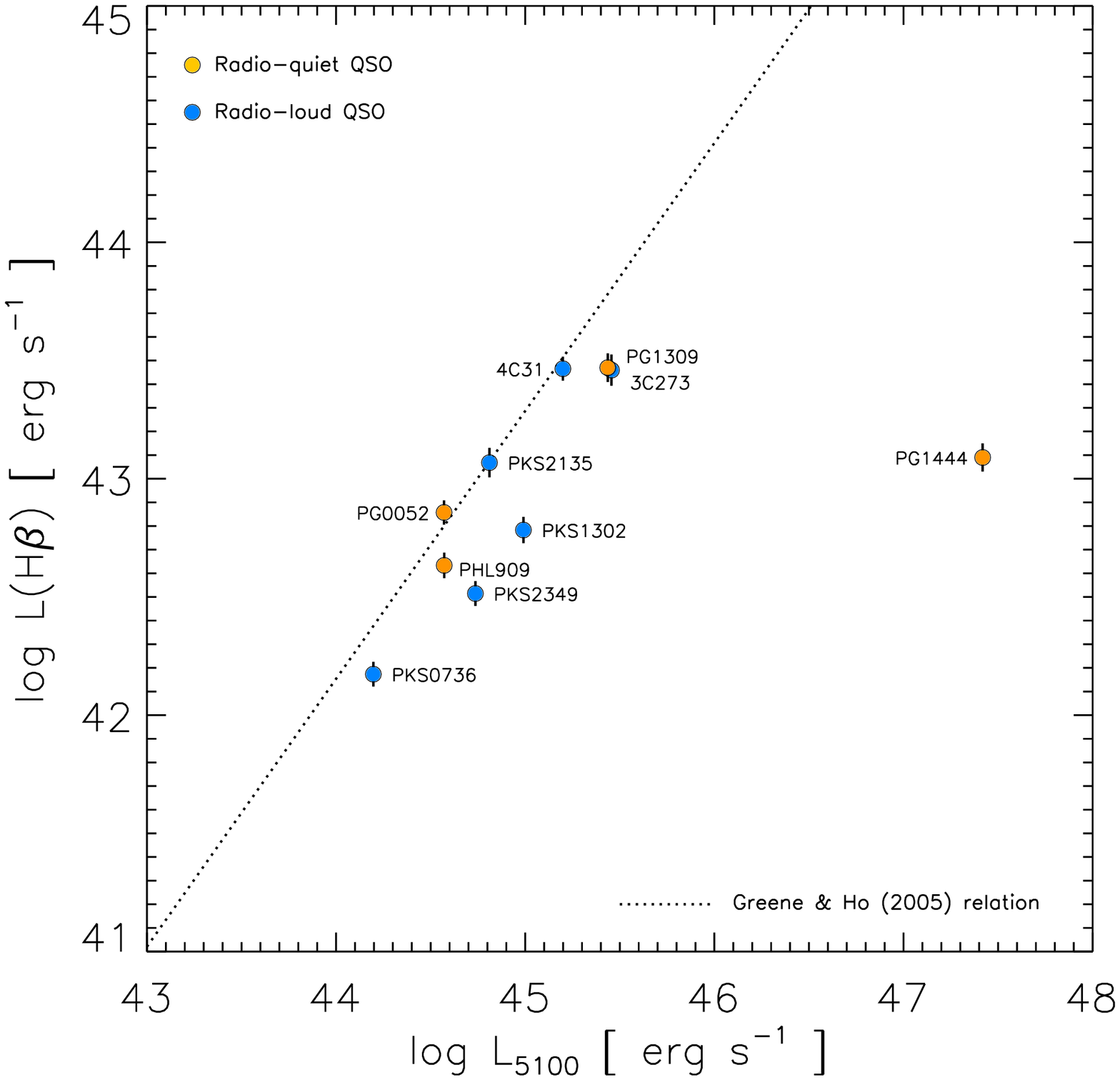} \hspace{0.1cm}
\includegraphics[scale=0.4,angle=0]{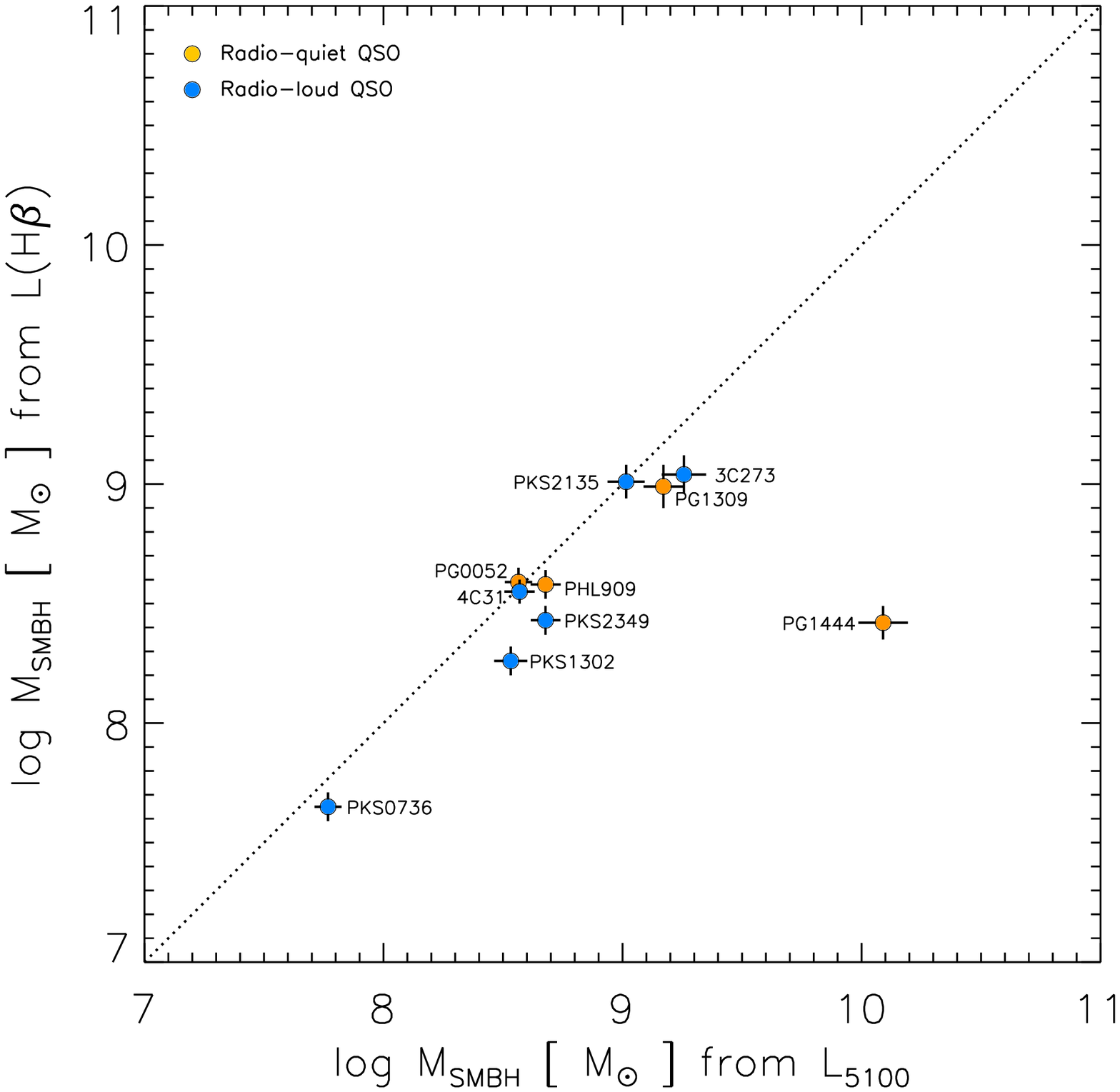}
 \caption{\small (Left) Comparison between the  continuum luminosity at rest frame wavelength 5100\,\AA, $\lambda L_{5100}$, and the broad \Hb\ luminosity derived from our fits. The dotted line plots the relationship between both quantities presented by \citet{GreeneHo05}.  (Right) Comparison between the SMBH mass derived using the fits to both the broad \Ha\ and \Hb\ emission lines and that derived from  $L_{5100}$  and FWHM$_{\rm H\beta}$ following Eq.~\ref{eqL5100}. The dotted line indicates $y=x$.
\label{lvfig} }
\end{figure*}

The only object in common with the sample analyzed in X rays by \citet{Gliozzi+11} is PG~0052+251 (see bottom panel of Fig.~\ref{masscomp}). These authors derived $\log (M_{\rm SMBH}/M_{\odot}$) = 8.48$\pm$0.26, which  matches within the errors with our result. This  also suggests that the value provided by \citet{Kim+08} using reverberation mapping techniques may be slightly underestimated. If this is also the case for 3C~273, it would explain why our $M_{\rm SMBH}$ estimation is higher than that derived by \citet{Peterson+04,Bentz+09a} and \citet{Kim+08} using reverberation mapping techniques. 

Finally, we also study how well the SMBH masses are recovered when using the continuum luminosity at 5100\,\AA\ instead of our fits to the emission line profiles. Left panel of Fig.~\ref{lvfig} compares the derived $\lambda L_{5100}$ luminosities with the broad  \Hb\ luminosity obtained in our fits. This panel includes the relationship provided by \citet{GreeneHo05}. Right panel of Fig.~\ref{lvfig} compares the SMBH mass computed using $\lambda L_{5100}$ and that obtained from our fits to the broad emission lines. As we see, all but one of our objects have a relatively good agreement between both SMBH estimations, although masses derived using  $\lambda L_{5100}$ seem to be systematically overestimated. Indeed, \citet{GreeneHo05} already pointed out this result, emphasizing that the effect is serious for relatively luminous, core-dominated sources and, in particular, radio-loud objects. These authors attributed the enhancement of the optical continuum to jet contamination, although they did not rule out that the enhancement is a consequence of changes in the ionizing spectrum or the covering factor of the ionized gas. Nevertheless there is a large difference in masses --almost 3 orders of magnitude-- for PG~1444+407, which is a radio-quiet quasar. 
We consider that the $M_{\rm SMBH}$ estimation derived using our fit to the \Hb\ emission line profile is much more robust than that obtained from the continuum luminosity. Indeed, if we assume the SMBH mass derived using the $\lambda L_{5100}$ luminosity as the correct one, PG~1444+407 will lie far away from the rest of the radio-quiet quasars in our subsequent analysis of the  $M_{\rm SMBH}$-$\sigma_{*}$ relation and  $L_{\rm 5\,GHz}$-$\sigma_{*}$ relation.

\begin{figure*}
\centering 
\includegraphics[scale=0.8,angle=0]{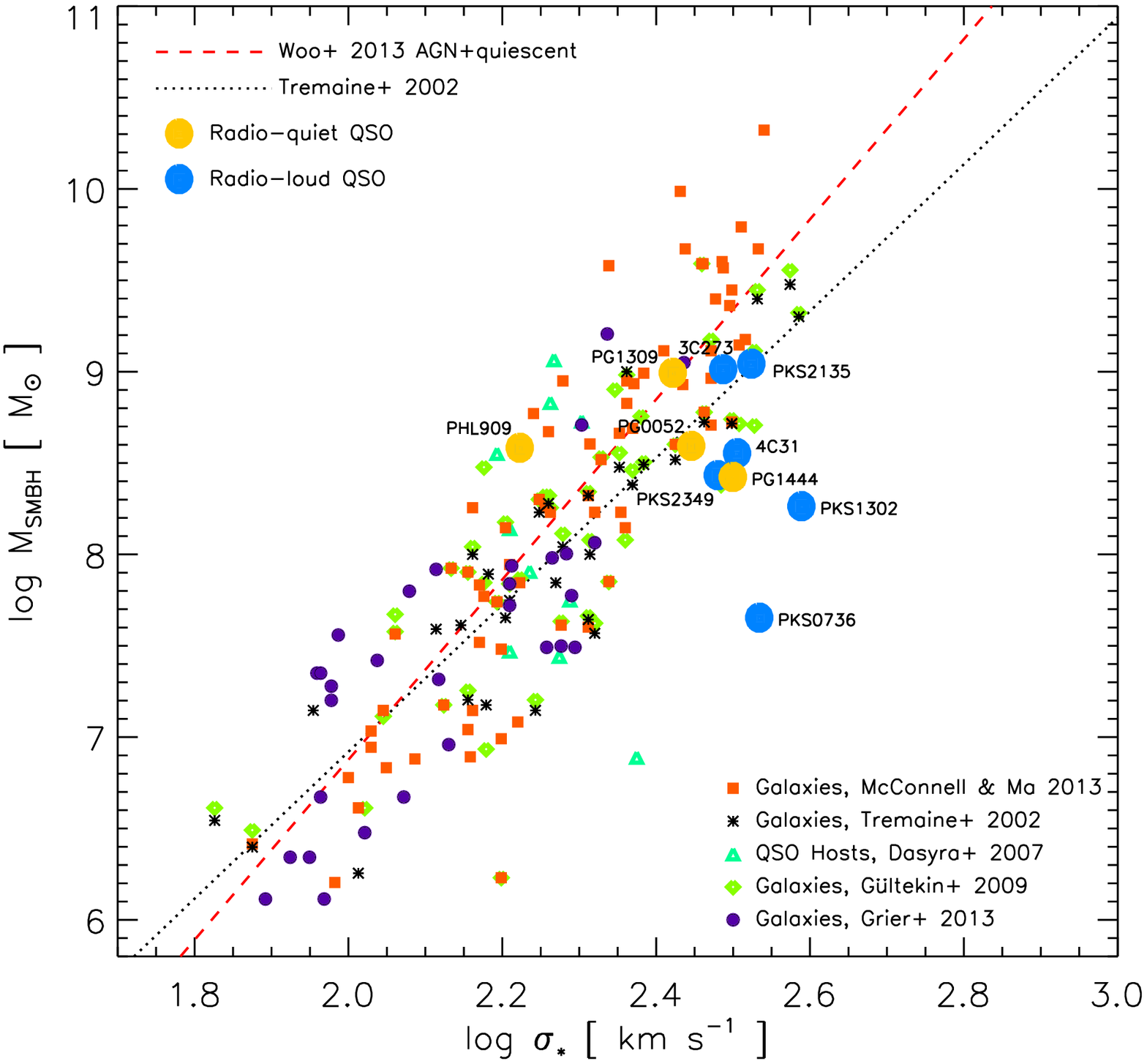}
 \caption{\small $M_{SMBH}$-$\sigma_{*}$ relation.
QSO host galaxies from this work are shown as
large circles (radio-loud are blue, radio-quiet are yellow). 
We also plot the data obtained for quiescent galaxies by \citet{Tremaine+02} --black stars--
and by \citet{McConnellMa13} --orange squares--,
%for AGN by \citet{Shields+03} --orange squares;  in this case, $\sigma_{*}$ was derived from the broad [\ion{O}{iii}]~$\lambda$5007 emission line--, 
for radio-quiet PG QSO hosts by \citet{Dasyra+07} --cyan triangles--, for
galaxies by \citet{Gultekin+09} --green diamonds--, and for galaxies by \citet{Grier+13} --purple circles--.
The  black dotted line is the relation derived by \citet{Tremaine+02}, and the red dashed line is the recent relation provided by \citet{Woo+13} considering both quiescent galaxies and AGN.
\label{mbhs_fig} }
\end{figure*}

\begin{figure*}[h!]
\centering 
\includegraphics[scale=0.4,angle=0]{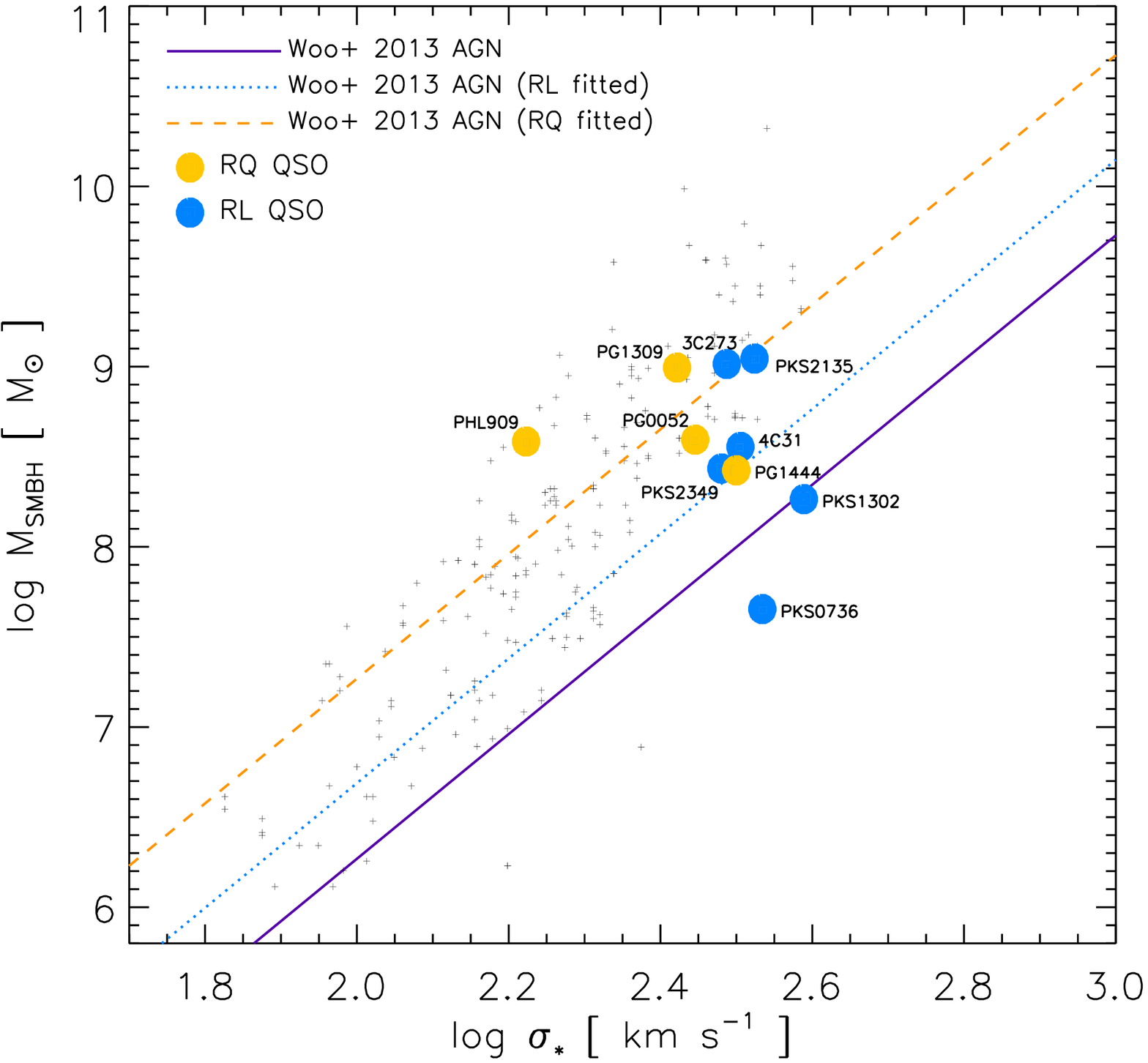} \hspace{0.1cm}
\includegraphics[scale=0.4,angle=0]{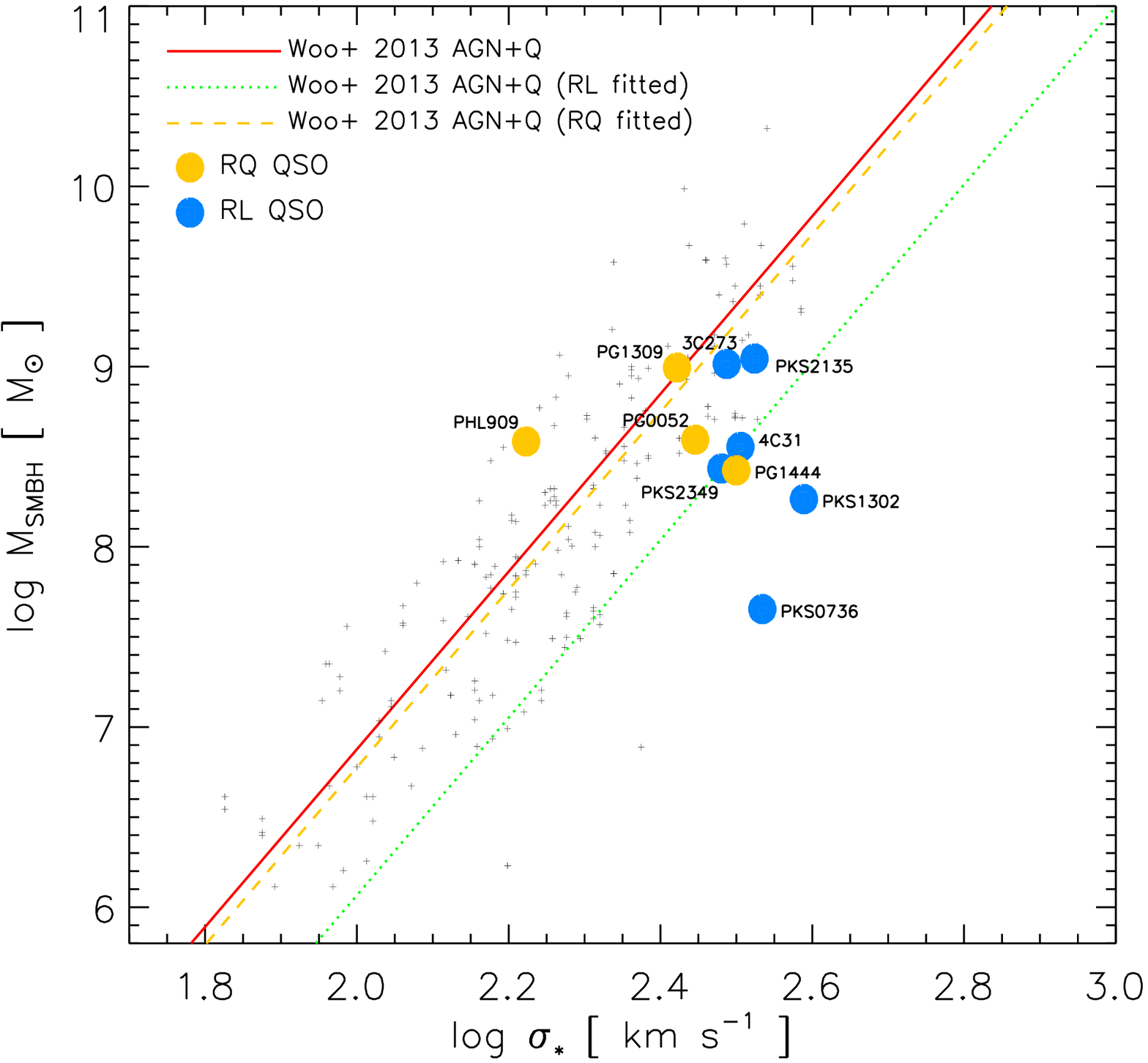}
 \caption{\small Comparison of our QSO sample datapoints with the new $M_{SMBH}$-$\sigma_{*}$ relations provided by \citet{Woo+13}. We consider the cases of only AGN (Left) and AGN + quiescent galaxies (Right) relations. QSO host galaxies from this work are shown as large circles (radio-loud are blue, radio-quiet are yellow). 
%For one object (PKS~2349-014) we plot the $M_{SMBH}$ values derived using both the fitting to the broad
%\Hb\ (blue) and \Ha\ (red hollow circle) emission lines (see text). Both values are connected by a red dotted line. 
The black crosses represent all the previous data included in Fig.~\ref{mbhs_fig}. The original $M_{SMBH}$-$\sigma_{*}$ relationships are plotted with a purple (AGN) or red (AGN+quiescent) continuos line. The best fits to the \citet{Woo+13} $M_{SMBH}$-$\sigma_{*}$ relation fixing the slope are plotted with a cyan dotted line (only for AGN and our RL QSO), an orange dashed line (only for AGN and our RQ QSO), a green dotted line (for AGN + quiescent galaxies and our RL QSO), and a dashed yellow line (for AGN + quiescent galaxies and our RQ QSO).
\label{SMBHSIGFIT} }
\end{figure*}

\subsection{The $M_{SMBH}$-$\sigma_{*}$ relation}

Figure~\ref{mbhs_fig} plots the $M_{SMBH}$-$\sigma_{*}$ relation found for our sample. Here we distinguish between RL (blue circles) and RQ (yellow circles) galaxies. 
%As explained before, for PKS~2349-014 we use the $M_{SMBH}$ values derived 
%both from the fitting to the broad \Hb\ (blue) and \Ha\ (red hollow circle) emission lines. 
%Both values are connected by a dotted red line. 
Figure~\ref{mbhs_fig} includes some data points found in the literature: purple stars plot quiescent galaxies data presented by \citet{Tremaine+02}, 
%orange squares indicate AGN data by \citet{Shields+03} --in this case, $\sigma_{*}$ was derived from the broad [\ion{O}{iii}]~$\lambda$5007 emission line, as we will further discuss below--, 
orange squares indicate galaxy data by \citet{McConnellMa13},
cyan triangles are the radio-quiet PG QSO hosts data by \citet{Dasyra+07}, the green diamonds represent the galaxy data by \citet{Gultekin+09}, and purple circles are galaxy data by \citet{Grier+13}. The red dashed line is the recent relation provided by \citet{Woo+13}, which considers data of both quiescent galaxies and AGN. The standard $M_{SMBH}$-$\sigma_{*}$ relation derived by \citet{Tremaine+02} is also shown with a  dotted black line.
As we see, the majority of our quasars seem to agree with these relations, although one object (PKS~0736+017) lies somehow below the \citet{Tremaine+02} relation.

It is also interesting to note the large variations in $M_{SMBH}$-$\sigma_{*}$ within the same $\sigma_{*}$ interval. For example, both PKS~0736+017 and 3C~273 (two radio-loud quasars) have a very similar velocity dispersion, $\log \sigma_{*}\sim2.52$, while their SMBH masses are differing by almost one and a half orders magnitude, $\log M_{SMBH}=7.65\pm0.06$ and $9.04\pm0.08$, respectively.

Some studies \citep[e.g.][and references within]{Grier+13} do not find a difference between AGNs and quiescent galaxies on the $M_{SMBH}$-$\sigma_{*}$ relation. However, \citet{Woo+13}  investigated the validity of the assumption that quiescent galaxies and active galaxies follow the same relation by updating observational data of both quiescent galaxies and AGN, in the latter case using all available reverberation-mapped AGN. 
Assuming the standard expression
\begin{eqnarray}
\log \Big[\frac{M_{SMBH}}{M_{\odot}} \Big]= \alpha + \beta\,\Big[\frac{\sigma_{*}}{\rm 200\,km\,s^{-1}}\Big],  
\end{eqnarray}
they derived these relations for the cases of (a) only AGN and (b) AGN + quiescent galaxies. They found that the $M_{SMBH}$-$\sigma_{*}$ of active galaxies (for which $\alpha=7.31\pm0.15$,  $\beta=3.46\pm0.61$, and the intrinsic scatter is  $\epsilon_0=0.41\pm0.05$) appears to be shallower than that of AGN+quiescent galaxies (for which $\alpha=8.36\pm0.05$,  $\beta=4.93\pm0.28$, and the intrinsic scatter is  $\epsilon_0=0.43\pm0.04$). However, \citet{Woo+13} also state that  the two relations are actually consistent with one another, as the two different relations are consistent with the same intrinsic $M_{SMBH}$-$\sigma_{*}$  relation due to selection effects.

\begin{figure*}[t!]
\centering 
\includegraphics[scale=0.35,angle=0]{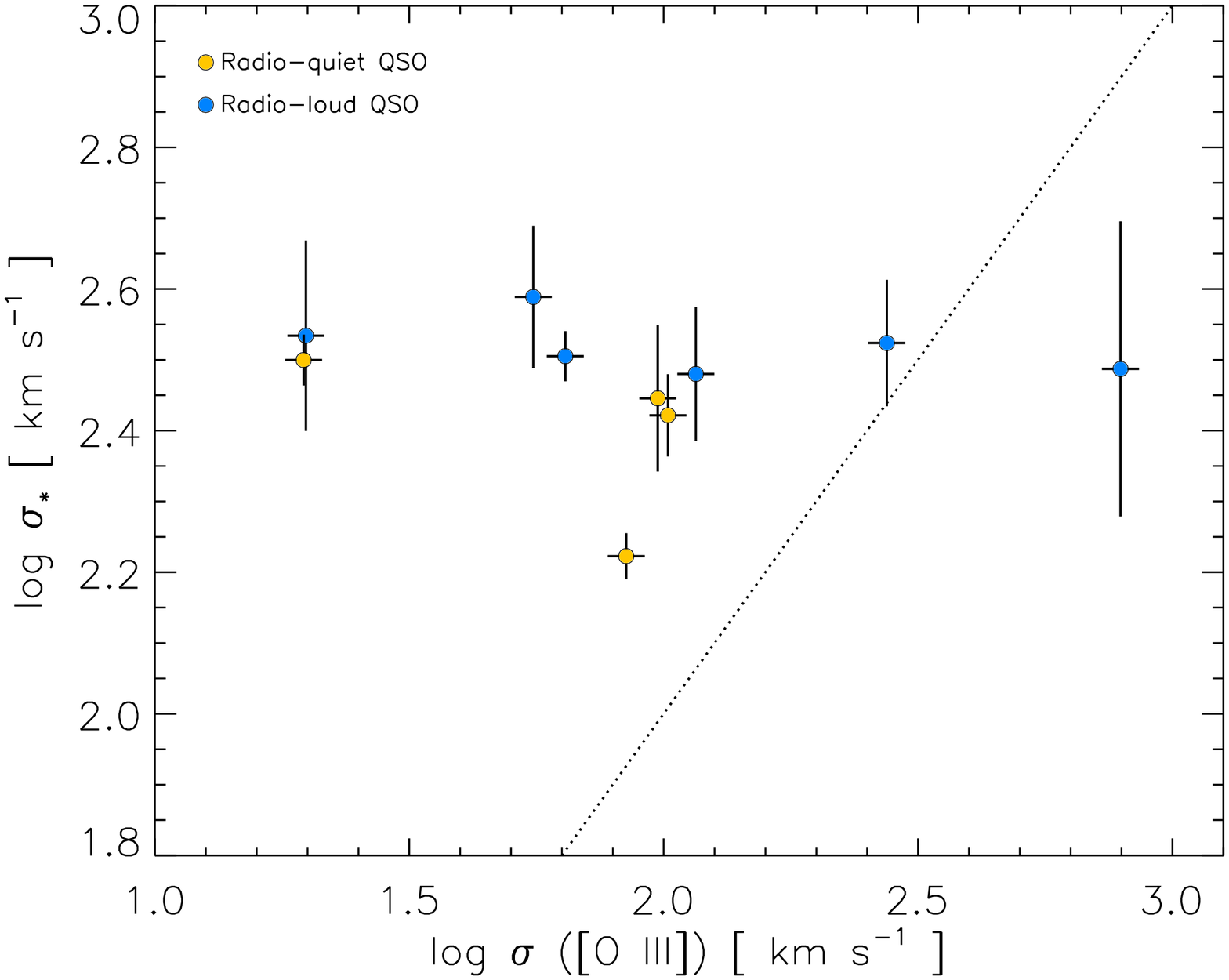} \hspace{0.4cm}
\includegraphics[scale=0.35,angle=0]{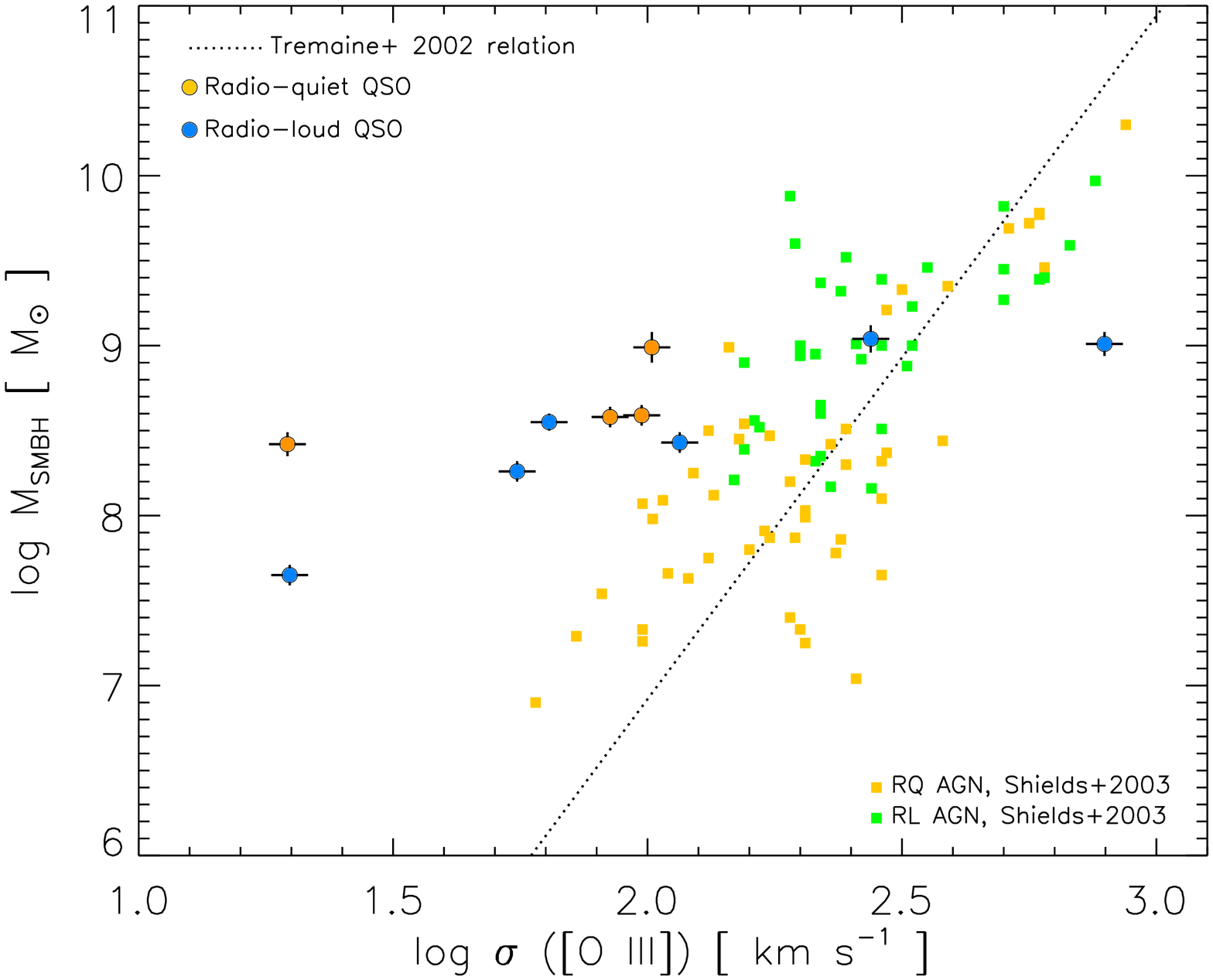}
 \caption{\small (Left) Comparison between the stellar velocity dispersion, $\sigma_{\star}$, given by \citet{WolfSheinis08}, with the $\sigma$ of the broad [\ion{O}{iii}]~$\lambda$5007 emission line derived in our fitting.  The dotted line represents $x=y$. (Right) The  $M_{SMBH}$-$\sigma$ relation using the $\sigma$ of the narrow [\ion{O}{iii}]~$\lambda$5007. AGN data points by \citet{Shields+03} are plotted with orange squares. In both panels radio-loud QSO are blue circles, radio-quiet QSO are yellow circles. 
\label{FWHMO3} }
\end{figure*}

\begin{table}[h!]
  \caption{\small  Results of the fittings of our data sample, subdivided in RL and RQ QSO, when compared with the  $M_{SMBH}$-$\sigma_{*}$ relationships given by \citet{Woo+13} for AGN and AGN + quiescent galaxies.   \label{WOOTABLE} }
\smallskip
  \scriptsize
  \centering
  \begin{tabular}{l@{\hspace{10pt}}     c@{\hspace{8pt}}c@{\hspace{8pt}}c  c@{\hspace{5pt}}  c@{\hspace{8pt}}c@{\hspace{8pt}}c }
\noalign{\smallskip}
\tableline
\noalign{\smallskip}
                &       \multicolumn{3}{c}{W13 AGN}  & &   \multicolumn{3}{c}{W13 AGN+Q}      \\
\cline{2-4} \cline{6-8}
\noalign{\smallskip}
  
           &    $\alpha$     &   $\beta$    & $\epsilon_0$ &&   $\alpha$     &   $\beta$    & $\epsilon_0$ \\
                 
\noalign{\smallskip}

\tableline

\noalign{\smallskip}				      

W13 original                       &   7.31  &  3.46  &  0.657 &&     8.36 &   4.93  &  0.870     \\
W13 fitted to RL                 &   7.73  &  3.46  &  0.218 &&     7.55 &   4.93  &  0.450    \\

\noalign{\smallskip}
\tableline
\noalign{\smallskip}	
 
W13 original                    &   7.31  &  3.46  &  1.144    &&    8.36 &   4.93  &  0.603     \\
W13 fitted to RQ             &   8.31  &  3.46  &  0.216     &&     8.26 &   4.93  &  0.511    \\

\noalign{\smallskip}
\tableline
 \end{tabular}
\end{table}

Figure~\ref{SMBHSIGFIT} compares our QSO datapoints with the \citet{Woo+13} relationships. The left panel considers their $M_{SMBH}$-$\sigma_{*}$ relation for AGN only (purple solid line), whist the right shows  their  $M_{SMBH}$-$\sigma_{*}$ relation for AGN and quiescent galaxies (red solid line). We distinguish between radio-loud quasars (RL QSO, blue circles) and radio-quiet quasars (RQ QSO, yellow circles). 
We now investigate which of these relations fits better to our RL and RQ QSO data when we fixed the slope, $\beta$, and leave the zero point, $\alpha$, free. We search for the $\alpha$ parameter which provides the lowest dispersion between the observational points and these modified relationships for each case, and derive also the intrinsic scatter. Table~\ref{WOOTABLE} compiles our results, which are also plotted in Fig,~\ref{SMBHSIGFIT}.
Interestingly, the $M_{SMBH}$-$\sigma_{*}$ relation for AGN and quiescent galaxies provided by \citet{Woo+13}  matches well our RQ QSO sample (both lines almost overlap in Fig,~\ref{SMBHSIGFIT}, as the best fit has $\alpha$=8.26, being the scatter $\epsilon_0=0.511$), while the RL QSO sample is in better agreement with the relation for only AGN reported by these authors. In this case, we derive that an $\alpha$=7.73 provides the best match between our observational points and the $M_{SMBH}$-$\sigma_{*}$ relation for only AGN (cyan dotted line).

\begin{figure*}[t!]
%\centering 
\includegraphics[scale=0.376,angle=0]{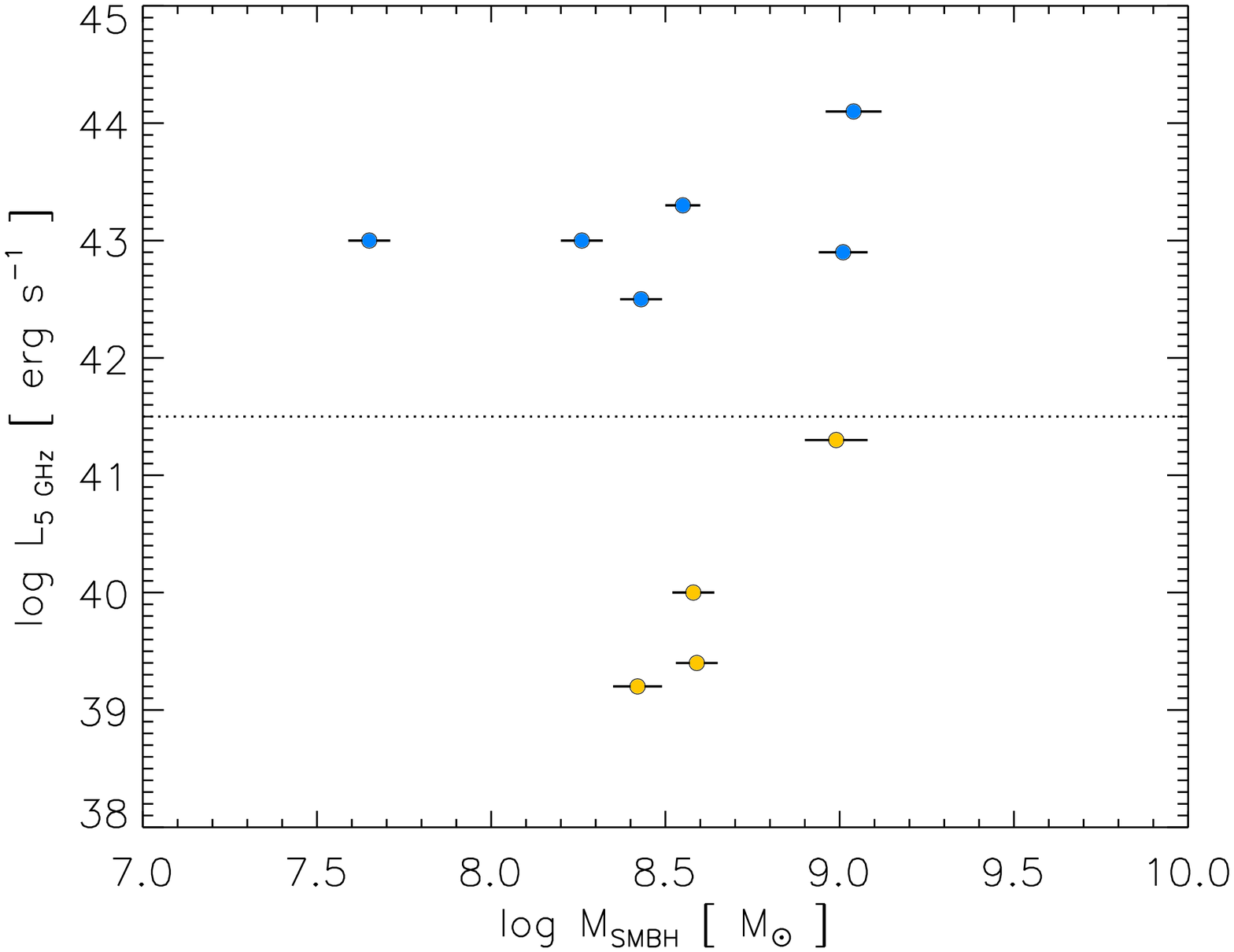}
\includegraphics[scale=0.376,angle=0]{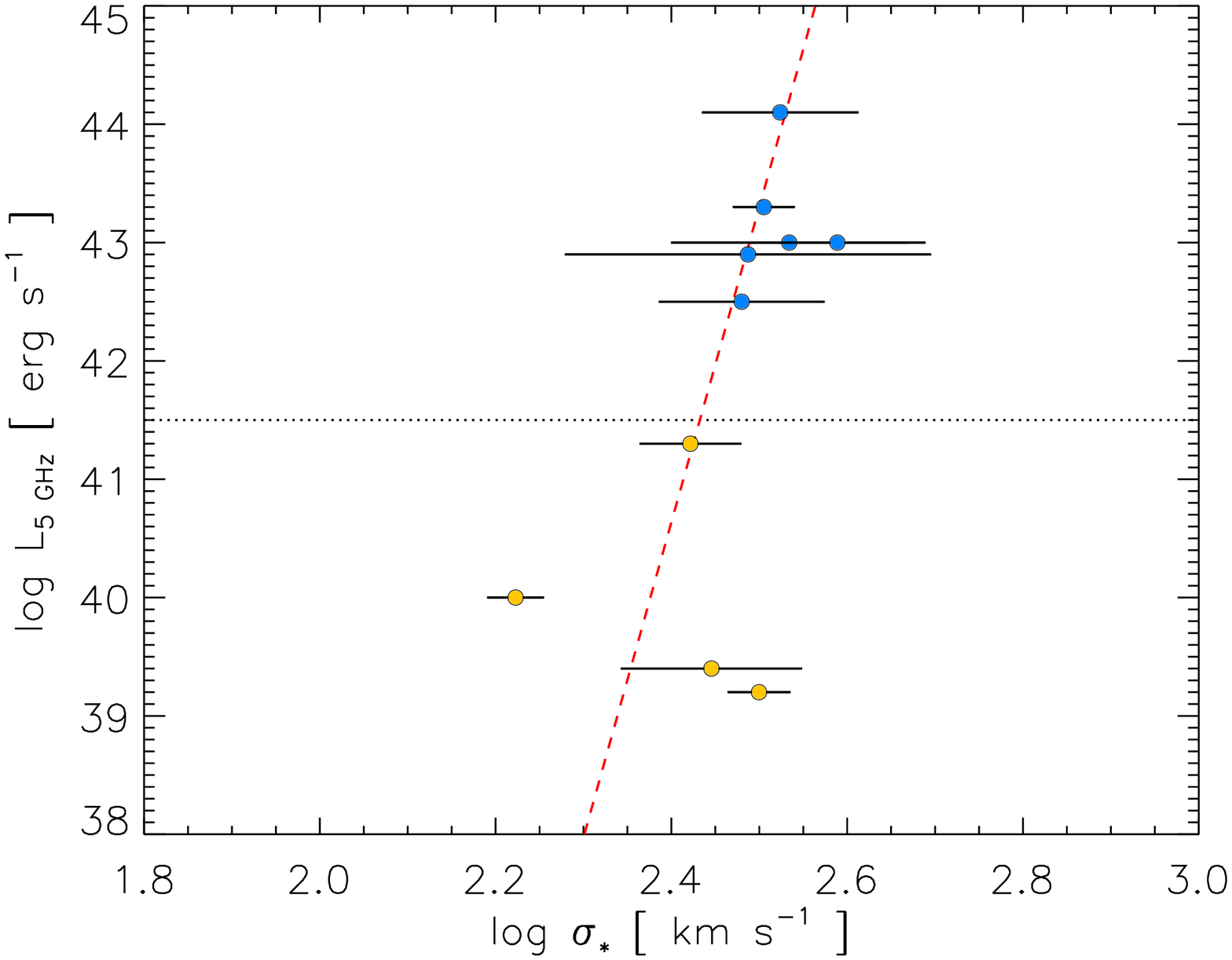}
 \caption{\small (Left panel) Comparison between the mass derived for the SMBH and the  5 GHz radio-luminosity of the host galaxy, as given by \citet{Wold+10}. (Right panel) Comparison between the stellar velocity dispersion of the galaxy hosting a QSO --given by  \citet{WolfSheinis08}-- and its 5 GHz radio-luminosity --given by \citet{Wold+10}--. In both panels, radio-loud QSO are blue circles, radio-quiet QSO are yellow circles. We include the limit between RL and RQ quasars as given by $L_{\rm 5\,GHz} \gtrsim 10^{41.5}$\,erg\,s$^{-1}$ (Wold et al. 2010). Red dashed line in right panel represents a linear fit to our data
\label{massL5GHz} }
\end{figure*}

\subsection{Comparing stellar velocity dispersions with FWHM of [\ion{O}{iii}]}

As pointed out in the introduction, the FWHM of the %broad
 [\ion{O}{iii}]~$\lambda$5007 emission line has been used by some authors as a surrogate for stellar velocity dispersions \citep{NelsonWhittle96}. In particular, \citet{Shields+03} used $\sigma$([\ion{O}{iii}]~$\lambda$5007) and not $\sigma_{*}$ to explore the $M_{SMBH}$-$\sigma$ relation, finding a relatively large scatter. 

We have compared the stellar velocity dispersions estimated for our sample quasars by  \citet{WolfSheinis08} with the velocity dispersion of the (narrow) %broad 
[\ion{O}{iii}]~$\lambda$5007 emission line derived from our fits (see Table~\ref{gaussianfits}). Left panel of Fig.~\ref{FWHMO3} compares both velocity dispersions. 
%As we see, only objects showing low [\ion{O}{iii}]~$\lambda$5007 velocity dispersions, $\log (\sigma$([\ion{O}{iii}]~$\lambda$5007))$\lesssim2.7$, seem to have a good agreement with $\log (\sigma_{*})$.
From this Figure, it is evident that the assumption of using the velocity dispersion of the narrow  [\ion{O}{iii}]~$\lambda$5007 emission line as surrogate for $\sigma_{*}$ is not satisfied. Actually, $\sigma$([\ion{O}{iii}]~$\lambda$5007) seems to be completely independent of $\sigma_{*}$, particularly for low values.
%For all quasars showing higher $\sigma$[\ion{O}{iii}]~$\lambda$5007 than that, the corresponding $\sigma_{*}$ 
% velocity dispersions that this value, the derived  $\sigma$[\ion{O}{iii}]~$\lambda$5007 is considerably lower than the adopted $\sigma_{*}$.

We further investigate this issue in right panel of Fig.~\ref{FWHMO3}, where we plot the $M_{SMBH}$-$\sigma$ relation using  $\sigma$([\ion{O}{iii}]~$\lambda$5007) instead of $\sigma_{*}$. This figure includes the data points compiled by \citet{Shields+03} --which are subdivided in RQ and RL AGN--, as well as the \citet{Tremaine+02} relation. We do not see any correlation at all %see a relatively poor correlation 
between the position of our data with the position of the other objects and the relation. Indeed, the \citet{Tremaine+02} relation is valid for only one object. Again, the situation is particularly poor 
%at the high-end of $\sigma$, when we observe a large scatter. 
 at the low-end of $\sigma$.
We therefore conclude that $\sigma$([\ion{O}{iii}]~$\lambda$5007) should not be used as a surrogate for stellar velocity dispersions for high-mass, high-luminosity quasars.

\subsection{Comparing QSO radio-luminosities\\ with SMBH masses}

\citet{NelsonWhittle96} reported a rather tight correlation between
$\sigma_{*}$ and radio luminosity in Seyfert galaxies, linking
directly observable AGN properties to host galaxy properties. Given
the subsequent \MSMBH --$\sigma_{*}$ correlation, one would expect to
see a similar correlation between \MSMBH\ and radio luminosity. Such
a correlation was reported by \citet*{Franceschini+98} for a small
sample of 8 nearby early-type galaxies with black hole masses
determined from stellar and gas dynamical studies. Later work with
more galaxies, both quiescent and active, have shown much larger
scatter or no correlation at all
\citep{Laor00,Ho02,Snellen+03,Woo+05}. In fact, a bimodality in radio
loudness as a function of $M_{SMBH}$ has been suggested \citep{Laor00}.

Left panel of Fig.~\ref{massL5GHz} compares the 5\,GHz radio-continuum luminosity, $L_{\rm 5\,GHz}$, of the quasar host galaxy with the derived SMBH masses. The radio data have been extracted from \citet{Wold+10}, who estimated $L_{\rm 5\,GHz}$ from previous literature data listed in the NED assuming a spectral index $\alpha=-5$.  The limit between RL and RQ quasars --as given by $L_{\rm 5\,GHz} \gtrsim 10^{41.5}$\,erg\,s$^{-1}$ following \citet{Wold+10}-- is shown in Fig.~\ref{massL5GHz} with a dotted line. Following this figure, we see that RL QSO are found in our entire mass range, including in the low-mass limit --PKS\,0736+017, which has $\log (M_{\rm SMBH}/M_{\odot}$) = 7.67$\pm$0.04--.
%However, following the dataset provided by \citet{Shields+03}, RL AGN hosts are found only when $M_{\rm SMBH}$ is greater than $\sim10^8\,M_{\odot}$.
As shown in the left panel of Fig.~\ref{massL5GHz} a correlation between radio loudness and $M_{SMBH}$ is not found in this mass range.

\begin{figure}[h!]
\centering
\includegraphics[width=\linewidth]{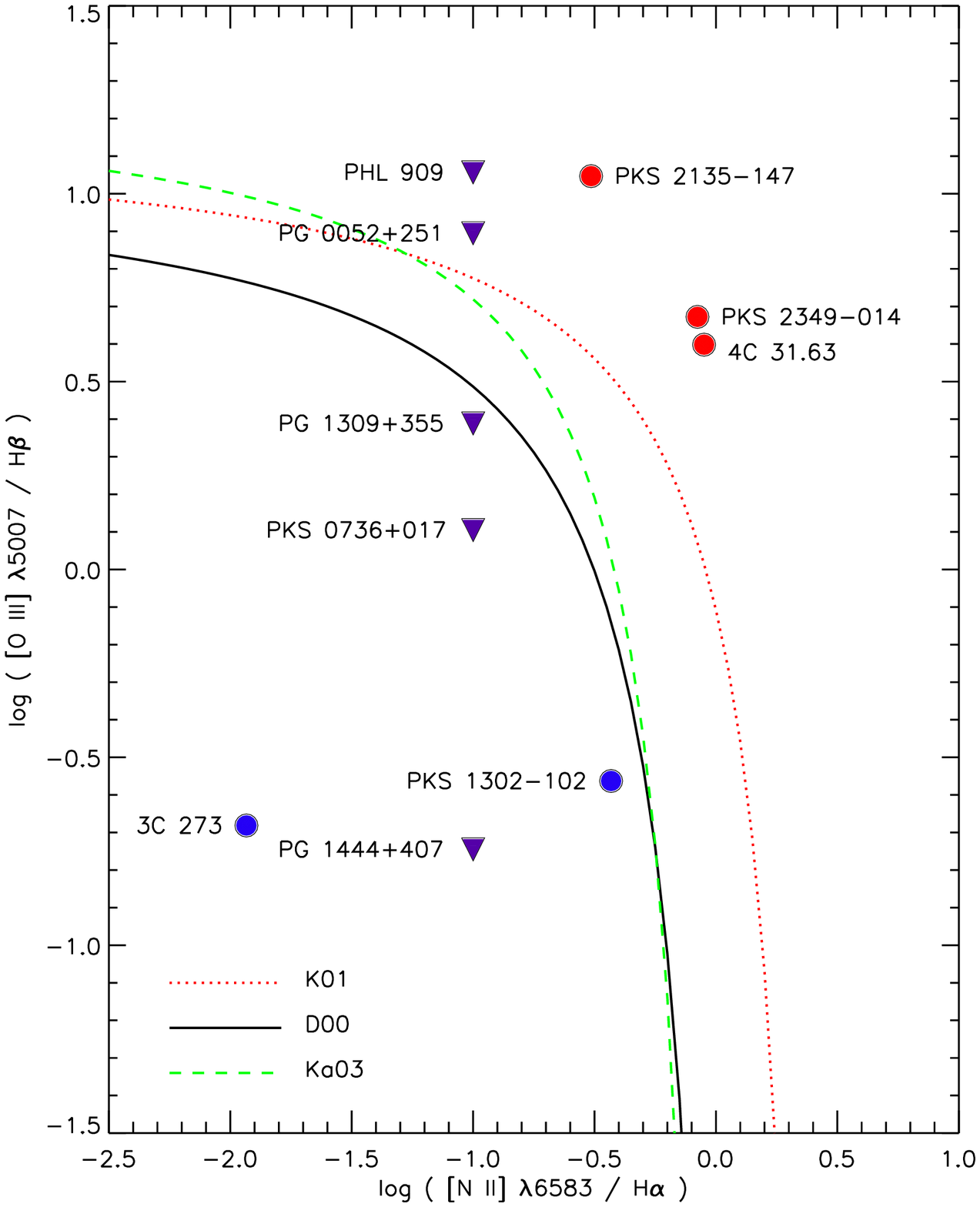}   
\caption{[\ion{O}{iii}] $\lambda$5007/H$\beta$ versus [\ion{N}{ii}] $\lambda$6583/H$\alpha$ diagnostic diagram for the ionized emission coming from the narrow components of our fits.  The loci of the \HII\ regions as proposed by 
\citet{Do00}, %Dopita et al. (2000), 
black continuous line (D00), \citet{KD01}, %Kewley et al. (2001), 
red dotted line (K01), and the empirical relation provided by \citet{Kauffmann+03b}, dashed green line (Ka03), are also plotted. 
Circles indicate quasars for which both [\ion{O}{iii}] $\lambda$5007/H$\beta$ and [\ion{N}{ii}] $\lambda$6583/H$\alpha$ have been measured.
In this case, only the narrow components of  3C~273 and PKS~1302-102 (blue circles) are expected to come from photoionization by massive stars, but an important shock contribution is expected in 4C~31.63, PKS~2135-147, and PK~2349-014 (red circles). However, the position of 3C~273 is too far for what expected from \HII\ regions.
%Green stars show the position of objects for which \mbox{[\ion{N}{ii}]~$\lambda$6583/H$\alpha$} 
%has been estimated assuming the oxygen abundance provided by \citet{P01a,P01b} and the calibrations given by \citet{PP04}. 
Purple triangles indicate the [\ion{O}{iii}] $\lambda$5007/H$\beta$ position of objects for which  [\ion{N}{ii}] $\lambda$6583/H$\alpha$  is unknown. 
%Note that the position of PG~0052+251 and PHL~909 is almost identical.
}
\label{diagnostic}
\end{figure}

Conversely, the right panel of of Fig.~\ref{massL5GHz} compares $L_{\rm 5\,GHz}$,  with the stellar velocity dispersion of the host galaxy. In this case, we observe a trend that galaxies with larger stellar velocity dispersion have larger 5\,GHz radio-continuum luminosities. The red dashed line in this panel represents a linear fit to our data, in the form $\log (L_{\rm 5\,GHz}) = A + B \log (\sigma_{*})$, for which we derive $A=-23.3\pm19.0$, $B=26.6\pm7.7$. The correlation coefficient is $r$=0.555.

%{\it Include more objects coming from other samples?}

%Leipski et al. 2006A&A...455..161L (PG 0052+251, 4.8 GHz, VLA)
%Condon et al. 1998AJ....115.1693C (PHL 909, 1.4 GHz VLA)

\begin{deluxetable}{c    cc    cc c      cccc}
\tabletypesize{\scriptsize}
%\rotate
\tablecaption{Oxygen abundances of the host galaxies using the narrow lines obtained in the fit of our spectra.   \label{oxygen} }
\tablewidth{0pt}
\tablehead{       Object  &      [\ion{N}{ii}]~$\lambda$6583/\Ha    &  [\ion{O}{iii}]~$\lambda$5007/\Hb       &    $N2$  &  $O3N2$ &  BPT  &       
  \multicolumn{4}{c}{12 + log (O/H) } \\ 
   \noalign{\smallskip}
\cline{7-10}
\noalign{\smallskip}
%        & &   &     && &        PP04   ($N2$)       &   PP04 ($O3N2$)   &  S-B98$_{\rm PP04}$ & Adopted 
        & &   &     && &        PP04   ($N2$)       &   PP04 ($O3N2$)   &  S-B98 & Adopted 
}
\startdata
PG~0052+251 & \nodata  & 7.85  &   \nodata & \nodata  & AGN? & \nodata & \nodata & \nodata & \nodata \\
PHL~909     & \nodata  & 11.4  &   \nodata & \nodata  & AGN? & \nodata & \nodata & \nodata & \nodata \\

PKS~0736+017 & \nodata & 1.27  & \nodata  &\nodata & \HII? &\nodata&\nodata&\nodata&\nodata\\

  3C~273     &  0.0116 &  0.208 & -1.936 &   1.254 & \HII? &  7.80 &  8.33   & 8.34 &   8.3$^a$ \\
PKS~1302-102 &  0.370  &  2.74 & -0.870 &  0.907 & \HII &  8.65 &  8.44   & 8.43 &   8.55 $\pm$ 0.10\\

PG~1309+355  & \nodata & 3.53   & \nodata & \nodata & \HII? & \nodata & \nodata & \nodata & 8.48?$^b$ \\ 
PG~1444+407  & \nodata & 0.180  & \nodata & \nodata & \HII? & \nodata & \nodata & \nodata &\nodata\\

PKS~2135-147 &  0.306 & 11.1  & -0.514 &  1.560 & AGN   &  8.61 &  8.23  & 8.57 & 8.3 -- 8.6   \\

 4C~31.63    &  0.893 &  3.97   & -0.0491 &  0.648 & AGN  & 8.87 &  8.52  & 8.56 &   8.5 -- 8.7 \\

PKS~2349-014 &  0.839 &  4.70   & -0.0762 &  0.748  & AGN &  8.86 &  8.49   & 8.56 &  8.5 -- 8.7 \\

\enddata
\tablenotetext{a}{The derived value for [\ion{N}{ii}]~$\lambda$6583 is extremely low. Although its position agrees with that expected for \HII\ regions, another ionization mechanism is in action. We then consider the oxygen abundance provided by S-B98 calibration as tentative value for the metallicity of this object. See text for details.} 
%This is a tentative value to the oxygen abundance in 3C~273, as the $\sigma$ of the narrow emission lines are relatively large to those expected in \HII\ regions. See text for details.
\tablenotetext{b}{Tentative value considering  the oxygen abundance given by the empirical calibration provided by \citet{P01a,P01b}, which considers the $R23$ and $P$ parameters.}
\end{deluxetable}

\subsection{Gas-phase metallicity of the host galaxies}

We now use the fitted narrow emission lines to estimate the gas-phase metallicity of the host galaxies. 
Caution must be taken when attempting to do this, as the physical conditions that are responsible of the ionization of the gas are very different in \HII\ regions and AGNs. 
The NLR spectrum is also a function of radiation pressure and, most important, the form of the extreme ultraviolet spectrum. 
Therefore photoionization models need to be considered to quantitatively derive metallicities in AGN hosts as the typical and very easy-to-apply empirical calibrations involving bright emission-lines used for computing metallicities in \HII\ regions cannot be applied \citep[e.g.,][]{Storchi-BergmannPastoriza89,Storchi-Bergmann+98,Hamann+02,GrovesHeckmanKauffman06,Husemann+14, Dopita+14}.

We first check that the nature of the ionization of the narrow lines is massive stars (i.e., the narrow lines are produced in star-forming regions) using the  [\ion{O}{iii}]~$\lambda$5007/H$\beta$ versus [\ion{N}{ii}]~$\lambda$6583/H$\alpha$ diagnostic diagram \citep{BPT81,VO87}. 
This diagram is plotted in Fig.~\ref{diagnostic}, and includes the analytic relations given by \citet{Do00} and \citet{KD01}, %Dopita et al. (2000) and Kewley et al. (2001) 
as well as the empirical relation provided by \citet{Kauffmann+03b}.
The narrow lines derived in 4C~31.63, PKS~2135-147, and PKS~2349-014 (red circles) %, see Table~\ref{gaussianfits}) 
are still too broad to be produced only for photoionization by massive stars, as these objects do not lie in the loci of the \HII\ regions in the diagnostic diagram. 
 Hence, other mechanisms, as doppler broadening, outflows, or jet interactions, are playing an important role in the ionization of the lines.
Thus,  as shown in Fig.~\ref{diagnostic} we expect that photoionization by massive stars is the main excitation mechanism of the narrow lines in 3C~273 and PKS~1302-102 (blue circles). 
However, the faintness of the [\ion{N}{ii}]~$\lambda$6583 line in 3C~273 
moves this datapoint too far for what expected for an \HII\ region, and hence other mechanisms are affecting the ionization of the narrow lines.

Purple triangles in Fig.~\ref{diagnostic}  indicate the [\ion{O}{iii}] $\lambda$5007/H$\beta$ position of objects for which  [\ion{N}{ii}] $\lambda$6583/H$\alpha$  is unknown (i.e., their true $x$-position on this diagram is not known). In the cases of PG~0052+251 and PHL~909, the high [\ion{O}{iii}] $\lambda$5007/H$\beta$ value suggests that these objects do not lie in the loci of \HII\ regions but in the AGN regime 
\citep[if not, their metallicities should be very low, see][]{Sanchez+14b}. On the other hand, the true position of PG~1444+407 --which has the lowest  [\ion{O}{iii}] $\lambda$5007/H$\beta$ ratio in our sample-- in the BPT diagram shown in Fig.~\ref{diagnostic}  may be close to the analytic relations plotted in this figure and therefore consistent with \HII\ regions. The true position of objects PG~1309+355 and PKS~0736+017 in the BPT diagram may be consistent with \HII\ regions (in the case $N2\lesssim-0.8$ and $-0.5$, respectively) or LINERs (low-ionization narrow-emission line regions).

As mentioned above, for PKS~1302-102  the ionization of the narrow lines is massive stars. Therefore we can derive its
%In those objects for which the ionization of the narrow lines is massive stars we compute their 
oxygen abundance -- in units of 12+log(O/H)-- 
assuming the standard empirical calibrations to derive the metallicity in \HII\ regions.
We use the narrow lines of our fits and apply the standard empirical calibrations obtained by \citet{PP04}, %Pettini \& Pagel (2004), 
which consider the $N2=\log$([\ion{N}{ii}]~$\lambda6583$/\Ha) and $O3N2=\log$\{([\ion{O}{iii}]~$\lambda5007$/\Hb) / ([\ion{N}{ii}]~$\lambda6583$/\Ha)\} parameters. 
Table~\ref{oxygen} compiles the results of the oxygen abundances obtained following these calibrations and the values of the $N2$ and $O3N2$ %, $R23$, and $P$ 
parameters when available.
However, we note that caution must be taken when using empirical calibrations to derive the oxygen abundance in galaxies \citep[see][for a recent review about this topic]{LSDK+12}. 
With this, we derive an oxygen abundance of  8.55$\pm$0.10 for PKS~1302-102. This value is the average oxygen abundance provided by these two empirical calibrations.

%The results  of the oxygen abundance obtained for 3C~273 and PKS~1302-102 are quite consistent following the empirical 
%calibrations based on the $N2$ and $O3N2$ parameters and the \citet{PP04} calibrations. 
%We derive an oxygen abundance of  8.62$\pm$0.08 and 8.52$\pm$0.08 for 3C~273 and PKS~1302-102, respectively. 
%
%However, we should note than, in the case of 3C~273, the velocity dispersion of the narrow lines are large 
%(e.g., $\sigma_{\rm [O\,III]~\lambda5007}$=9.452~\AA~$\sim$1365~\kms) in comparison to those typically found in \HII\ regions \
%citep[$\sigma_{NL}\lesssim$300~\kms, see][]{OsterbrockFerland06}, and hence the value we provide for the oxygen abundance 
%of this quasar should be taken as a tentative estimation of its true value. 

The spectrum of PG~1309+355 shows the narrow [\ion{O}{ii}]~$\lambda$3727 emission line. Assuming that the photoionization of the gas is coming from massive stars (as implied by Fig.~\ref{diagnostic}), we can apply
%Additionally, we measure the narrow [\ion{O}{ii}]~$\lambda$3727 emission line (if it is detected in the quasar spectrum) and apply 
the empirical calibrations given by \citet{P01a,P01b} --which consider the $R23$ and $P$ parameters, defined as
$R_3=  ([\textsc{O\,iii}]\,\lambda 4959 + [\textsc{O\,iii}]\,\lambda 5007) / {\rm H\beta}$,
$R_2=  [\textsc{O\,ii}]\,\lambda 3727 / {\rm H\beta}$, $R_{23} =  R_3 + R_2$, and  $P = R_3/R_{23}$-- to get an estimation of the oxygen abundance of the ionized gas. We consider the  \citet{CardelliClaytonMathis89} extinction law and assumed a reddening coefficient of $c$(\Hb)=0.4 to correct the  [\ion{O}{ii}]~$\lambda$3727 flux for reddening. With this, we derive $f$([\ion{O}{ii}]~$\lambda$3727)/$f$(\Hb) = 0.343, $R23$=46.383, $P$=0.722, and 12+log(O/H)=8.48 for this object.

\citet{Storchi-Bergmann+98} computed the oxygen abundances for an artificial grid of line ratios with the photoionization code CLOUDY \citep{Ferland+98} using an empirical AGN spectrum for the ionizing source. \citet{Storchi-Bergmann+98} also provided a calibration based on the ([\ion{O}{iii}] $\lambda\lambda$4959,5007)/H$\beta$ and ([\ion{N}{ii}] $\lambda\lambda$6548,6583)/H$\alpha$ line ratios by fitting a two-dimensional polynomial of second order to their grid of oxygen abundances. We use this calibration to get an estimation of the metallicity in 3C~273, 4C~31.63, PKS~2135-147, and \mbox{PKS~2349-014.} 
%As the \citet{Storchi-Bergmann+98} calibration is based on photoionization models \citep[which systematically overpredict in 0.2 -- 0.4~dex the 
%oxygen abundances derived using the direct \Te-method, see][and references within]{LSDK+12},  we scaled their results into the \citet{PP04} absolute scale applying the equations compiled by \citet{KE08}. 
The results %, \mbox{S-B98$_{\rm PP04}$,} 
are also listed in Table~\ref{oxygen}. In all four cases we found 12+log(O/H) values between 8.3 and 8.6.
As applying the \citet{PP04} $N2$ calibration in AGNs will provide higher oxygen abundances than their true value, their results can be considered upper limits to the real metallicity of these objects. Hence, 4C~31.63, PKS~2135-147, and PKS~2349-014 seem to have oxygen abundances between 8.3 and 8.7. %, i.e., metallicities between 0.5\Zo\ and \Zo. 
For the case of 3C~273, only the \citet{Storchi-Bergmann+98} equation can be used, we provide a tentative value of 12+log(O/H)$\sim$8.3 for its oxygen abundance. Although its real value may be slightly higher, the metallicity of 3C~273 is definitively subsolar.

These are the first attempt of deriving the gas-phase oxygen abundance in these quasars we are aware of. It is challenging to derive the oxygen abundances in active galaxies due to the strong contamination of the broad emission lines created by the AGN activity. However, detailed analyses indicate that most AGN tend to have solar --12+log(O/H)$_{\odot}$=8.69, \citet{Asplund09}-- to supersolar metallicities \citep{Storchi-BergmannPastoriza89,Storchi-Bergmann+98,Hamann+02}, reinforcing the hypothesis that AGN are usually found at high metallicities.
Recently, \citet{Husemann+14} used the narrow emission lines to estimate the oxygen abundances of 11 low redshift ($z<0.2$) quasar host galaxies, all of them ranging between 8.5 and 9.0 (in the $T_e$-based absolute scale).
Conversely, only few galaxies hosting AGN are found to have sub-solar metallicities, having in all cases masses lower than 10$^{10}$\,\Mo\ 
\citep*[e.g.,][]{GrovesHeckmanKauffman06,BarthGreeneHo08,Izotov+07}. %(e.g., Groves, Heckman \&  Kauffmann 2006; Barth, Greene \& Ho 2008; Izotov et al. 2007). 
In particular, out of a sample of $\sim$23000 Seyfert~2 galaxies, \citet{GrovesHeckmanKauffman06} found only $\sim$40 clear clear candidates for AGN with NLR abundances that are below solar. Low-metallicity AGN seem to be extremely rare \citep{IT08,Izotov+10}. 
Hence, it is interesting to note that  the oxygen abundances derived for our quasar sample are in the subsolar regime (metallicities between 0.4\Zo\ and \Zo).
%both the RL quasars 3C~273 and PKS~1302-102, with metallicities 0.85\Zo\ and 0.68\Zo\ respectively, are in the subsolar regime.
%
This may suggest that low-$z$, higher luminosity quasars have higher probability of being found in subsolar metallicity hosts than low-luminosity quasars. This is a reasonable argument, as brighter AGNs are expected to consume more gas, suggesting these AGNs are found in host galaxies with large gas reservoirs, and hence lower metallicity as the gas has not yet been primarily processed into stars.

\subsection{On the nature of the RL and RQ quasars}

Is there a true physical difference between RL and RQ quasars beyond their radio luminosity? %Concerning the SMBH mass, 
\citet{Laor00} found that virtually all of the RL quasars contain SMBH with masses $M_{\rm SMBH}>10^{9}$\Mo, whereas the majority of the RQ quasars host a SMBH with masses $M_{\rm SMBH}<3\times10^{8}$\Mo.
\citet{Lacy+01} established that a continuous variation of radio luminosity with BH mass existed and that the radio power also depends on the accretion rate relative to the Eddington limit.
Later, \citet{McLureDunlop02} found a similar result, quantifying that the median BH mass of the RL quasars are a factor 2 larger than that of their RQ counterparts.
However, \citet{Ho02} and \citet{JarvisMcLure02} found no clear relationship between radio power and black hole mass. 
\citet{McLureJarvis04} used a sample of more than 6000 quasars from SDSS to find that RL quasars are found to harbour systematically more massive black holes than are the RQ quasars. These authors also reported a strong correlation between radio luminosity and SMBH mass.

RQ quasars seem to be associated with mixed morphologies including spiral galaxies, as they tend to have lower $\sigma_{*}$  than RL quasars \citep{WolfSheinis08}.
Recently \citet{Kimball+11} used the Expanded Very Large Array (EVLA) to  observe at 6~GHz 179 quasars in the redshift range $0.2<z<0.3$,  detecting 97\% of them. 
These authors concluded that the radio luminosity function is consistent with the hypothesis that the radio emission in RL quasars is powered primary by AGNs, while the radio emission in RQ quasars is powered primarily by star-formation in their host galaxies (i.e., their host galaxies are spiral galaxies). 

Although we consider only the study of 10 objects, our analysis appears to agree with \citet{Kimball+11} conclusions, as it suggests that the $M_{SMBH}$-$\sigma_{*}$  relation of RQ quasars matches well with that found in AGN+quiescent galaxies (i.e., still star-forming galaxies) and that the $M_{SMBH}$-$\sigma_{*}$  relation of RL quasars follows that found only in AGN. 
%(i.e., non-star-forming galaxies). 
%
Indeed, stellar population synthesis analyses of quasars have shown that the radio-luminosity of RQ quasars is consistent with that found in quiescent star-forming galaxies
 \citep{Wold+10}. 
Following our analysis, PG~1309+355, a RQ quasar, also seems to show a relatively important star-formation activity. 
%However, even in RL quasars we may expect some contribution of star-formation, as it seems the case of 3C~273.
The importance of the star-formation activity in the quasar host galaxy may be just another consequence of whatever physical phenomenon is responsible for its RL or RQ nature. 

As presented in \citet{WolfSheinis08}, we hypothesize that RQ quasars may be formed by a
more secular merger scenario than RL quasars, in that many RL quasars show signs of recent or
ongoing interactions including major merger events \citep{Zakamska+06,Villar-Martin+11b,Villar-Martin+12,Bessiere+12,Villar-Martin+13}.
Low surface brightness fine structure indicate of past merger events can easily be missed if too shallow images are used \citep[e.g.][and references within]{Bennert+08}.
However 
%In particular, 
\citet{RamosAlmeida+12} recently found that $\sim$95\% of their radio-loud AGN host show interaction signatures, and suggested 
that radio-loud quasars are also triggered by interactions.
In the case or RQ quasars, the merger scenario
appears much more rare and may point to RL quasars likely occurring in harassment and
minor merger events (whose features can be difficult to detect even in deep images).  
If this is the case, it is also possible that the major
mergers are more likely to lead to binary black holes \citep{Begelman+80,Roos81,Khan+11}
as the BH mass scales with
galaxy mass  while RQ quasars events are characterize by minor mergers in which only a
single BH survives due to the mass disparity.  
The merger of two SMBH seems to be the most plausible scenario to explain the X-shaped morphology 
observed in some bright radio-galaxies \citep{MerrittEkers02,Hodges-Kluck+10b,GongLeZhang11,Mezcua+11,Mezcua+12}.
The scenario where the AGNÕs radio
luminosity is fueled by binary BH activity is consistent with our host galaxy
observations \citep[RL hosts tend to be more massive than RQ hosts, ][]{WolfSheinis08},
% \citep{WolfSheinis08}, 
with the observational result that RL quasars harbour systematically more massive black holes than the RQ quasars 
\citep{McLureDunlop02,McLureJarvis04}, 
as well as the observation by  \citet{Kimball+11} that show the radio
luminosity of the stellar component of the galaxy is a much smaller contribution to
the total radio luminosity on RL quasars.  These ideas require further exploration.

% SECTION 5: CONCLUSIONS  -----------------------------------------------------------------------------------------------

\section{Conclusions}

We have analyzed the emission line profiles of a sample of 10 quasars and derive their super-massive black hole masses following the virial method. 
Our sample consists in 6 radio-loud quasars and 4 radio-quiet quasars. We carefully fit a broad and narrow Gaussian component for each bright emission line in both the \Hb\ (10 objects) and \Ha\ regions (5 objects). Our main conclusions are:
\begin{enumerate}
\item[1.] We find a very good agreement of the derived SMBH masses, $M_{\rm SMBH}$, using the fitted broad \Hb\ and \Ha\ emission lines. 
We compare our  $M_{\rm SMBH}$ results with those found by previous studies using the reverberation mapping technique, the virial method and X-ray data, as well as those derived using the continuum luminosity at 5100~\AA.

\item[2.] We  study the relationship between the $M_{\rm SMBH}$ of the quasar and the stellar velocity dispersion, $\sigma_{*}$, of the host galaxy. 
We use the measured  $M_{\rm SMBH}$ and  $\sigma_{*}$ to investigate the $M_{\rm SMBH}$ -- $\sigma_{*}$ relation for both the radio-loud and radio-quiet subsamples. Besides the scatter and the low number of objects in our sample, we find a good agreement  between radio-quiet quasars and AGN+quiescent galaxies and between radio-loud quasars and AGN. The intercept in the latter case is 0.5~dex lower than in the first case.
Our results support the idea that both the star-formation phenomena in the host galaxy and the RL or RQ nature of the quasar are a consequence of a different merger scenario,  possibly being RL quasars triggered by a major merger event and RQ quasars ignited by a minor merger event.  Major mergers are more likely to lead to binary black holes, which inter-relationship may drastically enhance the radio activity.

\item[3.] Our analysis does not support the hypothesis of using $\sigma$([\ion{O}{iii}]~$\lambda$5007) as a surrogate for stellar velocity dispersions in high-mass, high-luminosity quasars.

\item[4.] We also investigate the relationship between the 5\,GHz radio-continuum luminosity, $L_{\rm 5\,GHz}$, of the quasar host galaxy with both $M_{\rm SMBH}$ and $\sigma_{*}$. We do not find any correlation between $L_{\rm 5\,GHz}$ and $M_{\rm SMBH}$, although we observe a trend that galaxies with larger stellar velocity dispersions have larger 5\,GHz radio-continuum luminosities. 

\item[5.] Using the results of our fitting for the narrow emission lines of [\ion{O}{iii}]~$\lambda$5007 and [\ion{N}{ii}]~$\lambda$6583 
we estimate the gas-phase oxygen abundance of six quasars, being sub-solar in all cases. These are the first attempt of deriving the gas-phase oxygen abundance in these quasars we know of. We suggest that low-$z$, higher luminosity quasars have higher probability of being found in subsolar metallicity hosts than low-luminosity quasars, as the latter have probably processed more gas into stars than high-luminosity quasars. %that are expected to still have large gas reservoirs.
% a the radio-loud quasars 3C~273 and PKS~1302-102, being sub-solar in both cases.
\end{enumerate}

%% If you wish to include an acknowledgments section in your paper,
%% separate it off from the body of the text using the \acknowledgments
%% command.

%% Included in this acknowledgments section are examples of the
%% AASTeX hypertext markup commands. Use \url without the optional [HREF]
%% argument when you want to print the url directly in the text. Otherwise,
%% use either \url or \anchor, with the HREF as the first argument and the
%% text to be printed in the second.

\acknowledgments

We are indebted to Marsha Wolf for her discussions and data reduction comments on the \mbox{$\sigma_{*}$ .}
We are grateful to Scott Croom, Paul Martini, C\'esar Esteban and Mike Dopita for providing their very valuable comments to this work 
and to Joe Miller who provided the original impetus for this research.
We also thank Louis Ho for the early discussions motivating this work.
This research has made use of the SAO/NASA Astrophysics Data System Bibliographic Services (ADS).
This research has made extensive use of the NASA/IPAC Extragalactic 
      Database (NED) which is operated by the Jet Propulsion Laboratory, 
      Caltech, under contract with the National Aeronautics and Space 
      Administration.

\bibliographystyle{apj}
{\scriptsize
\bibliography{/WORK/ARTICULOS/bibdesk_angelrls}
}

%% To help institutions obtain information on the effectiveness of their
%% telescopes, the AAS Journals has created a group of keywords for telescope
%% facilities. A common set of keywords will make these types of searches
%% significantly easier and more accurate. In addition, they will also be
%% useful in linking papers together which utilize the same telescopes
%% within the framework of the National Virtual Observatory.
%% See the AASTeX Web site at http://aastex.aas.org/
%% for information on obtaining the facility keywords.

%% After the acknowledgments section, use the following syntax and the
%% \facility{} macro to list the keywords of facilities used in the research
%% for the paper.  Each keyword will be checked against the master list during
%% copy editing.  Individual instruments or configurations can be provided 
%% in parentheses, after the keyword, but they will not be verified.

%{\it Facilities:} \facility{Nickel}, \facility{HST (STIS)}, \facility{CXO (ASIS)}.

%% Appendix material should be preceded with a single \appendix command.
%% There should be a \section command for each appendix. Mark appendix
%% subsections with the same markup you use in the main body of the paper.

%% Each Appendix (indicated with \section) will be lettered A, B, C, etc.
%% The equation counter will reset when it encounters the \appendix
%% command and will number appendix equations (A1), (A2), etc.

%\appendix

%\section{Appendix material}

\section{Appendix}

\begin{figure*}
\centering
\includegraphics[scale=0.6,angle=90]{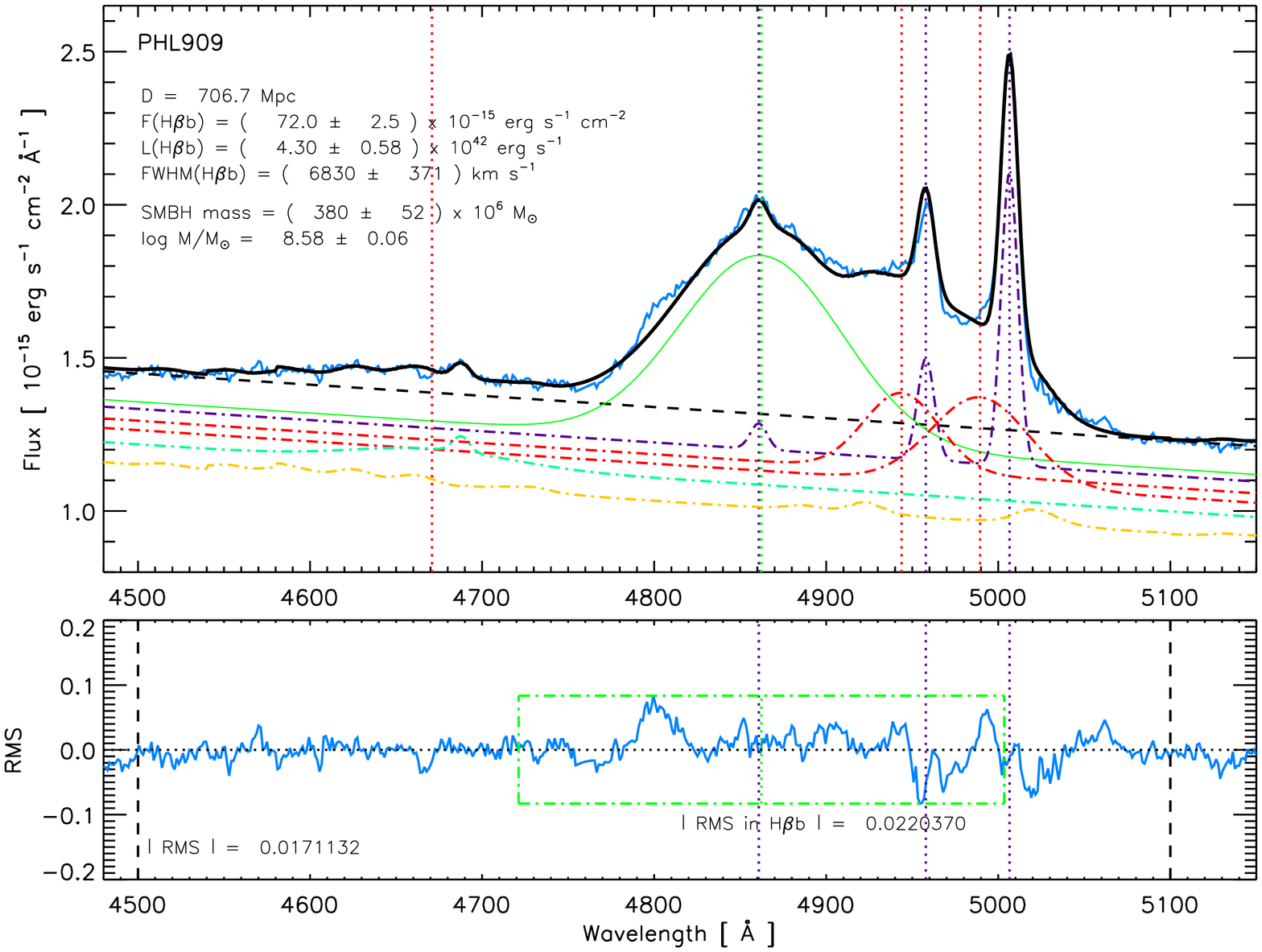}
\includegraphics[scale=0.6,angle=90]{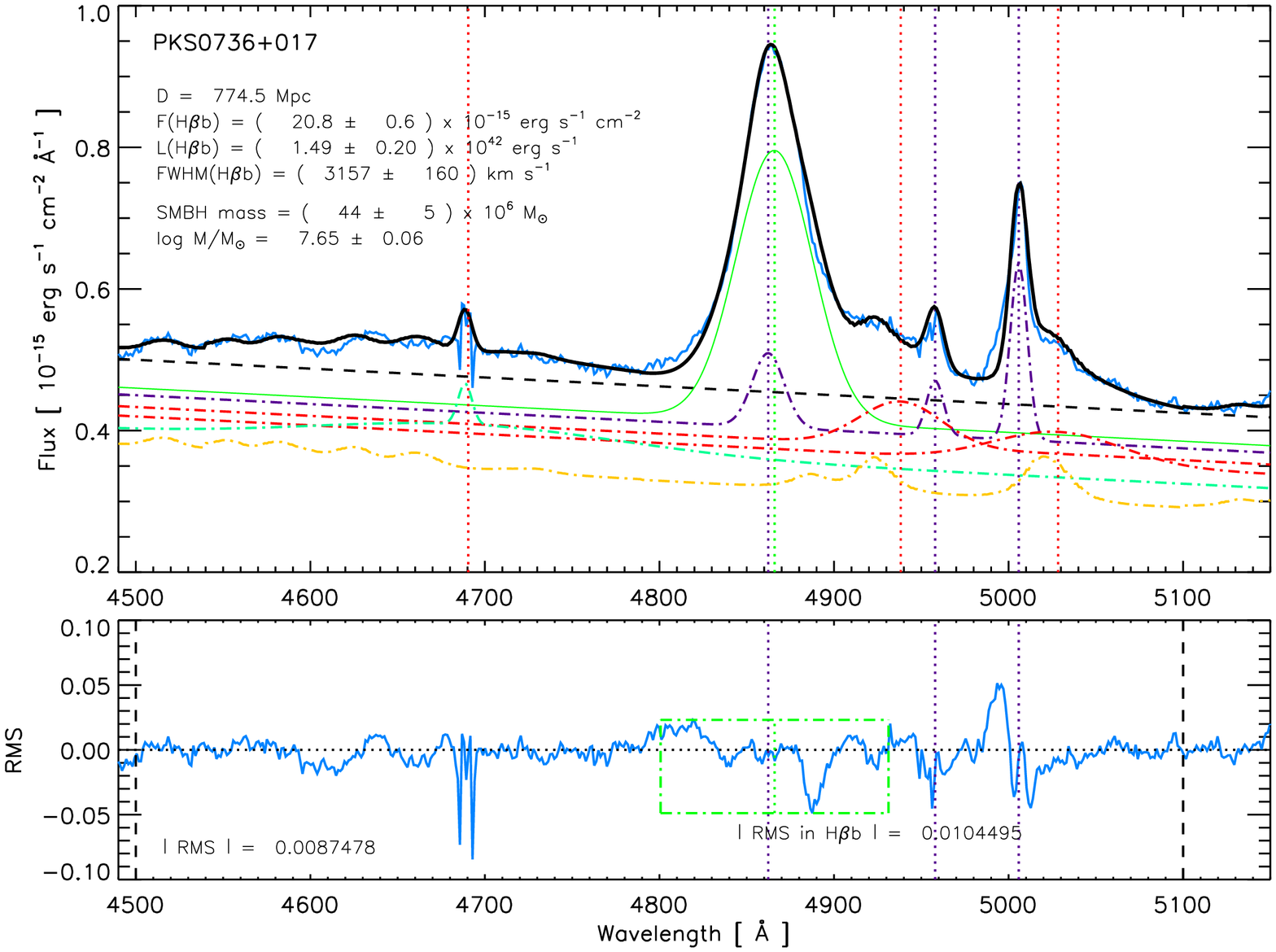} 
\caption{\small Fit to the emission line profiles of PHL~909 and PKS~0736+017 around the \Hb\ region.  The lines are the same described in Fig.~\ref{fits1}.}
\label{fits3}
\end{figure*}

\begin{figure*}
\centering
\includegraphics[scale=0.6,angle=90]{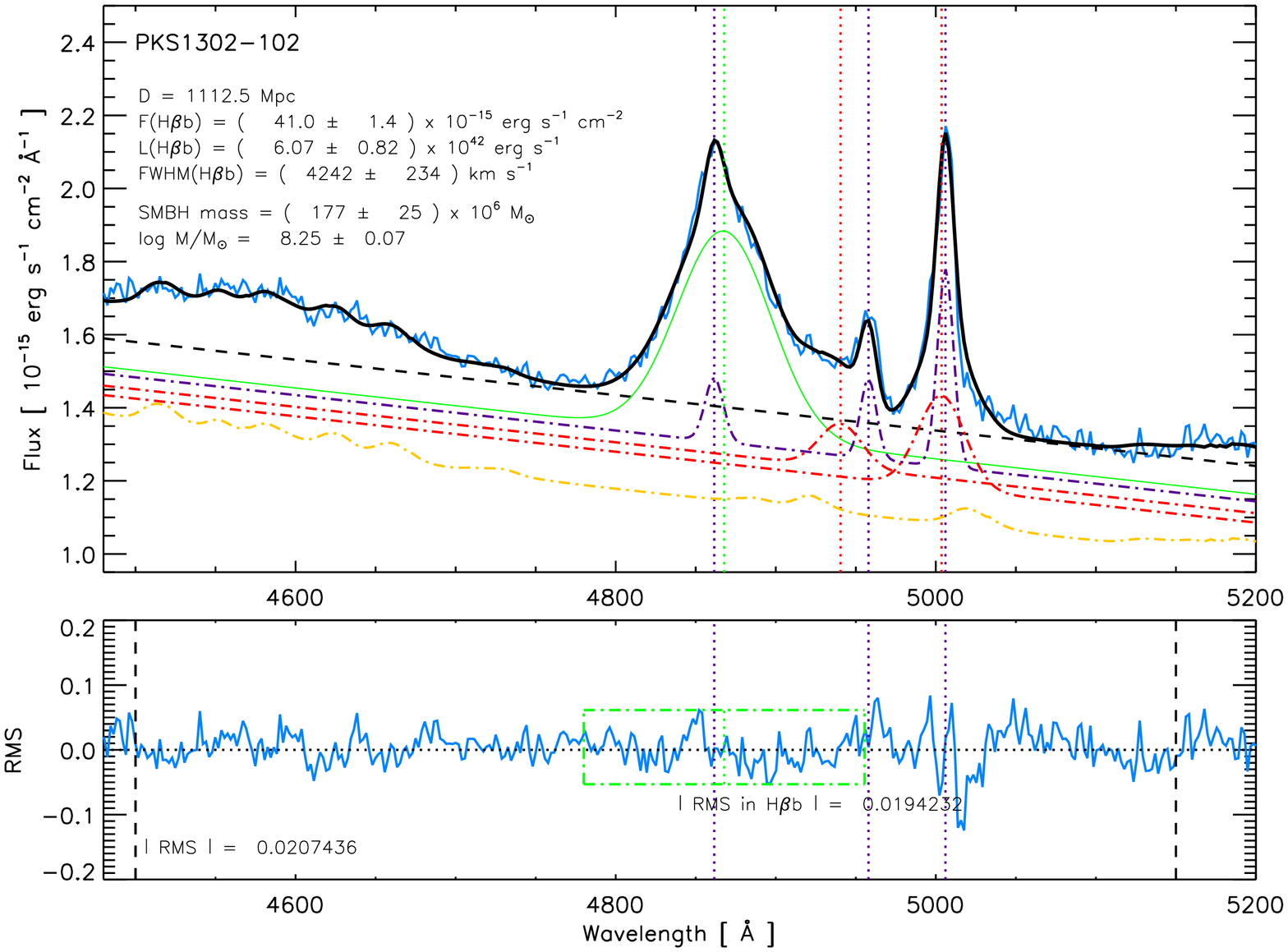}
\includegraphics[scale=0.6,angle=90]{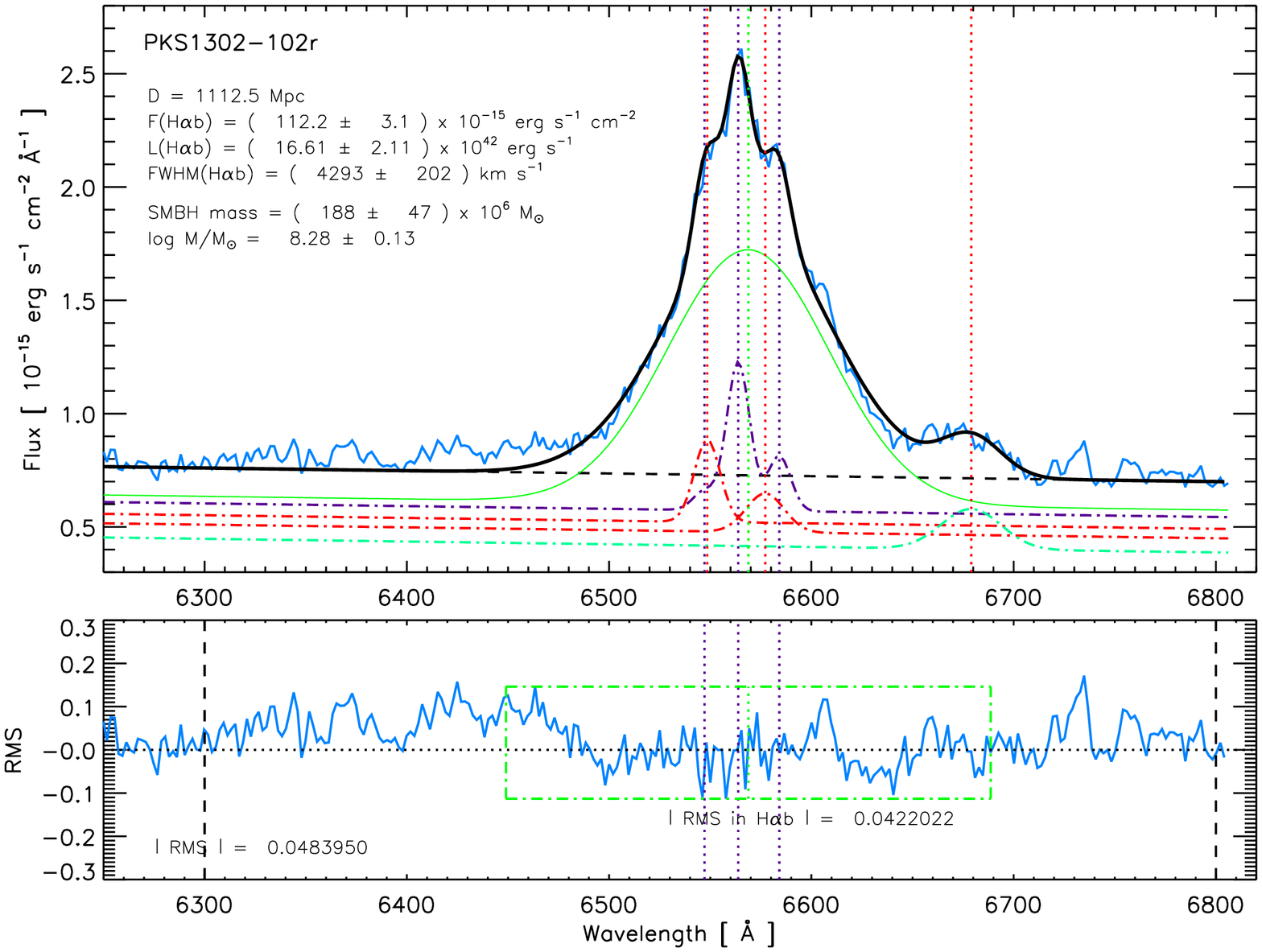} 
\caption{\small Fit to the emission line profiles of PKS~1302-102 around the \Hb\ region (top) and the \Ha\ region (bottom).  The lines are the same described in Fig.~\ref{fits1}, following the same procedure  for \Ha\ as described in Fig.~\ref{fits2}.}
\label{fits4}
\end{figure*}

\begin{figure*}
\centering
\includegraphics[scale=0.6,angle=90]{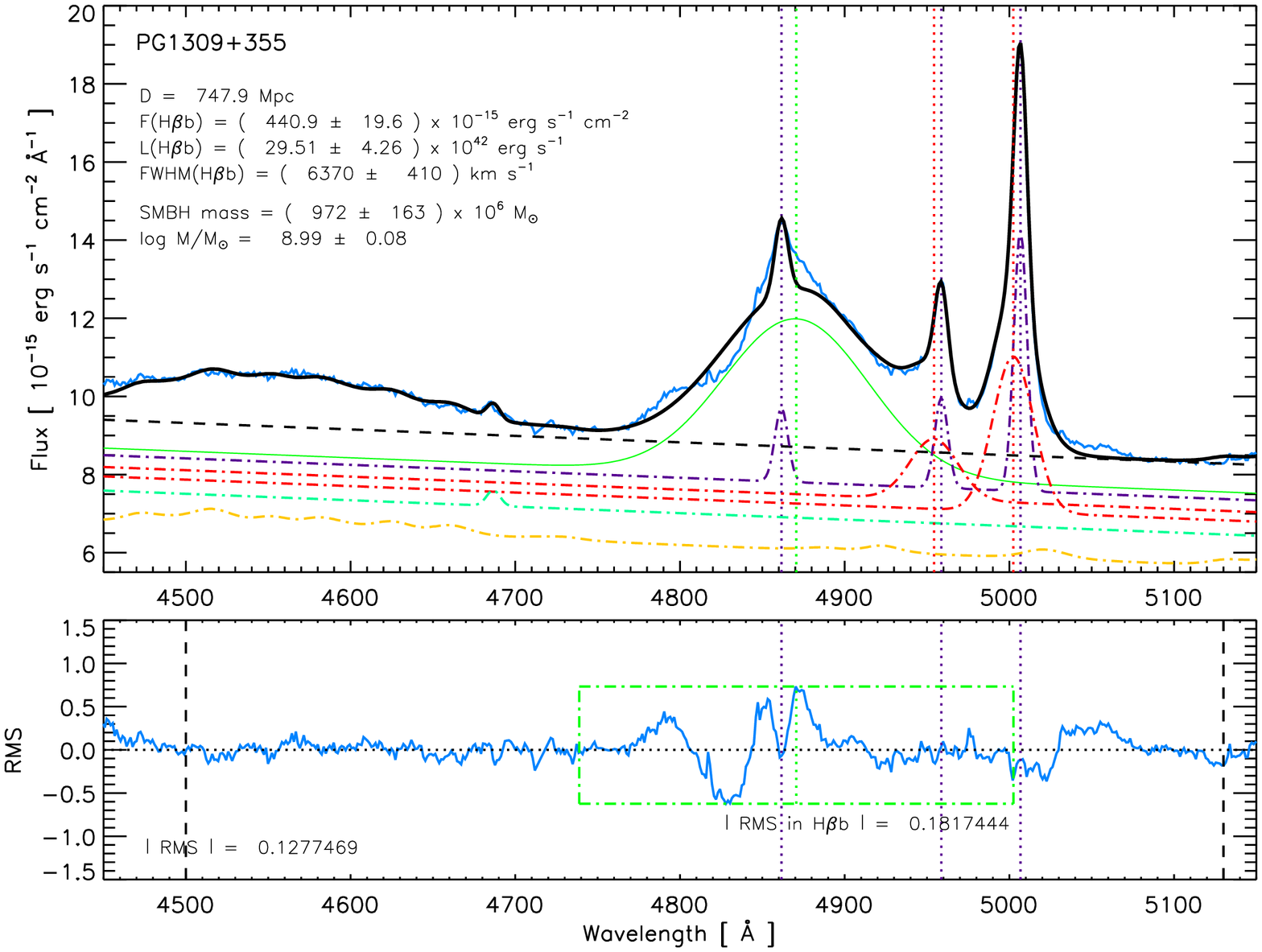}  
\includegraphics[scale=0.6,angle=90]{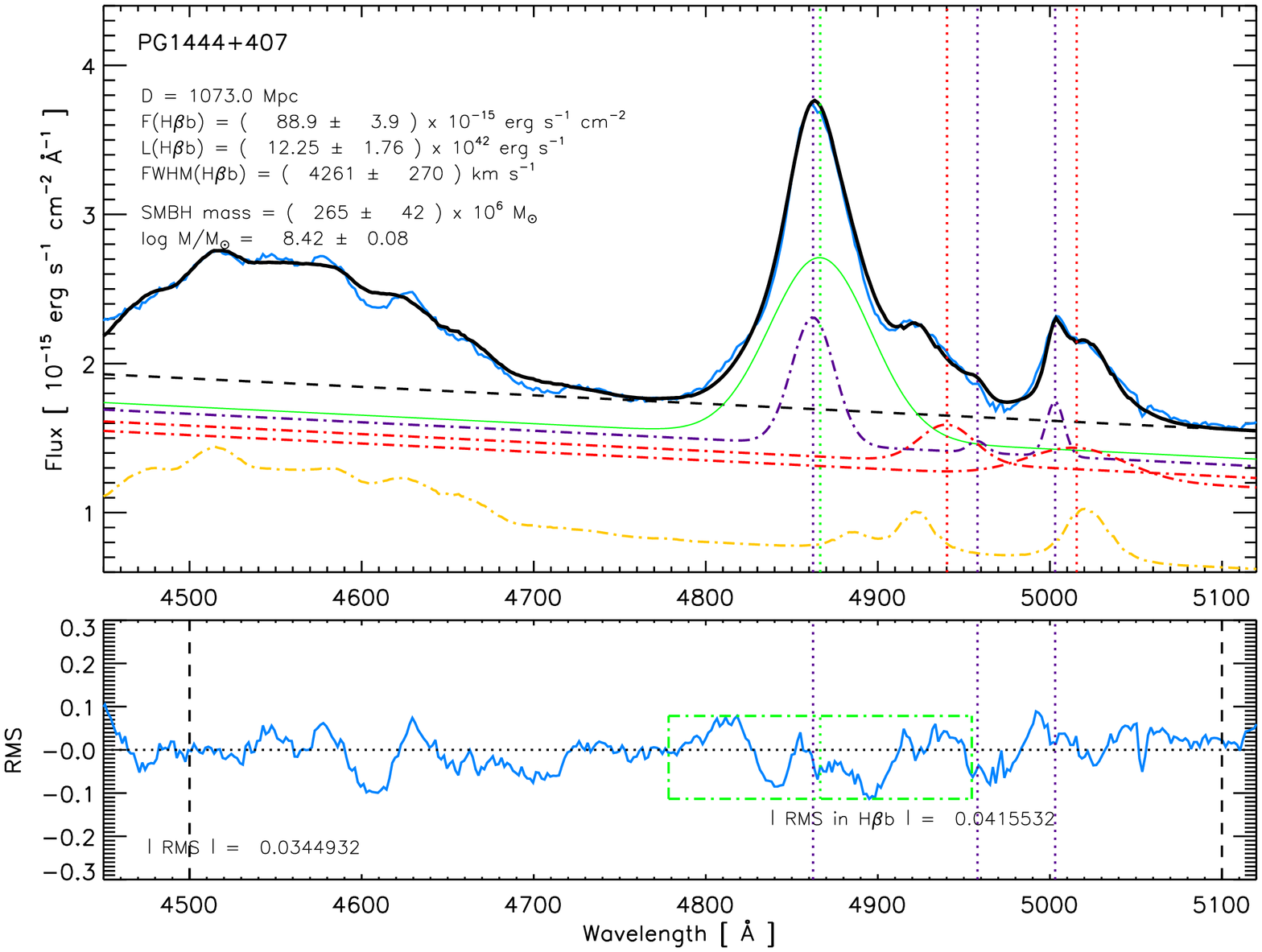}
\caption{\small Fit to the emission line profiles of PG~1309+355 and PG~1444+407 around the \Hb\ region.  The lines are the same described in Fig.~\ref{fits1}. }
\label{fits5}
\end{figure*}

\begin{figure*}
\centering
\includegraphics[scale=0.6,angle=90]{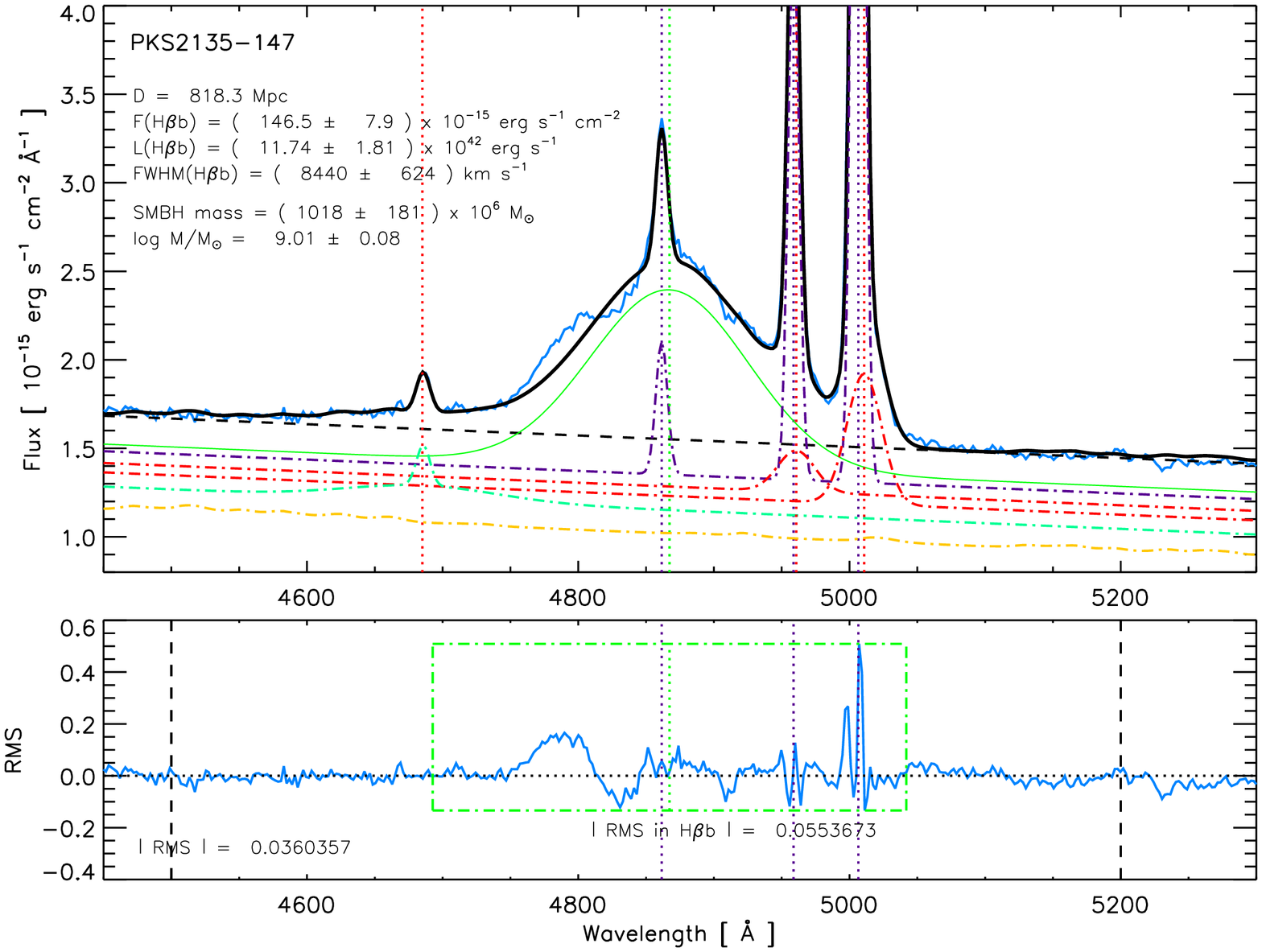} 
\includegraphics[scale=0.6,angle=90]{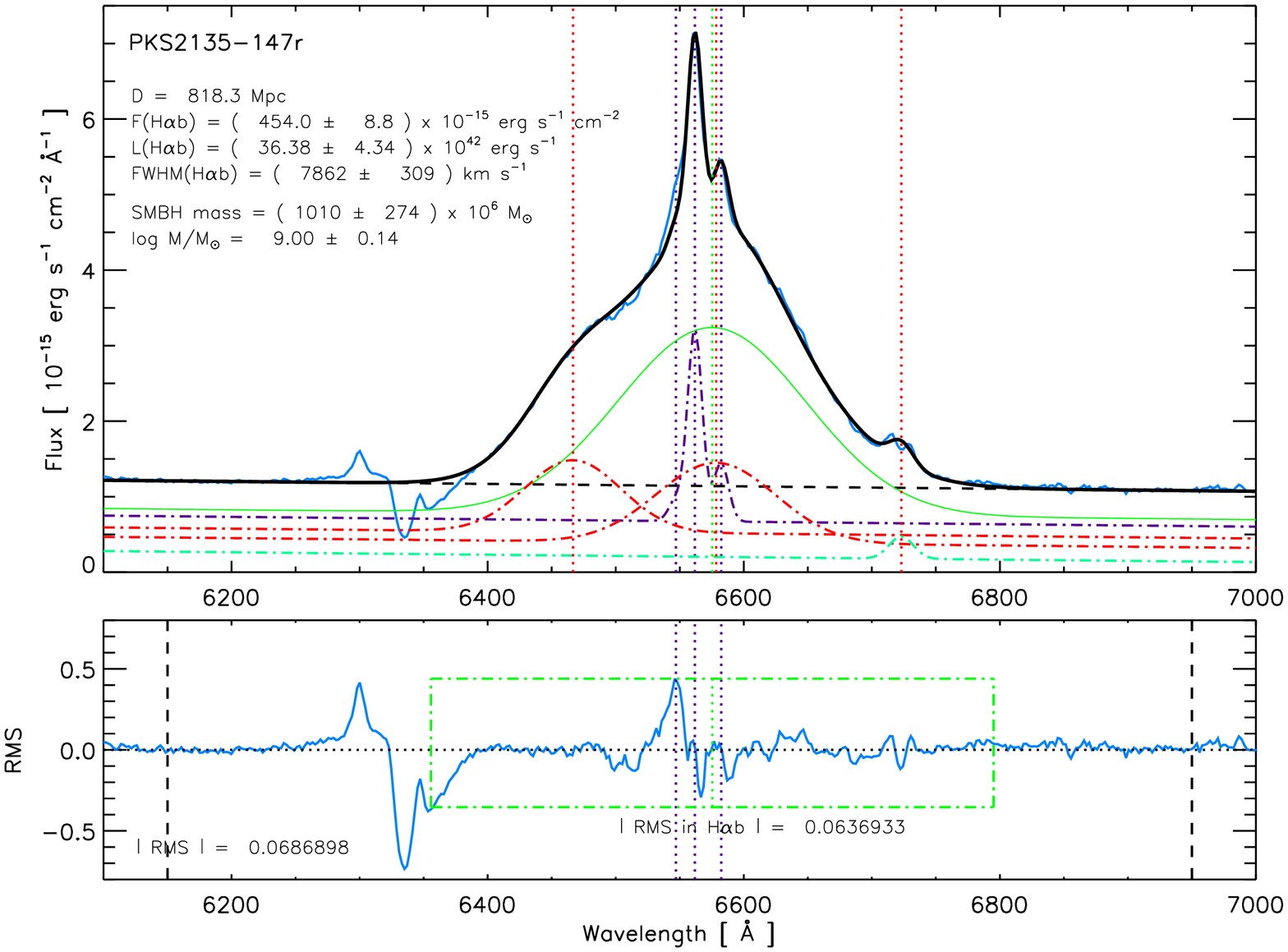}
\caption{\small Fit to the emission line profiles of PKS~2135-147 around the \Hb\ region (top) and the \Ha\ region (bottom).  The lines are the same described in Fig.~\ref{fits1}, following the same procedure  for \Ha\ as described in Fig.~\ref{fits2}. For the \Ha\ fit the dotted-dashed green line is a broad [\ion{S}{ii}]~$\lambda$6717+$\lambda$6731 line.}
\label{fits6}
\end{figure*}

\begin{figure*}
\centering
\includegraphics[scale=0.6,angle=90]{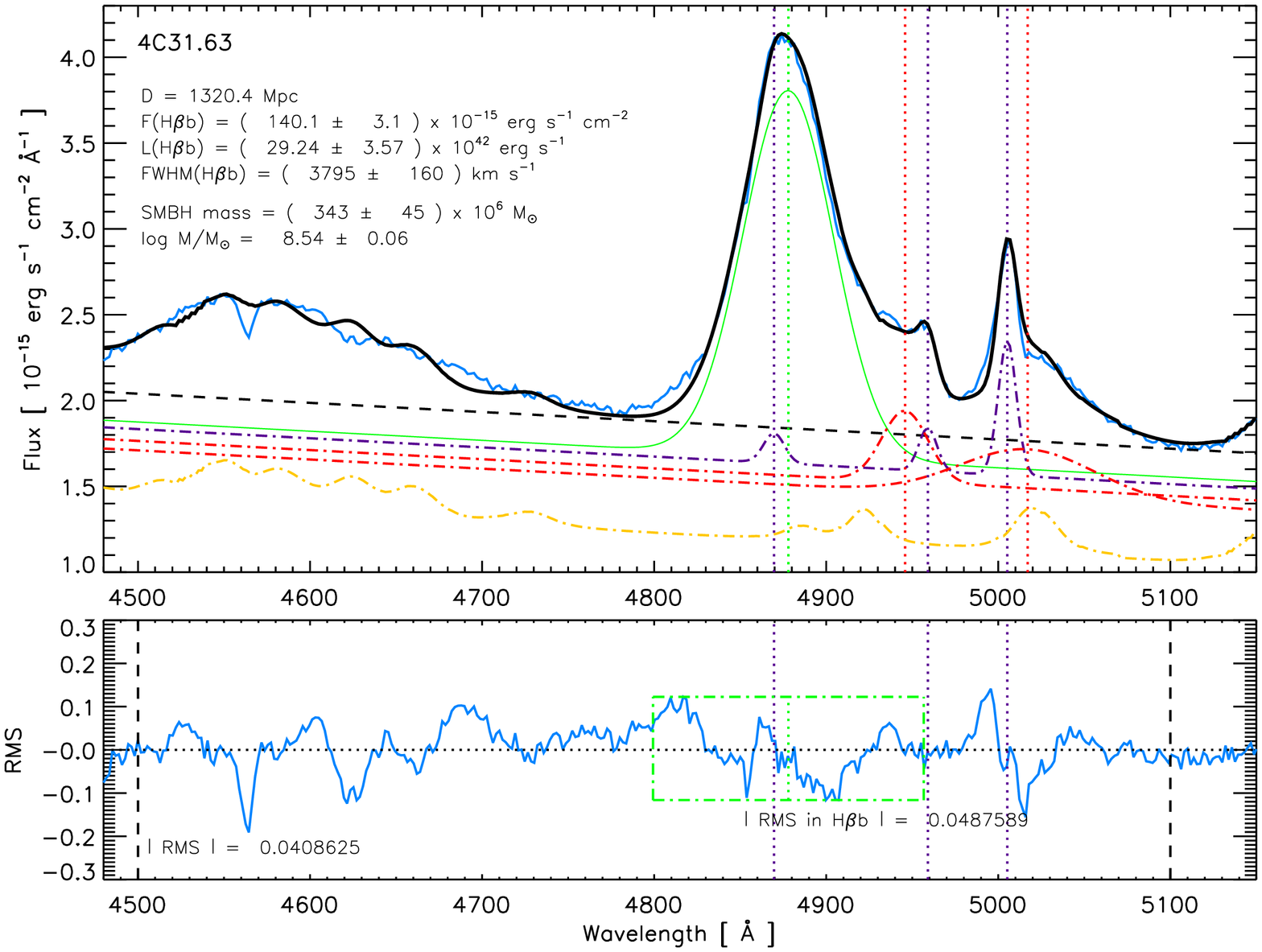}
\includegraphics[scale=0.6,angle=90]{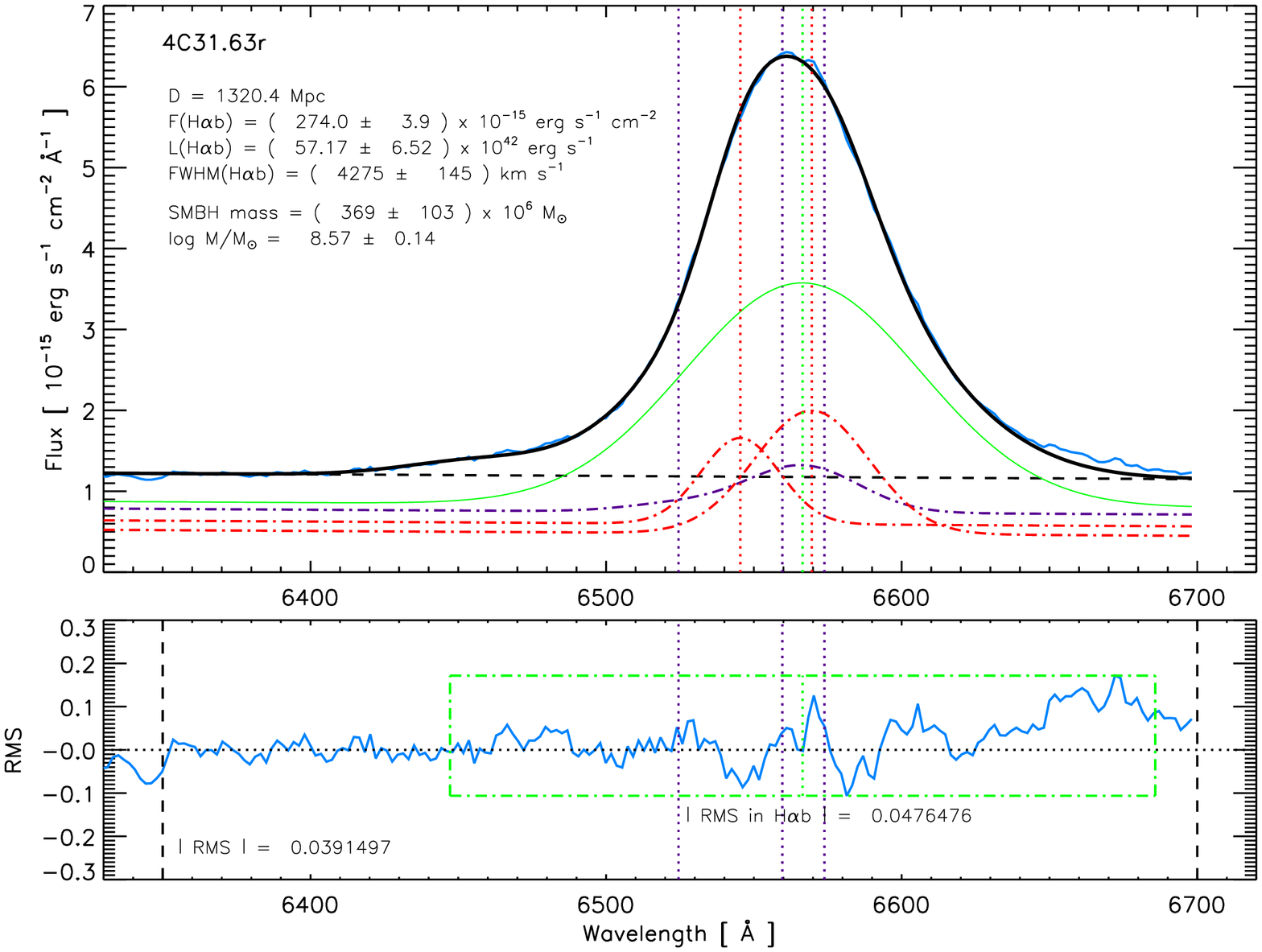}
\caption{\small Fit to the emission line profiles of 4C 31.63 around the \Hb\ region (top) and the \Ha\ region (bottom).  The lines are the same described in Fig.~\ref{fits1}, following the same procedure  for \Ha\ as described in Fig.~\ref{fits2}.}
\label{fits7}
\end{figure*}

\begin{figure*}
\centering
\includegraphics[scale=0.6,angle=90]{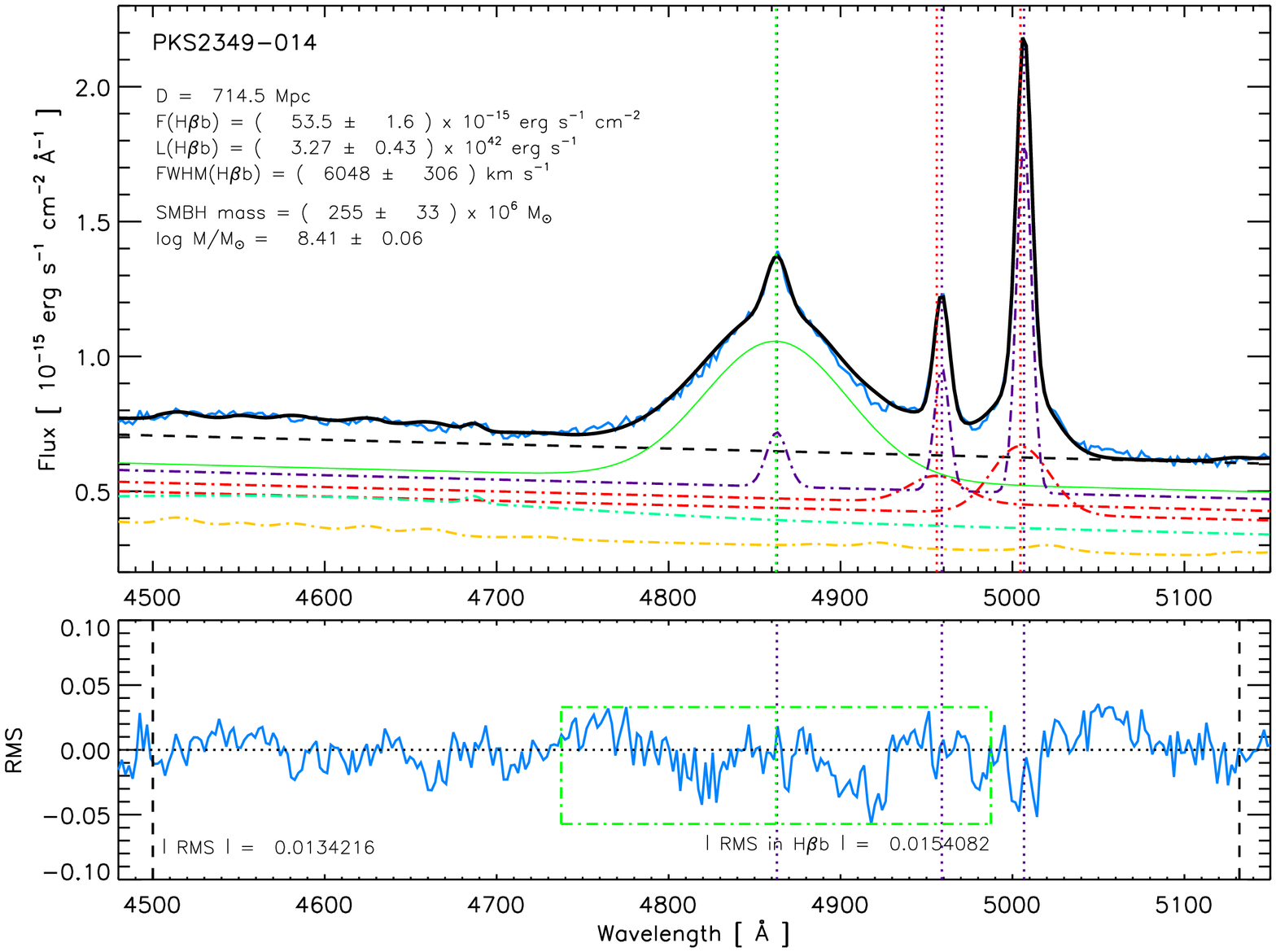}
\includegraphics[scale=0.6,angle=90]{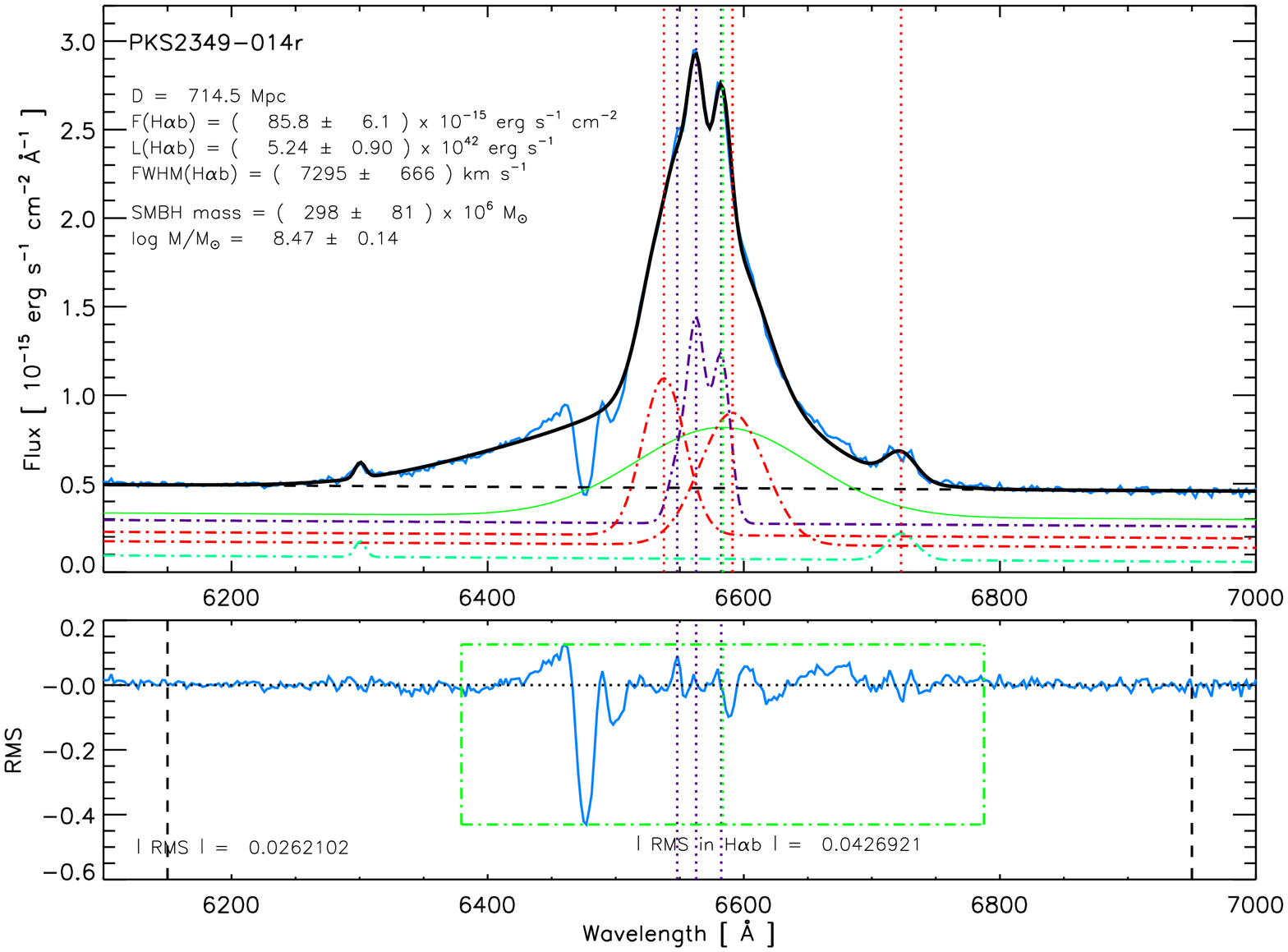} 
\caption{\small Fit to the emission line profiles of PKS~2349-014 around the \Hb\ region (top) and the \Ha\ region (bottom).  The lines are the same described in Fig.~\ref{fits1}, following the same procedure  for \Ha\  as described in Fig.~\ref{fits2}. For the \Ha\ fit the dotted-dashed green line is a narrow  [\ion{O}{i}]~$\lambda$6730 and a broad [\ion{S}{ii}]~$\lambda$6717+$\lambda$6731 line.}
\label{fits8}
\end{figure*}

Figures~\ref{fits3} to \ref{fits8} show the fit to the emission line profiles of the rest of our quasar sample, following the same conventions we specified in Figs.~\ref{fits1}~and~\ref{fits2}.

%% The reference list follows the main body and any appendices.
%% Use LaTeX's thebibliography environment to mark up your reference list.
%% Note \begin{thebibliography} is followed by an empty set of
%% curly braces.  If you forget this, LaTeX will generate the error
%% "Perhaps a missing \item?".
%%
%% thebibliography produces citations in the text using \bibitem-\cite
%% cross-referencing. Each reference is preceded by a
%% \bibitem command that defines in curly braces the KEY that corresponds
%% to the KEY in the \cite commands (see the first section above).
%% Make sure that you provide a unique KEY for every \bibitem or else the
%% paper will not LaTeX. The square brackets should contain
%% the citation text that LaTeX will insert in
%% place of the \cite commands.

%% We have used macros to produce journal name abbreviations.
%% AASTeX provides a number of these for the more frequently-cited journals.
%% See the Author Guide for a list of them.

%% Note that the style of the \bibitem labels (in []) is slightly
%% different from previous examples.  The natbib system solves a host
%% of citation expression problems, but it is necessary to clearly
%% delimit the year from the author name used in the citation.
%% See the natbib documentation for more details and options.

%% Use the figure environment and \plotone or \plottwo to include
%% figures and captions in your electronic submission.
%% To embed the sample graphics in
%% the file, uncomment the \plotone, \plottwo, and
%% \includegraphics commands
%%

%%%%%%%%%%%%% FIGURES IN APPENDIX

\clearpage

\end{document}